\documentclass[11pt,tightenlines,nofootinbib,superscriptaddress,aps]{revtex4-2}

\usepackage[utf8]{inputenc}
\usepackage[british]{babel}
\usepackage{amsthm,amscd,physics}
\usepackage{mathtools}
\usepackage{mathrsfs}
\usepackage{pgf}
\usepackage{mathrsfs}
\usepackage{enumerate}
\usepackage{dsfont}
\usepackage{paralist}
\usepackage{booktabs}
\usepackage{comment}
\usepackage{enumerate}
\usepackage{soul}
\usepackage[colorlinks=true, urlcolor=cyan,citecolor=cyan,anchorcolor=cyan, linkcolor=cyan]{hyperref}
\usepackage[margin=2.6cm,bmargin=3cm,tmargin=3cm]{geometry}
\usepackage[nameinlink,capitalise]{cleveref}
\usepackage{amsmath,amssymb}
\usepackage{xcolor}
\usepackage{tikz}
\usepackage{pgfplots}
\pgfplotsset{compat=1.18}
\usetikzlibrary{arrows}
\definecolor{htBlue}{HTML}{2E5FA3}
\definecolor{htPurple}{HTML}{7B52A8}
\definecolor{htOrange}{HTML}{E07B29}
\definecolor{htInk}{HTML}{2B2B33}
\pgfplotsset{colormap={htY}{rgb255(0cm)=(92,62,140) rgb255(3cm)=(146,114,194) rgb255(6cm)=(206,191,232) rgb255(10cm)=(236,230,247)}}
\pgfplotsset{colormap={htC}{rgb255=(150,178,218) rgb255=(205,219,240) rgb255=(236,242,250)}}
\pgfplotsset{colormap={htCop}{rgb255=(120,156,208) rgb255=(191,209,236) rgb255=(231,238,249)}}

\usetikzlibrary{arrows.meta,calc}
\makeatletter
\newcommand{\apptocfile}{atoc}
\let\oldappendix\appendix
\renewcommand{\appendix}{
  \oldappendix
  \let\oldaddtocontents\addtocontents
  \long\def\addtocontents##1##2{%
    \def\apptoc@ext{##1}%
    \def\apptoc@toc{toc}%
    \ifx\apptoc@ext\apptoc@toc
      \oldaddtocontents{\apptocfile}{##2}%
    \else
      \oldaddtocontents{##1}{##2}%
    \fi
  }%
}

\newcommand{\appendixheading}{%
  \begingroup
    \let\addcontentsline\@gobblethree
      {\huge\bfseries Appendices\par}
      \vspace{1.2em}
  \endgroup
}

\newcommand{\appendixtableofcontents}{%
  \begingroup
    \setcounter{tocdepth}{3}%
    \phantomsection
    \let\addcontentsline\@gobblethree
    \begin{center}
      {\large\bfseries Contents\par}
    \end{center}
    \vspace{1em}
    \@starttoc{\apptocfile}%
  \endgroup
}

\usetikzlibrary{decorations.pathreplacing,calc,positioning}
\definecolor{wblue}{HTML}{2E5FA3}
\definecolor{wpurple}{HTML}{7B52A8}
\definecolor{worange}{HTML}{E07B29}
\makeatother

\newcommand{\com}{\mathbb{C}}

\newcommand{\J}{\mathcal{J}}

\newcommand{\ii}{\mathrm i}

\newcommand{\EE}{\mathbb E}

\newcommand{\F}{{F_{\mathrm{th}}}}
\newcommand{\FF}{{F_{\mathrm{exp}}}}
\renewcommand\vec{\boldsymbol}
\def\r{{\vec{r}}}
\def\m{{\mathsf{m}}}

\def\d{{\vec{d}}} 
\def\b{B}

\renewcommand{\phi}{\varphi}
\renewcommand{\epsilon}{\varepsilon}

\renewcommand\vec{\boldsymbol}

\theoremstyle{plain}
\newtheorem{theorem}{Theorem}[section]
\newtheorem{corollary}[theorem]{Corollary}

\newtheorem{proposition}[theorem]{Proposition}

\theoremstyle{remark}

\newtheorem*{claim*}{Claim}
\newtheorem*{remark*}{Remark}
\newtheorem*{example*}{Example}
\newtheorem*{notation*}{Notation}
\numberwithin{equation}{section}

\begin{document}

\title{Learning Quantum Hamiltonians from the Dynamical Structure Factor}

\author{Nisarga Paul}
\email{npaul@caltech.edu}
\affiliation{Department of Physics and Institute for Quantum Information and
Matter, California Institute of Technology, Pasadena, CA 91125, USA}

\author{Thomas Schuster}
\affiliation{Department of Physics and Institute for Quantum Information and
Matter, California Institute of Technology, Pasadena, CA 91125, USA}
\affiliation{Google Quantum AI, Venice, CA 90291, USA}

\author{Jason Alicea}
\affiliation{Department of Physics and Institute for Quantum Information and
Matter, California Institute of Technology, Pasadena, CA 91125, USA}

\begin{abstract}
\noindent Identifying the microscopic interactions governing a quantum system is essential for developing predictive theories of materials and validating quantum technologies. Despite considerable progress, learning Hamiltonians from realistic experimental data remains a challenge. Motivated by neutron scattering and quantum simulations, we study the problem of learning a quantum Hamiltonian from the dynamical structure factor (DSF), which captures two-point correlation data in quantum magnets. We show that the DSF encodes sufficient information to efficiently reconstruct broad classes of quantum Hamiltonians. Our approach exploits simple relations between low-order spectral moments of the DSF and microscopic couplings, giving rise to three complementary Hamiltonian tomography protocols based on the high-temperature expansion, strong-field polarization, and the fluctuation-dissipation theorem, respectively. The framework we introduce establishes practical and robust Hamiltonian learning from common experimental data, with potentially immediate application to neutron scattering experiments. 
\end{abstract}

\maketitle

\section{Introduction}

\noindent Quantum systems are typically well-described by a local Hamiltonian, specified by a small number of couplings, from which rich collective phenomena such as symmetry breaking, topological order, and quantum criticality can emerge. Constraining and characterizing the Hamiltonian from experimental data presents a recurring challenge across physics, from particle phenomenology to quantum materials. This task of \emph{Hamiltonian learning} enables one to extract effective models governing the quantum system over energy and length scales of interest,  informing theoretical analyses and subsequent experimental refinements. In condensed matter, Hamiltonian learning facilitates the design of quantum materials with targeted properties~\cite{savary2016quantum,andrei2021marvels,geim2013van,RevModPhys.82.3045,RevModPhys.63.239}; in quantum simulation and device engineering, it is essential for benchmarking,  calibration, and building scalable platforms with verifiable performance~\cite{RevModPhys.86.153,PhysRevA.54.R4649,lee2002quantum,boulant2003robust,PhysRevA.101.062305,shulman2014suppressing,wiebe2014quantum,wang2017experimental}.
\par 
Hamiltonian learning has prompted many empirical and theoretical approaches, drawing on tools from quantum information and machine learning~\cite{WiebeGranadeFerrieCoryPRL,gu2024practical,burgarth2017evolution,BaireyAradLindnerPRL,AnshuArunachalamNatPhys,HaahKothariTangNatPhys,BakshiLiuMoitraTang,HuangTongFangSuPRL,zhang2015identification,hangleiter2024robustly,yu2023robust,stilck2024efficient,zubida2021optimal,bakshi2024learning}.  
A common empirical approach, applied successfully to certain quantum magnets~\cite{ColdeaHaydenAeppli2001PRL,ColdeaTennantHabicht2002PRL,ross2011quantum}, falls into the framework of model fitting, involving iteratively updating a model prediction and classical simulation until convergence to experiment.  For generic quantum systems, however, model fitting is computationally prohibitive to scale to large system size, leaving open the existence of a general, scalable approach under realistic experimental measurement constraints. More systematic theoretical approaches often presuppose global digital access to arbitrary measurements.  Experimental platforms nevertheless typically operate with far more limited control: Measurements in quantum materials are usually limited to observables characterizing responses to external probe fields, while in quantum simulators, access can be restricted to a few native local observables or correlation functions due to architectural and resource constraints.

\par 
Can one learn the Hamiltonian from \emph{limited, realistic} experimental outputs? This fundamentally and practically important question has drawn significant recent attention~\cite{ZhangSarovarPRL,SoneCappellaroPRA,PhysRevResearch.5.043284,ChenCotlerHuangQPT,gong2026out}. In this work, we address the problem for a broadly accessible and physically illuminating observable in quantum experiments: the \textit{dynamical structure factor} (DSF). This quantity [Eq.~\eqref{eq:dsf}] is directly measurable via inelastic neutron scattering in quantum magnetism and has been extracted in programmable quantum simulators using its simple relation to two-point correlation functions~\cite{chiesa2019quantum,lee2026benchmarking,granet2026dynamical,kokkas2026quantum,sun2026observation}. The DSF offers experimental fingerprints of complex quantum phases which can potentially enable the reconstruction of underlying Hamiltonian parameters, as perhaps best exemplified by the quest for Kitaev spin liquid materials~\cite{do2017majorana,banerjee2018excitations,hermanns2018physics,BanerjeeMajoranaTomography}. 
\par 
\begin{figure}
    \centering
    \includegraphics[width=\linewidth]{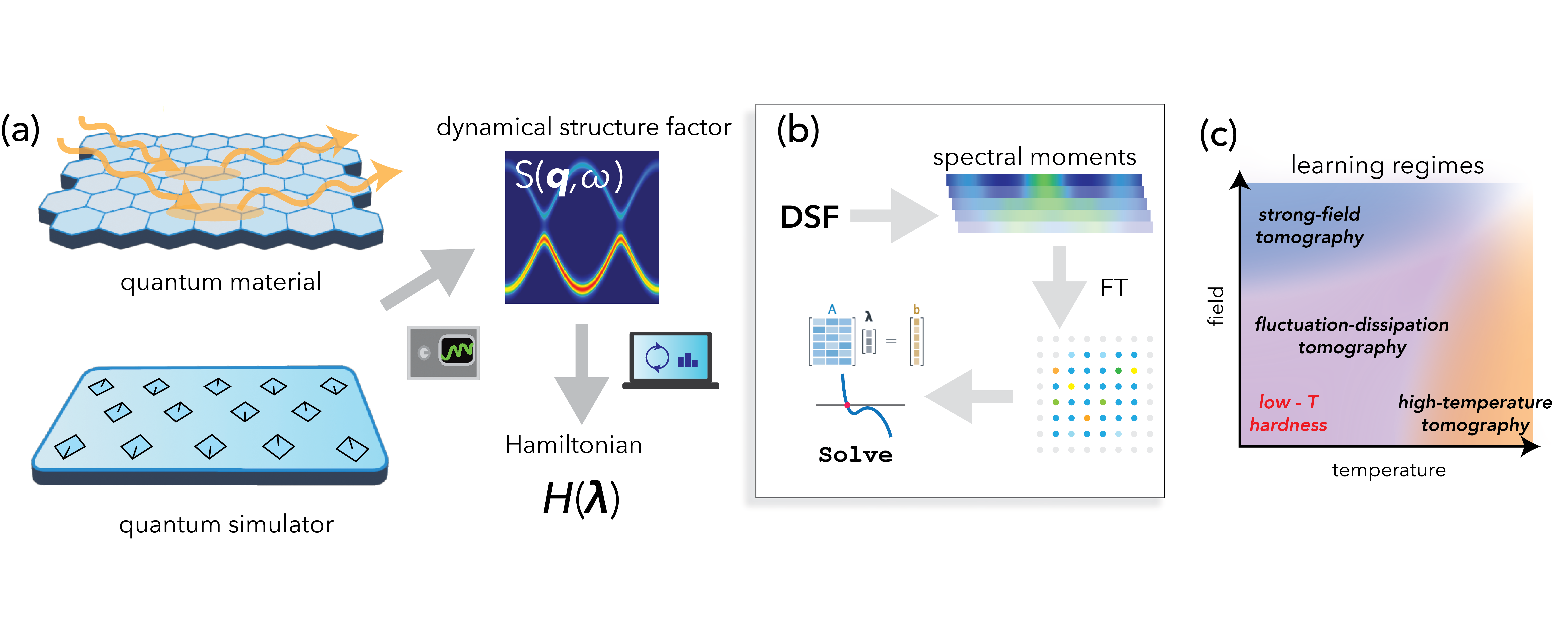}
    \caption{\textbf{Hamiltonian learning from DSF.} (a) Illustration of Hamiltonian learning from the dynamical structure factor (DSF). The DSF is experimentally accessible in quantum materials through neutron scattering or quantum simulators through two-point functions. Assuming access to the DSF at varying temperature and/or external global magnetic field, we study the learnability of a Hamiltonian $H(\vec \lambda)$ with unknown couplings $\vec\lambda$. (b) High-level idea of Hamiltonian learning protocols. We prove integrated frequency moments and Fourier harmonics of the DSF encode $\vec\lambda$ through efficiently solvable means. (c) Schematic of favorable regimes for protocols, in the temperature and magnetic field plane.}
    \label{fig:1}
\end{figure}
It is \textit{a priori} far from clear that the DSF alone provides sufficient information to recover the entire Hamiltonian. Indeed, while the DSF contains information on magnon dispersions and other spinful excitations, it encodes only a slice of two-point correlation data---a small portion of all observables. Although several machine-learning studies provide empirical evidence of this task's tractability, at least for specific models with a few free parameters~\cite{ChitturiINRDSF,samarakoon2020machine}, the well-posedness of this inverse problem has not been theoretically established. 
Indeed, building on known hardness bounds for learning from Gibbs states~\cite{HaahKothariTangNatPhys}, we rigorously demonstrate that it can be ill-posed: We identify continuous families of few-body Hamiltonians that exhibit identical DSFs. Moreover, even when the inverse problem is in principle well-posed, the DSF potentially depends on Hamiltonian couplings in a highly complex fashion that can in practice stymie finding the inversion needed to learn the Hamiltonian.

\par 
In this work, we develop a versatile framework
for Hamiltonian tomography from the DSF [Fig.~\ref{fig:1}a]. Our framework does not require simulation of the Hamiltonian
and comes with rigorous guarantees for general classes of Hamiltonians across several broad experimentally relevant regimes, thus providing strong evidence that the aforementioned challenges do not impede practical learnability. Our central insight is that the microscopic Hamiltonian parameters can be extracted from explicit relations, or sum rules, among integrated frequency moments of the DSF [Fig.~\ref{fig:1}b]. In particular, we provide three scalable Hamiltonian learning algorithms that we dub \textit{high-temperature}, \textit{strong-field}, and \textit{fluctuation-dissipation} protocols---each taming the complex dependence of the DSF on Hamiltonian parameters in complementary regimes [Fig.~\ref{fig:1}c; see Table~\ref{tab:protocols} for an executive summary]. These protocols respectively rely on a high-temperature expansion, perturbation theory around a strong polarizing Zeeman field, and nontrivial exact identities satisfied by the DSF due to stationarity of thermal states and the fluctuation-dissipation theorem. For the concrete case of translation-invariant
spin-$\frac12$ Hamiltonians in arbitrary dimensions with one- and two-body interactions,
we establish end-to-end learning guarantees and in several instances reduce Hamiltonian learning to solving a simple linear system. Extensive numerical tests across
one- and two-dimensional Hamiltonian families demonstrate appreciable robustness of our protocols to measurement noise and experimental imperfections. Together, these results establish a direct, practical, and scalable route to learning Hamiltonians from limited, physically relevant output of quantum experiments.

\begin{table*}[t]
\centering
\caption{\textbf{Summary of Hamiltonian learning protocols.}
Overview of protocols and their experimental
requirements in quantum materials and programmable quantum simulators.}
\label{tab:protocols}

\begingroup
\fontsize{8}{9.0}\selectfont
\setlength{\tabcolsep}{4pt}
\renewcommand{\arraystretch}{1.2}

\begin{tabular}{@{}
p{0.13\textwidth}
p{0.26\textwidth}
p{0.27\textwidth}
p{0.28\textwidth}@{}}

\toprule
\textbf{Protocol}
&
\textbf{High-temperature (HT)}
&
\textbf{Strong-field (SF)}
&
\textbf{Fluctuation--dissipation (FD)}
\tabularnewline
\midrule

\textbf{Executive summary}
&
Leverages a high-$T$ expansion of DSF spectral moments
to construct a low-degree polynomial system for the
Hamiltonian couplings [Sec.~\ref{sec:hightempmain}].
&
Leverages strong-field polarization and a $1/B$
expansion of DSF spectral moments to construct a
low-degree polynomial system for the couplings
[Sec.~\ref{sec:sft}].
&
Uses exact DSF identities derived from FD theorem to construct a linear system for the couplings, without requiring
high temperatures or strong fields
[Sec.~\ref{sec:fdt}].
\tabularnewline[5pt]
\midrule

\textbf{Experimental requirements}
&
\textit{Quantum materials:}
DSF measurements (e.g., neutron scattering)
at multiple high temperatures
($T\gg$ couplings).

\par\smallskip
\textit{Quantum simulators:}
Tunable high-$T$ Gibbs state preparation and
short-time two-point correlation data.
&
\textit{Quantum materials:}
DSF measurements at strong fields
($B\gg T$, couplings), with tunable
field strength \& orientation.

\par\smallskip
\textit{Quantum simulators:}
Polarized state preparation with a tunable global
field, short-time two-point correlation data.
&
\textit{Quantum materials:}
DSF measurements without requiring high
temperatures or strong applied fields, one- and two-body Hamiltonian. 

\par\smallskip
\textit{Quantum simulators:}
Short-time two-point correlation data in
a stationary state (e.g., Gibbs state), one- and two-body Hamiltonian.
\tabularnewline
\bottomrule
\end{tabular}
\endgroup
\end{table*}

\section{Hamiltonians and spectral moments}

\noindent Consider a translation-invariant quantum spin system (see \textsection\ref{sec:end} for an extension to the disordered case lacking translation symmetry). On a lattice $\Lambda$ of $N$ sites, the Hamiltonian reads
\begin{equation}
    H(\vec\lambda) = \sum_{j=1}^p \lambda^j H^j.
\end{equation}
Here each $H^j = \sum_{\vec r\in \Lambda} h^j_{\vec r}$ is an extensive interaction term---e.g., involving exchange or Zeeman fields---arising from a sum of translated local operators $h_{\vec r}^j$ with $O(1)$ magnitudes $\|h_{\vec r}^j\|_\infty$. We take $\vec\lambda \in [-1,1]^p$ to be the unknown parameters to be learned. The number of independent couplings $p$ and the interaction range are fixed as $N$ increases. For a local spin operator 
$O^a_{\vec r}$, define
$O^a_{\vec q}=N^{-1/2}\sum_{\vec r\in \Lambda}e^{-\ii\vec q\cdot\mathbf r}O^a_{\vec r}$. Below, $a,b,c$ indices run over $x,y,z$  with repeated indices summed. The DSF at inverse temperature $\beta$ is then
\begin{equation}\label{eq:dsf}
    S^{ab}(\vec q,\omega) = \int \mathrm{d}t \, e^{\ii\omega t} \langle O^a_{\vec q}(t) O^b_{-\vec q}\rangle_\beta,
\end{equation}
where $\langle A\rangle_\beta \coloneqq \Tr(e^{-\beta H}A)/\Tr(e^{-\beta H})$ indicates a thermal quantum average. At later stages, we distinguish access to the full tensor $S^{ab}$ from \textit{diagonal access} to $S^{aa}$. In either case, the task is to recover $\vec \lambda$ from the DSF at known temperatures in a manner robust to noise and scaling efficiently in $N$ and $\beta$.

\par

Next, define DSF spectral moments 
\begin{equation}
\label{eq:DSFspecmom}
    m_n^{ab}(\vec q)= \int_{-\infty}^{\infty}\dd\omega \, \omega^n \, S^{ab}(\vec q,\omega),
\end{equation}
which are equivalent to short-time derivatives of the two-point spin correlation function: $m_n^{ab}(\vec q)=2\pi\, \ii^n\partial_t^n \langle O_{\vec q}^a(t)O_{-\vec q}^b\rangle_{\beta}|_{t=0}$, with $m_0^{ab}(\vec q)/2\pi$ the static spin structure factor. Additionally Fourier transforming back to real space yields the particularly useful quantities 
\begin{equation}
\label{eq:DSFspecmomharm}
    m_{n,\vec d}^{ab} = \int_{\mathrm{BZ}} \frac{\dd \vec q}{\mathrm{vol(BZ)}}e^{-\ii\vec q\cdot \vec d}m_n^{ab}(\vec q) = 2\pi\, \ii^n\partial_t^n\langle O_{\vec 0}^a(t)O_{\vec d}^b\rangle_\beta|_{t=0}.
\end{equation}
Equivalently, $m_{n,\vec d}^{ab} = 2\pi \left\langle [[\cdots [O_{\vec 0}^a,H],H],\cdots,H]O_{\vec d}^b\right\rangle_\beta$, with the right-hand comprising $n$ commutators.
For a finite-range Hamiltonian, $m_{n,\vec d}^{ab}$ at fixed $n$ depends only on interactions within a finite neighborhood of $\vec 0$. Each time derivative introduces one factor of $H$, and the remaining dependence on $\vec \lambda$ lies in the thermal expectation value, which in general does not take a simple form. Our three protocols will render the dependence of the $m_{n,\vec d}^{ab}$ on $\vec\lambda$ learnable in three distinct but complementary ways, as illustrated in Fig.~\ref{fig:1}c.

\section{High-temperature tomography}
\label{sec:hightempmain}

\noindent Our \textit{high-temperature (HT) protocol} leverages the fact that Eq.~\eqref{eq:DSFspecmomharm} depends more mildly on Hamiltonian parameters as temperature increases and the Gibbs state approaches a featureless infinite-temperature mixed state. This trend forms the basis of the high-temperature series expansion in quantum lattice models~\cite{oitmaa2006series} and existing digital quantum algorithms for learning Hamiltonians from high-temperature Gibbs states~\cite{HaahKothariTangNatPhys}, and here allows us to tame the complexity of Hamiltonian learning. In particular, a high-temperature expansion allows us to derive low-degree polynomial relations between Hamiltonian couplings and DSF spectral moments.
At small $\beta$, we expand Eq.~\eqref{eq:DSFspecmomharm} as
\begin{equation}
\label{eq:mabndexpansion}
    m_{n,\vec d}^{ab}(\beta,\vec\lambda) = \sum_{i=0}^{\infty} \beta^i \m_{n,\vec d,i}^{ab}(\vec \lambda),
\end{equation}
with each nonzero coefficient $\m_{n,\mathbf d,i}^{ab}(\vec \lambda)$ a homogeneous polynomial of degree $n+i$ in the couplings. For our choice of bounded, finite-range interactions, truncating Eq.~\eqref{eq:mabndexpansion} at order $R$ incurs a controlled $O\left((\beta/\beta_{\mathsf c})^{R+1}\right)$ error as $\beta/\beta_{\mathsf c}\to0$, where $\beta_{\mathsf c}>0$ is an $O(1)$ threshold determined by Hamiltonian details. The $\vec \lambda$-dependent coefficients on the right side of Eq.~\eqref{eq:mabndexpansion} can be evaluated efficiently using local operator algebra without full many-body simulation [App.~\ref{sec:hightemp}]. 
\par 
We collect a finite set of $n_f$ theoretically derived coefficients $\m_{n,\vec d,i}^{ab}(\vec \lambda)$ and denote the resulting collection by $\F(\vec\lambda)$, which comprises different spin channels ($ab$), moment orders ($n$), spatial harmonics $(\vec d)$, and high-$T$ expansion orders ($i$). We refer to $\F(\vec\lambda)$ as the \textit{feature map}. The experimental data corresponding to the feature map, which we denote by 
$\FF$, is extracted from the measured DSF, assumed to be available across multiple high temperatures and the full Brillouin zone.
The HT protocol entails solving the polynomial system $\FF = \F(\vec\lambda)$. Crucially, for fixed feature map, interaction range and parameter count, the size and degree of this polynomial system does not scale with $N$. Hence the computational cost of the HT protocol is independent of system size, in contrast to model fitting approaches requiring repeated simulations of the DSF~\cite{samarakoon2022extraction}.\par 

The remaining question is whether the polynomial system $\FF = \F(\vec\lambda)$ uniquely determines the Hamiltonian couplings. We identify three possible scenarios analyzed in detail in App.~\ref{sec:hightemp}. First, the system may be underdetermined---e.g., if the feature map includes fewer than $p$ components---and admit a continuously infinite set of solutions. Expanding the feature set can remedy this non-solvability. Second, the system may admit a discrete set of solutions---a case we call \textit{locally solvable}. Local solvability can arise from spurious solutions or unavoidable degeneracy that we term \textit{measurement equivalence} arising from distinct Hamiltonians which are indistinguishable under $\F(\vec\lambda)$. A powerful property of polynomial feature maps is that local solvability can be certified for almost all couplings by checking that the Jacobian of $\F(\vec\lambda)$ has full rank at a \textit{random} $\vec\lambda$~\cite{ChenCotlerHuangQPT}. Third, the system may have a unique solution, in which case we say it is \textit{globally solvable}. Instances of both global solvability and local solvability with measurement equivalence are discussed below.
\begin{figure}
    \centering
    \includegraphics[width=\linewidth]{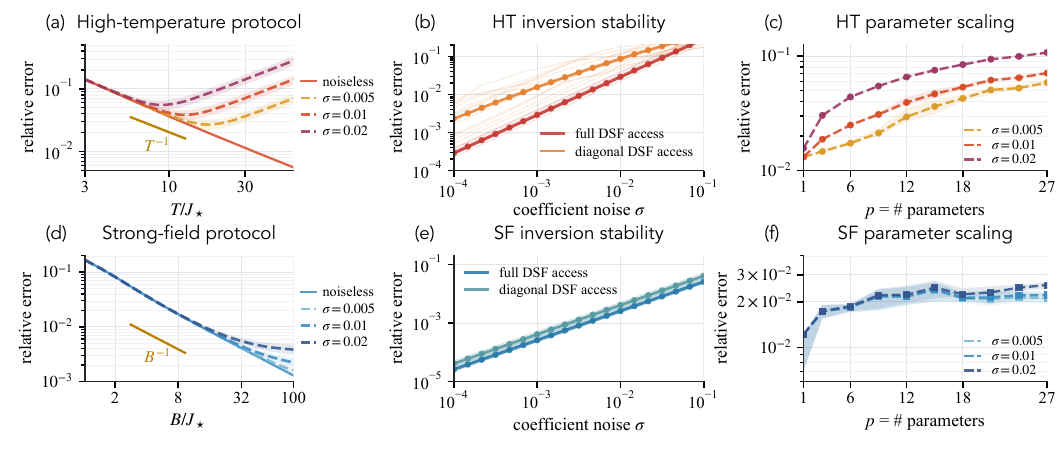}
    \caption{\textbf{High-temperature and strong-field tomography.} (a,d) Relative coupling error for the $J$–$K$–$\Gamma$ model, $(J,K,\Gamma)=(0.4,-0.7,0.2)J_\star$, on a periodic 12-spin honeycomb cluster. Solid curves are noiseless; dashed curves include Gaussian moment noise $\sigma$. (b,e) Noise dependence for 1,000 twelve-parameter chains with full or diagonal DSF access, where $\sigma$ denotes noise added to exact HT or SF expansion coefficients. (c,f) Performance versus the number of unknown couplings at $\beta J_\star=0.2$ for HT and $B/J_\star=32$ for SF. Relative errors are $|\boldsymbol\lambda_{\mathrm{pred}}-\boldsymbol\lambda_{\mathrm{true}}|/|\boldsymbol\lambda_{\mathrm{true}}|$. Curves show medians with central $25\%$ ribbons, except in (b,e), which show ensemble means (thick) and example trials (thin).}
    \label{fig:HTSF}
\end{figure}

To illustrate the HT protocol concretely, let us specialize to spin-$\frac12$ Hamiltonians with arbitrary one- and two-body interactions in any dimension. Consider, for a finite set of bond vectors $\mathcal{R}$,
\begin{equation}
\label{eq:12body}
    H(\vec \lambda) = \sum_{\vec r}b^a\sigma_{\vec r}^a + \sum_{\vec r,\vec d\in \mathcal{R}} \lambda_{\vec d}^{ab}\sigma^a_{\vec r}\sigma^{b}_{\vec r+\vec d},
\end{equation}
where $\sigma^a$ are Pauli matrices. Direct calculation of the feature map $F_{\rm th}$ reveals $b^a=\frac{\ii}{4\pi}\epsilon_{abc}\m_{0,\vec0,1}^{bc}$ and $\lambda_{\vec d}^{ab}=-\frac{1}{2\pi}\m_{0,\vec d,1}^{ab}$. Therefore, spatial harmonics of the static structure factor ($n=0$) alone, expanded to first order in $\beta$, contain the information of \emph{all} one- and two-qubit couplings. 
\par

Frequently in neutron scattering experiments, access is limited to unpolarized neutron beams, whereby only the diagonal DSF is available~\cite{squires1996introduction}. In such cases, we may still learn diagonal exchange interactions using $\lambda_{\vec d}^{aa}=-\frac{1}{2\pi}\m^{aa}_{0,\vec d,1}$. However, the off-diagonal exchanges $\lambda^{ab}_{\vec d}$ and fields $b^a$ must be inferred from higher DSF moments. For instance, direct calculation shows $\frac{1}{2\pi}\m^{aa}_{2,\vec d,0} = -8\left(\lambda_{\vec d}^{bb}\lambda_{\vec d}^{cc}-\lambda_{\vec d}^{bc}\lambda_{\vec d}^{cb}\right)$, which allows one to further determine the products $\lambda_{\vec d}^{bc}\lambda_{\vec d}^{cb}$. We show in App.~\ref{sec:hightemp} that the feature map $F_{\mathrm{th}}^{\mathrm{diag}}= \left\{\m^{aa}_{0,\vec d,1},\, \m^{aa}_{2,\vec d,0},\, \m^{aa}_{2,\vec 0,0},\Re \m^{aa}_{2,\vec d,1}\right\}_{a, \vec d}$ defines a locally solvable polynomial system for all couplings when only the diagonal DSF are accessible. In particular, arbitrary one- and two-body couplings are generically locally solvable from diagonal DSF data alone, using polynomial equations of degree at most three. However, global solvability requires additional care in this case. In particular, uniform $\pi$ spin rotations about coordinate axes can reverse signs of $\lambda^{ab}_{\vec d}$ and $b^a$ while preserving diagonal DSF data, an example of a measurement equivalence. See App.~\ref{sec:hightemp} for a rigorous proof.

\par 
We now validate the practical performance of the HT protocol in the presence of measurement noise. We consider three settings, each represented by Eq.~\eqref{eq:12body} with different restrictions: a honeycomb
$J$--$K$--$\Gamma$ model~\cite{kitaev2006anyons,balents2010spin} [Fig.~\ref{fig:HTSF}a], generic spin chains with twelve independent
one- and two-body nearest-neighbor couplings [Fig.~\ref{fig:HTSF}b], and honeycomb exchange Hamiltonians with up to 27 independent nearest-neighbor couplings [Fig.~\ref{fig:HTSF}c]. The numerics affirms the HT protocol's generally strong performance and versatility. Contrary to na\"ive expectation, performance is optimal at a finite, noise-dependent $T$ rather than arbitrarily high $T$ [Fig.~\ref{fig:HTSF}a], reflecting competition between the magnification of errors by noise at large $T$ and HT expansion error at small $T$. Figure~\ref{fig:HTSF}b confirms general expectations that lower-order polynomial systems $\FF = \F(\vec\lambda)$ ease learning, while Fig.~\ref{fig:HTSF}c reveals a mild growth of error with the number of learned parameters $p$.

\par

Though we obtained exact results on local and global solvability for Eq.~\eqref{eq:12body}, the HT protocol is more general. For three-body and higher-order interactions, higher-degree polynomial systems generally appear, but the HT protocol still applies. We demonstrate explicit end-to-end reconstruction for Hamiltonians with three-body interactions and nineteen parameters in App.~\ref{sec:supp_tfim_end2end}.

\section{Low-temperature hardness}
\noindent Though versatile and scalable, the HT protocol may face hurdles in practical settings, such as in validating quantum simulations at low temperatures, or in materials when low-energy descriptions break down at elevated temperatures due to thermal activation of additional degrees of freedom. Consequently, it is desirable to find protocols that also operate at low temperatures. We prove, however, that \textit{Hamiltonian learning from the DSF can be exponentially difficult in $\beta$}, even for simple systems. This extends results from static Gibbs states to a dynamical setting~\cite{HaahKothariTangNatPhys}. As an illuminating example, consider the Hamiltonian 
\begin{equation}
    \label{eq:lowThardmain}
    H = \frac h2(\sigma^z_1+\sigma^z_2) + \frac U4(1+\sigma^z_1)(1+\sigma^z_2),
\end{equation}
acting on two spins, with $h,U>0$. Its ground state $\ket{\downarrow\downarrow}$ is separated by energy $h$ from two excited states $\ket{\uparrow\downarrow},\ket{\downarrow\uparrow}$ and by energy $2h+U$ from the remaining state $\ket{\uparrow\uparrow}$. The latter shares no single-spin matrix element with the ground state: $\braket{ \downarrow\downarrow}{O^a_i|\uparrow\uparrow} = 0$. Therefore, the $T=0$ DSF is \textit{strictly independent} of $U$. At low finite temperature, the DSF depends on $U$ only through a thermal occupation factor, and $\pdv{U}S^{ab} = O\left(e^{-\beta h}\right)$ is exponentially small. It follows that learning $U$ at any fixed precision in a noisy experiment requires a number of observations growing exponentially in $\beta h$ [App.~\ref{sec:supphardness}]. 

Although this example provides a valuable worst-case caution, Eq.~\eqref{eq:lowThardmain} need not be representative of typical instances. 
Indeed, we will establish that, in many physically relevant cases, Hamiltonian learning from the DSF at low temperature is generically feasible.

We develop two routes: $(i)$ applying external experimentally motivated perturbations to $H$ [Sec.~\ref{sec:sft}] and $(ii)$ using nontrivial identities derived from the fluctuation-dissipation theorem that relate spectral moments at different orders [Sec.~\ref{sec:fdt}]. 

\section{Strong-field tomography}
\label{sec:sft}
\noindent As our first route to practical Hamiltonian learning at low temperatures, we study Hamiltonians $H(\vec\lambda)$ perturbed by strong global Zeeman fields. Our motivation is twofold. First, a Zeeman field along $\hat x$ overcomes the low-temperature hardness of Eq.~\eqref{eq:lowThardmain}, since the leading excited states contributing to the DSF are energy-split by $U/2$, rendering $U$ straightforwardly learnable. Second, polarizing Zeeman fields greatly simplify the structure of ground states and excitations; consequently, they have long been profitably used to extract interactions via model fitting in quantum magnets~\cite{ColdeaTennantHabicht2002PRL}, and will similarly streamline our systematic extraction of Hamiltonian parameters from DSF moments. 

\par 
In the \textit{strong-field (SF) protocol}, we specifically measure the DSF of
\begin{equation}
\label{eq:strongfieldmainH}
H_\b^{(\hat n)}(\boldsymbol{\lambda}) = H(\boldsymbol{\lambda}) -\b\sum_{\boldsymbol r} \hat n\cdot\boldsymbol{\sigma}_{\boldsymbol r},
\end{equation}
where $\b$ is the strength of a magnetic field oriented along $\hat n$ (both assumed to be known) and $H(\vec \lambda)$ is the Hamiltonian to be learned. We will consider $B$ large relative to the scale of $\vec\lambda$; in this limit, the ground state of $H^{(\hat n)}_\b$ is approximately a nondegenerate polarized product state with gap $\Delta =2\b+O(\vec\lambda)$. \par  

Similar in spirit to the HT protocol, the SF protocol proceeds via computation of DSF spectral moment harmonics, which here admit a controlled perturbative expansion in $1/\b$~\cite{yarotsky2005uniqueness,yarotsky2006ground}. This approach is suited to the low-temperature regime $T\ll\Delta$, where thermal corrections are exponentially suppressed in $\Delta/T$. We therefore work at $T=0$, writing 
\begin{equation}
\label{eq:SFexpansion}
    \b^{-n}m^{ab}_{n,\vec d}(1/\b,\vec\lambda) = \sum_{i=0}^\infty (1/\b)^i \mathsf{m}^{ab}_{n,\vec d,i}(\vec\lambda),
\end{equation}
leaving $\hat n$ dependence implicit. Similar to Eq.~\eqref{eq:mabndexpansion}, at large fields one can truncate the expansion at a given desired order with controlled error. Each nonzero coefficient $\mathsf{m}_{n,\vec d,i}^{ab}(\vec\lambda)$ is a homogeneous polynomial of degree $i$ in the couplings, efficiently computable using perturbation theory. The solvability analysis of Sec.~\ref{sec:hightempmain} then applies, with different field directions now incorporated into the feature map. For spin-$\frac12$ Hamiltonians with arbitrary one- and two-body interactions [Eq.~\eqref{eq:12body}], we establish global solvability at sufficiently large $\b$ using only the coefficients $\mathsf m_{0,\vec d,1}^{ab}$ and $\mathsf m_{1,\vec d,1}^{ab}$, which determine all couplings linearly. Remarkably, fields along just three directions, $\hat x,\hat y,\hat z$, suffice with full DSF access, while three additional tilted directions, $(\hat x-\hat y)/\sqrt2$, $(\hat x-\hat z)/\sqrt2$, and $(\hat y-\hat z)/\sqrt2$ suffice with only diagonal access. Additional spin symmetries can substantially reduce these requirements: For SU(2)-symmetric Hamiltonians, a single arbitrary field direction suffices, while for U(1)-symmetric Hamiltonians, any field direction unaligned with the symmetry axis suffices for learning all couplings. Proofs of these statements are left to App.~\ref{sec:suppprobefield}.

\par 
We evaluate the SF protocol at finite $B$ in the presence of measurement noise with the benchmark models used previously for the HT protocol: a honeycomb $J$-$K$-$\Gamma$ model [Fig.~\ref{fig:HTSF}d], spin chains with twelve independent couplings [Fig.~\ref{fig:HTSF}e], and honeycomb exchange Hamiltonians with up to 27 independent couplings [Fig.~\ref{fig:HTSF}f], all instances of Eq.~\eqref{eq:12body}. We observe performance trends qualitatively similar to those of the HT protocol, reflecting the controlled small-parameter expansions underlying both protocols. However, the SF protocol exhibits appreciably greater robustness to measurement noise, particularly with diagonal DSF access, and a weaker dependence of error on the number of unknown couplings. This quantitative difference originates from the persistence at large $B$ of magnon bands in the DSF split by $O(\vec\lambda)$ [cf. Fig.~\ref{fig:sftomography}]; by contrast, the $O(\vec\lambda)$ data in HT protocol vanishes as $O(\beta)$. Together, these results demonstrate the potential of strong-field measurements for robust Hamiltonian learning at low temperatures, even with restricted DSF access and many unknown couplings.

\section{Fluctuation–dissipation tomography}
\label{sec:fdt}

\begin{figure}
    \centering
    \includegraphics[width=\linewidth]{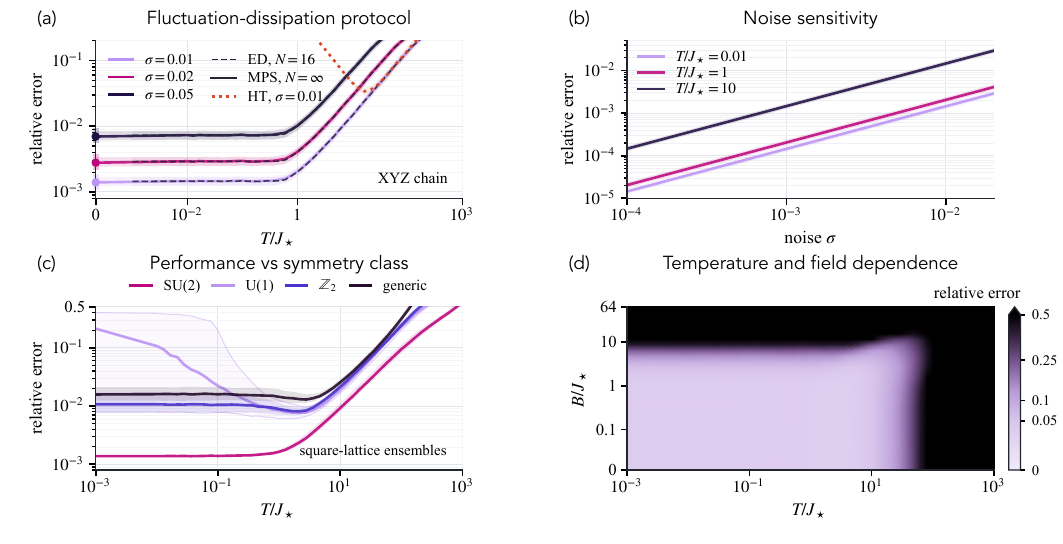}
    \caption{\textbf{Fluctuation-dissipation tomography.}
    (a) Relative coupling error for the XYZ chain at $(J_x,J_y,J_z)=(0.4,0.2,-1)J_\star$, using $N=16$ exact diagonalization (dashed) and infinite matrix product states (solid). (b) Noise dependence at three temperatures. (c) Temperature dependence for 1,000 Hamiltonians per symmetry class on a $4\times3$ square lattice ($\sigma=0.01$). (d) Median relative error for 1,000 generic Hamiltonians ($\sigma=0.02$) as field and temperature are varied. Throughout, relative error is $|\boldsymbol\lambda_{\mathrm{pred}}-\boldsymbol\lambda_{\mathrm{true}}|/|\boldsymbol\lambda_{\mathrm{true}}|$ and curves are medians with $25\%$ error ribbons.}
    \label{fig:fdt}
\end{figure}

\noindent Both the HT and SF protocols exploit small-parameter expansions to relate measured DSF data to low-degree polynomials in the couplings.  Suitable high-temperature/strong-field regimes may not be available in practice, however, especially if the Hamiltonian couplings to be learned are themselves `large'.  Moving beyond these frameworks, we introduce the fundamentally different \textit{fluctuation-dissipation (FD) protocol} that avoids the preceding requirements. The FD protocol instead leverages nontrivial relations satisfied by DSF spectral moments to construct a linear system $\mathsf{A}\vec\lambda = \mathsf{b}$, where both $\mathsf{A}$ and $\mathsf{b}$ are extracted from the DSF, and from which $\vec\lambda$ is linearly solved. \par

Our starting point is the fluctuation-dissipation theorem,
\begin{equation}
    \label{eq:fdt}
2\chi^{\prime\prime\,ab}(\boldsymbol q,\omega)
=(1-e^{-\beta\omega})S^{ab}(\boldsymbol q,\omega)
\end{equation}
where $\chi^{\prime\prime\,ab}=(\chi^{ab}-\chi^{ba*})/(2\mathrm i)$
is the dissipative part of the retarded susceptibility
($\chi^{ab}(\boldsymbol q,\omega)
=\mathrm i\int_0^\infty\mathrm dt\,e^{\mathrm i\omega t}
\langle[O^a_{\boldsymbol q}(t),O^b_{-\boldsymbol q}]\rangle_\beta$). Applying detailed balance in the form $S^{ba}(-\vec q,-\omega) = e^{-\beta \omega}S^{ab}(\vec q,\omega)$ and taking the first frequency moment of Eq.~\eqref{eq:fdt} yields
\begin{equation}
\label{eq:FDmain}
\sum_{j=1}^{p}\mathsf A_{\boldsymbol d,j}^{ab}\lambda^j=\mathsf{b}^{ab}_{\vec d},\quad\mathsf A_{\boldsymbol d,j}^{ab}
\coloneqq
\left\langle
[O_{\vec d}^b,[H^j,O_{\vec 0}^a]]
\right\rangle_\beta, \quad \mathsf{b}^{ab}_{\vec d}\coloneqq\frac{1}{2\pi}\left(
m_{1,\boldsymbol d}^{ab}+m_{1,-\boldsymbol d}^{ba}\right);
\end{equation}
see App.~\ref{sec:suppfdt} for detailed derivation. Equation~\eqref{eq:FDmain} is completely general and valid in thermal equilibrium at any temperature. In fact, it holds more generally in any stationary state, i.e., whose density matrix commutes with the Hamiltonian (App.~\ref{sec:suppfdt}). While $\mathsf{b}$ manifestly involves DSF moments, dependence of $\mathsf{A}$ on the DSF is in general obscured.  For one- and two-body Hamiltonians, a key simplification occurs: each entry of $\mathsf{A}$ may be expressed in terms of $m_{0,\boldsymbol d}^{ab}$. Both $\mathsf{A}$ and $\mathsf{b}$ are then available directly from the DSF, and we may learn $\vec\lambda$ by simply solving $\mathsf{A}\vec\lambda = \mathsf{b}$, provided a solution exists. \par 

As an instructive example, consider the nearest-neighbor XYZ chain
$H_{\mathrm{XYZ}}=\sum_{j,a}J_a\sigma_j^{a}\sigma_{j+1}^a$. Equation~\eqref{eq:FDmain} yields the relations $m_{1,1}^{aa} + m_{1,-1}^{aa} = 4(J_b m_{0,1}^{cc} + J_c m_{0,1}^{bb})$ with $a,b,c=x,y,z$ distinct, which can be solved for $J_{x},J_y,J_z$ unless $m_{0,1}^{aa}$ vanishes for some $a$. Notably, for $J_x=J_y=J_z$ these relations reduce to the
Hohenberg--Brinkman sum rule~\cite{hohenbergbrinkman}, of which Eq.~\eqref{eq:FDmain} is a vast generalization. \par

More generally, for Hamiltonians with arbitrary one- and
two-body interactions [Eq.~\eqref{eq:12body}], we prove
that full DSF access uniquely determines all exchange
couplings $\lambda_{\vec d}^{ab}$ whenever
\begin{equation}
\label{eq:fdtcondition}
\det\,m_{0,\boldsymbol d}^{ab} \neq0
\end{equation}
for each $\vec d$. If, in addition, the system has nonzero magnetization,
$\langle\boldsymbol\sigma_{\boldsymbol0}\rangle_\beta
\neq\boldsymbol0$, all one-body fields $b^a$ can also be learned, and hence
the Hamiltonian is globally solvable. Rigorous proofs and explicit solutions for $\lambda_{\vec d}^{ab}$ and $b^a$ are left to App.~\ref{sec:suppfdt}, where we also prove a corollary bounding errors by $O(\varepsilon/\gamma)$ for error $\varepsilon$ in the spectral moments, provided $\gamma \equiv \det\, m_{0,\vec d}^{ab} >0$.

\par

Figure~\ref{fig:fdt}a,b numerically demonstrates the stability of the FD protocol for $H_{\mathrm{XYZ}}$ down to $T=0$ in the presence of noise. In Fig.~\ref{fig:fdt}a, the FD protocol is shown to reach relative errors of $\sim 0.1$--$1\%$ with noise of standard deviation $\sigma\leq 0.05$ at low temperature, outperforming the HT protocol at comparable $\sigma$ by about an order of magnitude. Errors are found to scale linearly with $\sigma$ [Fig.~\ref{fig:fdt}b], as expected for a linear least-squares fit. In App.~\ref{sec:suppfdt}, we prove using a variational argument directly in the thermodynamic limit that the FD protocol for $H_{\mathrm{XYZ}}$ is robust down to $T=0$.

\par 

Having established stability of the FD protocol for $H_{\mathrm{XYZ}}$, it is natural to ask how well the method extends to other models. To this end, we study the typical performance of the FD protocol across Hamiltonian ensembles classified by different symmetries.  Specifically, we take $H(\vec\lambda)$ as in Eq.~\eqref{eq:12body} with imposed SU(2), U(1), $\mathbb{Z}_2$, or no on-site symmetry and solve Eq.~\eqref{eq:FDmain} with noisy DSF data; see Fig.~\ref{fig:fdt}c. We find generally strong performance, similar to the case of $H_{\mathrm{XYZ}}$. A notable exception is the U(1)-symmetric ensemble at low temperature, where the relative error can become substantial. This behavior descends from the presence of a conserved magnetization, taken to be $M^z=\sum_{\boldsymbol r}\sigma^z_{\boldsymbol r}$ without loss of generality.  Nontrivial components of $\mathsf{A}$ associated with the coupling $b^z$ are, according to Eq.~\eqref{eq:FDmain}, proportional to $\langle M^z\rangle_\beta$.  At zero temperature, the ground state occupies a definite magnetization sector.  When the ground state has $M^z=0$, $\mathsf{A}$ becomes non-invertible and the FD protocol fails to learn $b^z$. At finite $T$, thermal occupation of additional magnetization sectors can restore the invertibility of $\mathsf{A}$, explaining why reconstruction can initially improve upon warming. The U(1)-symmetric ensemble thus identifies conserved quantities as potential impediments for Hamiltonian learning---albeit one that can be ameliorated with (even weak) explicit symmetry-breaking perturbations; we discuss this  further in App.~\ref{sec:suppfdt}. 
\par 
To compare FD tomography with the preceding protocols, in Fig.~\ref{fig:fdt}d we examine the
high-temperature limit and the consequences of a strong applied field. At infinite temperature,
each entry of $\mathsf A$ vanishes by cyclicity of the trace, and $\mathsf A=O(\beta)$ at high temperature. Upon application of a strong Zeeman field, correlations approximately decompose via
$\langle O_{\vec 0}^aO_{\vec d}^b\rangle
\to \langle O_{\vec 0}^a\rangle\langle O_{\vec d}^b\rangle$. In both cases, $m^{ab}_{0,\vec d}$  [Eq.~\eqref{eq:fdtcondition}] becomes increasingly ill-conditioned and hence the matrix inversion becomes increasingly sensitive to noise and experimental imperfection [App.~\ref{sec:temp_and_field_fdt}]. Thus the regime in which FD tomography operates well is complementary to those of the HT and SF protocols.

\begin{figure*}
\centering
\definecolor{dsfHT}{HTML}{E07B29}\definecolor{dsfSF}{HTML}{2E5FA3}\definecolor{dsfFD}{HTML}{7B52A8}
\def\bd{\boldsymbol d}\def\bq{\boldsymbol q}\def\br{\boldsymbol r}\def\bl{\boldsymbol\lambda}\def\sm{\mathsf m}
\def\dsfF{\fontsize{7}{8.3}\selectfont}
\long\def\dsfc#1{\begin{minipage}{143pt}\dsfF\raggedright\hyphenpenalty10000\exhyphenpenalty10000
  \lineskiplimit-.6pt\lineskip.4pt\relax#1\end{minipage}}
\long\def\dsfh#1#2{\setbox0\hbox{\dsfc{#1}}\edef#2{\the\dimexpr\ht0+\dp0}}
\def\dsfe#1{\par\vspace{1pt}\centerline{$#1$}\par\vspace{1pt}}
\def\dsfi#1{\par\hangindent1em\hangafter1\noindent\hbox to1em{#1.\hss}\ignorespaces}
\def\dsfl#1{\par\vspace{1.2pt}\noindent\textit{#1:} }
\def\dsfHs{\dsfi1 Measure $S^{ab}(\bq,\omega)$ at several high $T$.
  \dsfi2 Integrate data to moments $m^{ab}_{n,\bd}$.
  \dsfi3 Extract by finite differences in $\beta$:
  \dsfe{m^{ab}_{n,\bd}(\beta)=\sum_{i\geq0}\beta^i\,\sm^{ab}_{n,\bd,i}.}
  \dsfi4 Use operator algebra to derive $\sm^{ab}_{n,\bd,i}(\bl)$ from proposed Hamiltonian.
  \dsfi5 Equate and solve for $\bl$.}
\def\dsfSs{\dsfi1 Measure $S^{ab}(\bq,\omega)$ in strong fields $B\hat n$.
  \dsfi2 Integrate data to moments $m^{ab}_{n,\bd}$.
  \dsfi3 Extract by finite differences in $1/B$:
  \dsfe{B^{-n}m^{ab}_{n,\bd}(B)=\sum_{i\geq0}B^{-i}\,\sm^{ab}_{n,\bd,i}.}
  \dsfi4 Use perturbation theory to derive $\sm^{ab}_{n,\bd,i}(\bl)$ from proposed Hamiltonian.
  \dsfi5 Equate and solve for $\bl$.}
\def\dsfFs{\dsfi1 Measure $S^{ab}(\bq,\omega)$ at a single $T$.
  \dsfi2 Integrate data to $m^{ab}_{0,\bd}$ and $m^{ab}_{1,\bd}$.
  \dsfi3 Use measured data to form $\mathsf A$ and $\mathsf b$:
  \dsfe{\jot.6pt\begin{aligned}\mathsf A^{ab}_{\bd,j}&=\langle[O^b_{\bd},[H^j,O^a_{\boldsymbol0}]]\rangle_\beta = f(m_{0,\vec d}^{ab}),\\
    \mathsf b^{ab}_{\bd}&=(m^{ab}_{1,\bd}+m^{ba}_{1,-\bd})/2\pi.\end{aligned}}
  \dsfi4 Solve $\sum_j\mathsf A^{ab}_{\bd,j}\,\lambda^j=\mathsf b^{ab}_{\bd}$.}
\def\dsfHg{\dsfl{${\leq}2$-body guarantee}full $S^{ab}$ gives all couplings linearly:
  $\lambda^{ab}_{\bd}=-\sm^{ab}_{0,\bd,1}/2\pi$, $b^a=\mathrm i\varepsilon^{abc}\sm^{bc}_{0,\boldsymbol0,1}/4\pi$.
  \dsfl{Extensions}diagonal $S^{aa}$, higher-body $H$, spatial disorder}
\def\dsfSg{\dsfl{${\leq}2$-body guarantee}full $S^{ab}$ at $\hat n=\hat x,\hat y,\hat z$ gives all
  couplings linearly.
  \dsfl{Extensions}diagonal $S^{aa}$, higher body $H$, angular tomography.}
\def\dsfFg{\dsfl{${\leq}2$-body guarantee}full $S^{ab}$ gives all couplings exactly and linearly if $\det\,m^{ab}_{0,\bd} \neq0$ , $\langle\boldsymbol\sigma\rangle_\beta\neq\boldsymbol0$. \par 
\textit{Extensions:} any stationary state}
\dsfh\dsfHs\dsfpa\dsfh\dsfSs\dsfpb\dsfh\dsfFs\dsfpc\dsfh\dsfHg\dsfqa\dsfh\dsfSg\dsfqb\dsfh\dsfFg\dsfqc
\pgfmathsetmacro\dsfH{max(\dsfpa+\dsfqa,\dsfpb+\dsfqb,\dsfpc+\dsfqc)+13}
\resizebox{\textwidth}{!}{%
\begin{tikzpicture}[x=1pt,y=1pt,every node/.style={inner sep=0pt},
  dsfbox/.style={draw=black!80,line width=.35pt,rounded corners=1.6pt}]
\foreach \x/\c/\t/\s/\g/\p in {0/dsfHT/High-temperature/\dsfHs/\dsfHg/\dsfpa,
    157.5/dsfSF/Strong-field/\dsfSs/\dsfSg/\dsfpb,315/dsfFD/Fluctuation--dissipation/\dsfFs/\dsfFg/\dsfpc}{
  \node[anchor=base,font=\dsfF\bfseries] at (\x+75.5,3.5) {\t};
  \path[dsfbox,fill=\c!4] (\x,0) rectangle ++(151,-\dsfH);
  \node[anchor=north] at (\x+75.5,-3.5) {\dsfc{\s}};
  \draw[black!35,line width=.3pt] (\x+4,-\p-6.5) -- ++(143,0);
  \node[anchor=north] at (\x+75.5,-\p-9.5) {\dsfc{\g}};}
\matrix[anchor=north,column sep=10pt,ampersand replacement=\&,nodes={dsfbox,align=center,
  font=\dsfF,inner xsep=4.5pt,inner ysep=2.6pt}] at (233,-\dsfH-5) {
  \node{\textit{DSF spectral moments}\\[.6pt]$m^{ab}_{n,\bd}=\mathrm{vol(BZ)}^{-1}\!\int_{\mathrm{BZ}}
    \mathrm d\bq\,e^{-\mathrm i\bq\cdot\bd}\int\mathrm d\omega\,\omega^nS^{ab}(\bq,\omega)$}; \&
  \node{\textit{One- and two-body Hamiltonians}\\[.6pt]$H(\bl)=\sum_{\br}b^a\sigma^a_{\br}
    +\sum_{\br,\,\bd\in\mathcal R}\lambda^{ab}_{\bd}\,\sigma^a_{\br}\sigma^b_{\br+\bd}$};\\};
\end{tikzpicture}}
\caption{\textbf{Technical summary.} Steps, learning guarantees, and extensions of the protocols.}
\label{fig:technicalsummary}
\end{figure*}

\section{Outlook}
\label{sec:end}

\noindent We have studied an inverse problem motivated by both longstanding experimental objectives in characterizing quantum magnets with neutron scattering and emerging theoretical frameworks for Hamiltonian learning. Our suite of simple, scalable protocols reveal complementary regimes of learnability and lend themselves to general application in quantum experiments [Fig.~\ref{fig:technicalsummary}]. Below, we discuss extensions, further experimental considerations, and compelling open questions.

\begin{figure}
\centering
\includegraphics[width=\linewidth]{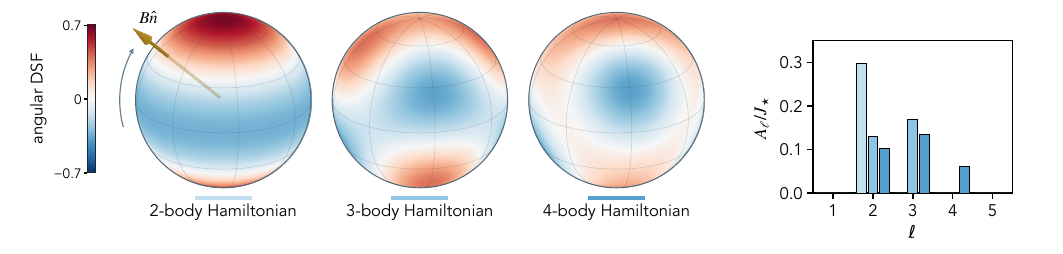}
\caption{\textbf{Angular tomography at strong field.} Spheres show the angular
DSF $\Delta\mathcal C_{\vec d}(\hat n)/J_\star$, where
$\Delta\mathcal C_{\vec d}=\mathcal C_{\vec d}
-\overline{\mathcal C}_{\vec d}$, $\mathcal C_{\vec d}$ is defined in Eq.~\eqref{eq:angularmain}, and $\overline{\mathcal C}_{\vec d}$ is its angular mean. Bars show $A_\ell/J_\star$, with
$A_\ell=[(4\pi)^{-1}\sum_m|\mathcal C_{\vec d;\ell m}|^2]^{1/2}$
for spherical-harmonic coefficients $\mathcal C_{\vec d;\ell m}$. Harmonics with angular momentum $\ell$ certify $\ell$-spin interactions in the Hamiltonian ($B\to\infty$,
$T=0$).}
    \label{fig:ATmain}
\end{figure}

\smallbreak
\noindent \textbf{Detecting multispin interactions.} In modeling many-body Hamiltonians, a useful notion is the largest $k$ such that the Hamiltonian is well-described, to good accuracy, by a sum of $k$-body interactions. In App.~\ref{sec:suppprobefield} we introduce an extension of the SF protocol, dubbed \textit{angular tomography}, which uses the full dependence of the DSF as a function of applied Zeeman field direction $\hat n$ to provide a lower bound on $k$. In particular, defining
\begin{equation}
\label{eq:angularmain}
\mathcal C_{\boldsymbol d}(\hat n)
=
\lim_{B\to\infty}\frac{1}{4\pi}
\sum_{a=x,y,z}
\left[m_{1,\boldsymbol d}^{aa}-2B\,m_{0,\boldsymbol d}^{aa}\right],
\end{equation}
we prove that nonzero spherical harmonics of $\mathcal{C}_{\vec d}(\hat n)$ of angular momentum $\ell$ indicate the presence of $\ell$-body interactions. For example, the $\ell=2,3,4$ components in
Fig.~\ref{fig:ATmain} indicate multispin interactions involving
at least $\ell$ spins, respectively, and their amplitudes lower-bound the associated multispin
interaction strength.\par 

\smallbreak\noindent \textbf{Hamiltonian learning with spatial disorder.} A practical extension of our framework is to the case of couplings $\vec\lambda_{\vec r}$ with spatial variation arising, e.g., from disorder. In App.~\ref{sec:spatialdisorder}, we prove that DSF data allows the characterization of statistical properties of spatially varying couplings, resembling the appearance of the Debye-Waller factor due to thermal motion in crystals. For broad classes of Hamiltonians, when high-temperature data is accessible, we prove that disorder-averaged DSF spectral moments can characterize the moments of a disorder distribution from which $\vec\lambda_{\vec r}$ arises, allowing the latter moments to be learned in a hierarchical fashion. \par

\smallbreak
\noindent \textbf{Comparison with model fitting and supervised ML.}
Conventional model fitting requires repeated simulations of the DSF as the Hamiltonian couplings are optimized, while supervised ML shifts much of this computational cost to generating a training set. Our protocols bypass both bottlenecks, leveraging linear or low-degree polynomial equations satisfied by DSF spectral moments. In App.~\ref{sec:litreview}, we benchmark against implicit neural representations~\cite{ChitturiINRDSF}, a supervised ML method, finding that reaching $\sim5\%$ relative error requires roughly $10^3$ training Hamiltonians on finite-size $J$--$K$--$\Gamma$ models where our protocols achieve sub-percent errors without training in their respective regimes.

\smallbreak
\noindent \textbf{Experimental implementation.}
Quantum magnets studied by neutron scattering, such as
Cs$_2$CuCl$_4$, provide natural benchmarks for our protocols~\cite{ColdeaTennantHabicht2002PRL}. Our work emphasizes the role of DSF spectral moments, which enjoy a degree of robustness to finite energy resolution or broadening. Moreover, the energy regimes relevant to HT and SF tomography are readily available especially in weak-exchange quantum magnets, while FD tomography enjoys application over several energetic orders of magnitude. Even when the assumed Hamiltonian omits interactions terms that are appreciable in an experimental system, HT tomography can identify the microscopic couplings of the retained terms, while SF and FD may instead learn effective couplings [App.~\ref{sec:furtherexperimental}]. Enriching the feature map and enlarging the Hamiltonian family can provide means of validating and refining the learned Hamiltonian. Our analyses support feasibility of implementation in existing quantum experiments.

\smallbreak
\noindent \textbf{Robustness of zero temperature learning.} Which physical systems admit stable Hamiltonian learning from the DSF down to zero temperature, for instance by the FD protocol? Can average-case guarantees at $T=0$ be established for broad Hamiltonian families beyond the present work's numerical evidence? \par

\begin{acknowledgments}
\noindent We thank Arnab Banerjee, Sitan Chen, Gabor Halasz, Hsin-Yuan Huang, Andrew Ma, Paul Neves, and Alan Tennant for helpful discussions. TS and NP acknowledge support from the Walter Burke Institute for Theoretical Physics and the Institute for Quantum Information and Matter, an NSF Physics Frontiers Center (NSF Grant PHY-2317110). The U.S. Department of Energy, Office of Science, National Quantum Information Science Research Centers, Quantum Science Center partially supported the analytical portions of this work. TS acknowledges support from the  U.S. Department of Energy, Office of Science, National Quantum Information Science Research Centers, Quantum Systems Accelerator. NP acknowledges AI (Claude, Codex, ChatGPT $\leq 6.1$) for coding + plotting assistance, helpful discussions, and checking theoretical results, and credits ChatGPT 5.6 for part of Thm. B.3. We provide code for the tomography protocols and data for reproducing selected figures at \href{https://github.com/nisargapaul/LearningDSF}{\texttt{github.com/nisargapaul/LearningDSF}}.
\end{acknowledgments}

\bibliography{refs}

\clearpage
\appendix

\appendixheading
\appendixtableofcontents

\section{Comparison with model fitting and supervised learning }
\label{sec:litreview}

\subsection{Model fitting}
\begin{table*}[t]
\centering
\scriptsize
\renewcommand{\arraystretch}{1.15}
\setlength{\tabcolsep}{5pt}
\caption{\textbf{Model fitting.} Conventional methods for learning
Hamiltonians from neutron scattering data.}
\label{tab:forward_methods}
\begin{tabular}{
    p{0.2\textwidth}
    p{0.25\textwidth}
    p{0.25\textwidth}
    p{0.18\textwidth}}
\toprule
\textbf{Method}
& \textbf{Typical regime}
& \textbf{Main limitation}
& \textbf{Example} \\
\midrule

Linear spin-wave theory
&
Ordered or field-polarized magnets with well-defined magnons
&
Unreliable for unordered magnets, strong fluctuations, \& broad magnon continua
&
Cs$_2$CuCl$_4$~\cite{ColdeaTennantHabicht2002PRL}
\\

Self-consistent Gaussian approximation
&
Paramagnets, diffuse scattering
&
Gaussian approximation to spin fluctuations, typically restricted to static correlations
&
NaCaNi$_2$F$_7$~\cite{plumb2019continuum}
\\

(Semi)classical spin dynamics
&
Finite-$T$, semiclassical
&
Misses quantum effects&
Zn$_2$VO(PO$_4$)$_2$~\cite{park2024quantum}
\\

Exact diagonalization
&
Strongly quantum magnets on small clusters
&
Exponential cost in system size 
&
CoNb$_2$O$_6$~\cite{woodland2023tuning}
\\

DMRG 
&
Strongly quantum 1D/quasi-1D systems
&
Entanglement/dimension-limited
&
$(\mathrm{C}_7\mathrm{H}_{10}\mathrm{N})_2\mathrm{CuBr}_4$~\cite{schmidiger2012spectral}
\\

Supervised ML & General, $\lesssim 10$ param. models
& Generating training data with one of the above methods 
& Dy$_2$Ti$_2$O$_7$\cite{samarakoon2020machine}

\\
\bottomrule
\end{tabular}
\end{table*}

\noindent A common existing approach to Hamiltonian learning from neutron scattering data involves what we refer to as \textit{model fitting}: $(i)$ specifying a few-parameter family of interactions informed by crystal symmetries and other physically motivated constraints, $(ii)$ calculating the DSF for the resulting model, and $(iii)$ adjusting the couplings until convergence is achieved with measured data. A mature toolbox of such methods, with varying domains of validity and limitations, is available. Linear spin-wave theory (LSWT), as implemented in packages such as \texttt{SpinW}, is widely used for ordered magnets with weak fluctuations~\cite{toth2015linear}. Classical and semiclassical spin dynamics provide complementary approaches, including
generalizations based on SU($N$) coherent states in
\texttt{Sunny.jl}~\cite{dahlbom2025sunny}. Exact diagonalization
and dynamical DMRG present fully quantum methods suitable to small clusters or low-dimensional geometries. Table~\ref{tab:forward_methods} summarizes representative
applications and limitations.

A number of challenges are encountered in model fitting. First, any classical $H\mapsto$ DSF simulation must inevitably make some set of approximations or incur an exponential complexity in general. For instance, among the only unbiased methods for $H\mapsto$ DSF simulation is exact diagonalization or quantum Monte Carlo, which scale exponentially in system size or can suffer from a sign problem. In practice, some set of approximations is typically made, with each method having its corresponding domain of validity. Moreover, model fitting based on classical simulation can involve an iterative optimization (simulate the DSF, compare with data, update $H$, repeat) with few convergence guarantees. 
\par 

In light of these challenges, our focus is on learning protocols with accuracy guarantees and that do not require any many-body simulation that scales with the system size. We note the latter is a realistic target due to our assumption of translation invariance, and this rules out protocols scaling as $\mathrm{poly}(N)$, which can still be prohibitively expensive for macroscopic materials. Our protocols avoid many-body simulation by exploiting analytical relations between couplings and spectral moments. These relations are obtained from local operator algebra, using controlled high-temperature expansions, strong-field expansions, or fluctuation-dissipation identities. \par 

Analytical constraints on spectral moments in terms of couplings have a history in neutron scattering. For example, the Hohenberg--Brinkman sum rules relate DSF spectral moments to exchange interactions in isotropic Heisenberg chains~\cite{hohenbergbrinkman}. Furthermore, inference of Hamiltonian couplings using a strong Zeeman field has experimental precedent in Cs$_2$CuCl$_4$, where exchange couplings were extracted directly from the magnon dispersion in the polarized magnet~\cite{ColdeaTennantHabicht2002PRL}. Some elements of our protocols can be regarded as substantial extensions of these precursors to more general Hamiltonians, put on rigorous footing.\par

Perhaps the most compelling advantage of our Hamiltonian learning protocols over conventional model fitting is in application to quantum magnets without long-range order or whose elementary excitations are not magnons. In long-range ordered magnets, infinite-lifetime quasiparticles have well-defined mode energies whose intensities can be fit with a Hamiltonian ansatz, a practice with experimental precedent (see e.g. \cite{PhysRevB.105.104402} for an example). In contrast, quantum magnets lacking conventional order can have broad quasiparticle continua, e.g. the two-spinon continuum in the DSF of a quantum spin liquid, in which case Hamiltonian learning can prove subtle. Our work provides systematic protocols for general quantum systems which are in principle impervious to these difficulties. 
 
\subsection{Supervised machine learning}

 \begin{figure}
    \centering    \includegraphics[width=\linewidth]{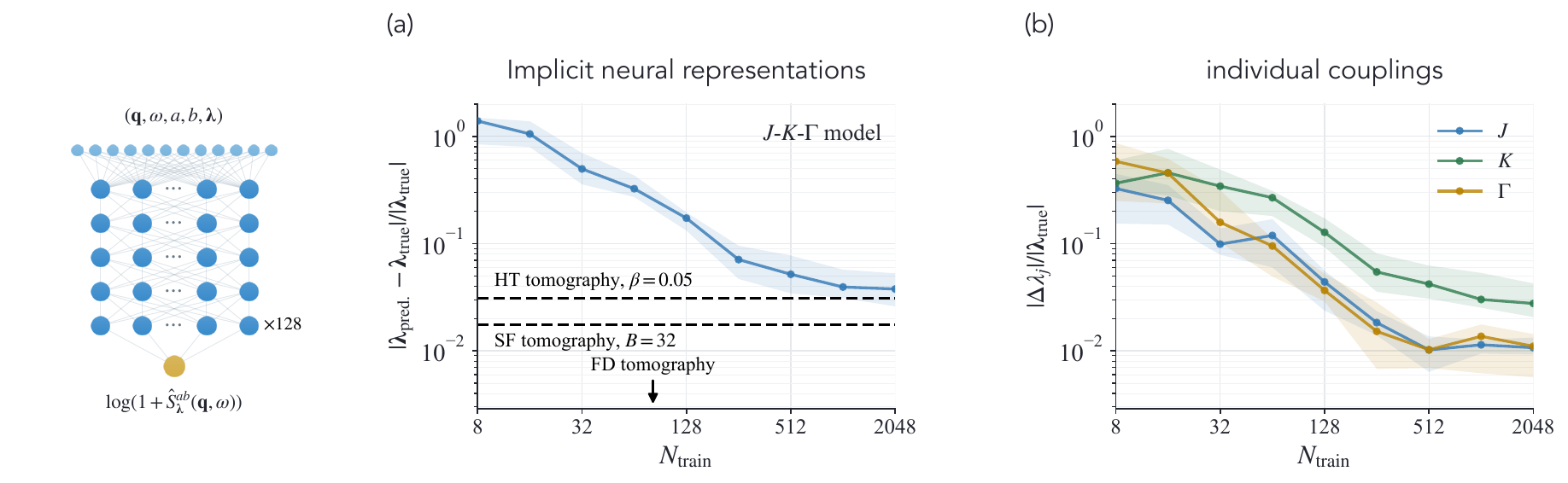}
    \caption{\textbf{Implicit neural representations.} We compare the supervised learning approach introduced in Ref.~\cite{ChitturiINRDSF} with our training-free Hamiltonian learning protocols for the $J-K-\Gamma$ model on a periodic 12-spin honeycomb cluster. Left, the implicit neural representation (INR) learns the DSF as a function of the couplings, using a depth-five width-128 neural network with sine activations. New couplings are then inferred by fitting the predicted DSF to test data. (a) Relative coupling error $|\boldsymbol{\lambda}_{\mathrm{pred}}-\boldsymbol{\lambda}_{\mathrm{true}}|/|\boldsymbol{\lambda}_{\mathrm{true}}|$ versus the size $N_{\mathrm{train}}$ of the training set, with INR evaluated at zero field and temperature. Dashed lines show the strong-field (SF) protocol ($\b=32$, $T=0$) and the high-temperature (HT) protocol ($\beta=0.05$, $\b=0$). All points use 16 test Hamiltonians and noiseless data. (b) Individual errors in $\lambda_j = (J,K,\Gamma)_j$ normalized by $|\boldsymbol{\lambda}_{\mathrm{true}}|$. INR curves show medians over test Hamiltonians with $25\%$ error ribbons. }
    \label{fig:MLstuff}
\end{figure}

\noindent A modern model fitting approach, which can be combined with previous ones, is supervised machine learning (ML)~\cite{ChitturiINRDSF,samarakoon2020machine,samarakoon2022extraction}. In these cases, many-body simulation is used to generate a labeled training set which is fit to a general DSF representation $(\vec q,\omega,a,b,\vec\lambda) \mapsto S^{ab}(\vec q,\omega;\vec\lambda)$ using neural networks. Ref.~\cite{ChitturiINRDSF} developed an ``implicit neural representation" (INR) architecture, trained it on LSWT-simulated data, and fitted to experimental neutron scattering data to extract exchange couplings. Their framework profitably employs gradient-based optimization to fit model parameters and accurately recovered the Hamiltonian of a two-parameter 2D antiferromagnetic Hamiltonian. The main drawback with supervised ML, as with other model fitting methods, is the reliance on expensive and/or approximate classical simulation for the DSF, in this case to generate a training set. \par

In Fig.~\ref{fig:MLstuff}, we compare an adapted INR algorithm with our training-free strong-field (SF), high-temperature (HT), and fluctuation-dissipation (FD) protocols. We refer the reader to Ref.~\cite{ChitturiINRDSF} for further details on INR. We apply our protocols and INR to the $J$-$K$-$\Gamma$ model~\cite{rau2014generic} with random couplings,
\begin{equation}
\begin{aligned}
H(\boldsymbol{\lambda}) = \sum_{\langle \vec r\vec r'\rangle_\gamma} \left[ J\,\boldsymbol{\sigma}_{\vec r}\cdot\boldsymbol{\sigma}_{\vec r'} +K\sigma_{\vec r}^\gamma\sigma_{\vec r'}^\gamma +\Gamma\left(\sigma_{\vec r}^\alpha\sigma_{\vec r'}^\beta +\sigma_{\vec r}^\beta\sigma_{\vec r'}^\alpha\right) \right], \end{aligned}
\end{equation}
where $\boldsymbol{\lambda}=(J,K,\Gamma)$ and $\alpha,\beta$ specify two axes other than $\gamma$. 
\par 
Our numerical results display two regimes for the INR: a steady improvement for $N_{\mathrm{train}}\leq 256$ with median coupling error falling from $\sim 1.4$ to $\sim 0.1$, and a plateau from $N_{\mathrm{train}}\geq 512$ during which the error only changes on the order of $\sim 0.02$ and improvement slows. Hence the INR requires roughly $10^3$ training Hamiltonians to reach a median error of $5\%$, while SF and HT tomography reach about $1-5\%$ \textit{without} training. Meanwhile, the FD protocol was exact to machine precision. This benchmark shows that, our protocols achieve comparable or better accuracy without requiring a large simulated training set. An interesting future direction would be to incorporate our protocols into supervised learning frameworks to combine their respective advantages.
\par 
\textit{Numerical details: } We generate DSFs for nested training sets of $N_{\mathrm{train}}=8,16,\ldots,2048$ Hamiltonians (with 128 validation and 16 test Hamiltonians) and broaden along $\omega$ with a Gaussian of standard deviation $0.15$ on a grid of spacing $0.05$ and total range $\approx 100$. The underlying neural network is a larger version of the SIREN architecture used in Ref.~\cite{ChitturiINRDSF}: it has 5 width-128 layers and sine nonlinearities. The neural network takes couplings, momentum, frequency, and polarization as inputs and outputs a nonnegative number,  $\widehat S$. It is trained on the aforementioned data using a (generalized) KL divergence $D(S\Vert\widehat S) =\sum_i\left[\widehat S_i-S_i+S_i\log(S_i/\widehat S_i)\right]$ as the loss function, where $S_i$ is the spectral weight in bin $i$ of the training data. We randomly sample Hamiltonians and polarization channels (i.e. $\sum_{ab}p_a^*p_bS^{ab}$). We use batches of 8192 samples and optimize with Adam at initial learning rate $2\times10^{-4}$ and a cosine decay training schedule. We train for 8000 optimization steps for $N_{\mathrm{train}}$ up to 128 and 16000 optimization steps at larger $N_{\mathrm{train}}$. We retain the lowest achieved validation loss across training, freeze the network weights and, denoting it $\widehat S_{\boldsymbol{\lambda}}$, produce an estimator $\vec\lambda_{\mathrm{pred}}$ for any test DSF $S$ by computing $\vec\lambda_{\mathrm{pred}} = \arg \min_{\vec\lambda} D(S\| \widehat S_{\boldsymbol{\lambda}})$, searching over $[-1,1]^3$.

\section{High-temperature Hamiltonian tomography}
\label{sec:hightemp}

\noindent In this section, we give an in-depth treatment of Hamiltonian learning from the dynamical structure factor (DSF) using the high-temperature (HT) expansion.  We are considering a family of local, translation-invariant Hamiltonians on the lattice $\Lambda$, defined
\begin{equation}
\label{eq:suppHTeqHam} H(\vec\lambda) = \sum_{a=1}^p \lambda^aH^a,\qquad H^a = \sum_{\vec r\in \Lambda} h_{\vec r}^a
\end{equation}
with $\vec \lambda \in [-1,1]^p$ and with $p$ a constant independent of the system size $N=|\Lambda|$. For definiteness, take $\Lambda$ to be a periodic Bravais lattice, with a fixed finite-dimensional spin Hilbert space at each site. The operators $h_{\vec r}^a$ are translated instances of a fixed local interaction term $h_{\vec 0}^a$. We exclude linear redundancies in Eq.~\eqref{eq:suppHTeqHam} and assume $\tr H=0$. We assume there exists a metric $d_\Lambda(\vec r,\vec r')$ on $\Lambda$ and define $\mathrm{supp}(A)$ to be the set of sites on which $A$ acts nontrivially. We define 
\begin{equation}
\mathrm{diam}(A) = \max_{\vec r,\vec r'\in \mathrm{supp}(A)}d_\Lambda(\vec r,\vec r').
\end{equation}
We assume each local term is bounded in the sense $ \|h_{\vec r}^a\|_\infty \leq h$ and $\mathrm{diam}(h_{\vec r}^a)\leq R$ for $O(1)$ constants $h, R$. We write 
\begin{equation}
    \langle A\rangle_\beta = \frac{\Tr(Ae^{-\beta H})}{\Tr(e^{-\beta H})},\qquad \langle A\rangle_0 = \frac{\Tr(A)}{\Tr(1)},
\end{equation}
with $\hbar = k_B=1$. For fixed local Hermitian spin operators $O_{\vec r}^a$, we define $O_{\vec q}^{a} = \frac{1}{\sqrt{N}} \sum_{\vec r\in\Lambda}e^{-\ii\vec q\cdot \vec r}O_{\vec r}^a$ and associate to it the dynamical structure factor (DSF) and its spectral moments:
\begin{equation}
    S^{ab}(\vec q,\omega) = \int_{-\infty}^{\infty} \dd t \, e^{\ii\omega t} \left\langle O_{\vec q}^a(t)O_{-\vec q}^b\right\rangle_\beta,\quad m_n^{ab}(\vec q) = \int_{-\infty}^{\infty} \dd \omega \,\omega^n S^{ab}(\vec q,\omega).
\end{equation}
Here, $n\in\mathbb{Z}_{\geq 0}$ and $O(t) = e^{\ii Ht}Oe^{-\ii Ht}$. We distinguish full access, where all entries of $S^{ab}$ are available, from diagonal access, in which only $S^{aa}$ is available. $S^{ab}(\vec q,\omega)$ is simply related to intensities $I(\vec q,\omega)$ measured by neutron scattering through $I \propto \sum_{ab}(\delta_{ab}-\hat q_a\hat q_b)S^{ab}$~\cite{squires1996introduction}. 

\par 

Using $\int \mathrm{d}\omega \, \omega^n e^{-\ii\omega t} = 2\pi \ii^n\delta^{(n)}(t)$ and 
$ \partial_t^nO(t)\left.\right|_{t=0} = (-\ii)^n \mathrm{ad}_H^nO$ we have
\begin{equation}
    \frac{m_n^{ab}(\vec q)}{2\pi} = \left\langle \mathrm{ad}_H^n\left(O_{\vec q}^a\right)\, O^b_{-\vec q}\right\rangle_\beta,
\end{equation}
where $\mathrm{ad}_HO \coloneqq [O,H]$. We now define spatial harmonics of the spectral moments: letting $\Lambda^\star$ denote the set of $N$ allowed momenta in the first Brillouin zone of $\Lambda$, we write
\begin{equation}
    m^{ab}_{n,\vec d} = \frac1N\sum_{\vec q\in \Lambda^\star} e^{-\ii \vec q\cdot \vec d}m_n^{ab}(\vec q).
\end{equation}
In the limit $N\to\infty$, the summation becomes identified with an integral, and we may use these interchangeably. Using $\frac1N\sum_{\vec q \in \Lambda^\star} e^{\ii\vec q\cdot(\vec r'-\vec r-\vec d)}=\delta_{\vec r',\vec r+\vec d}$, we obtain
\begin{equation}\label{eq:mnd_appendix}  \frac{m^{ab}_{n,\vec d}}{2\pi} =  \left\langle\mathrm{ad}_H^n \left(O_{\vec 0}^a\right)O_{\vec d}^b\right\rangle_\beta,
\end{equation}
where we have used translation invariance. Next, for a set of sites $X$, let
\begin{equation} \mathcal N_\ell(X) = \{\vec r\in\Lambda:d_\Lambda(\vec r,X)\leq\ell\} 
\end{equation}
denote its $\ell$-neighborhood. A commutator of the form $[A,H]$ receives contributions only from interaction terms whose support intersects $\operatorname{supp}(A)$. Consequently, $\operatorname{supp} \bigl(\operatorname{ad}_H^{\,n}O^a_{\vec0}\bigr) \subseteq \mathcal N_{nR}\bigl(\operatorname{supp}(O^a_{\vec0})\bigr).$ Thus $A_{n,\vec d}^{ab}(\vec\lambda) \coloneqq \operatorname{ad}_{H(\vec\lambda)}^{\,n}\bigl(O^a_{\vec0}\bigr) O^b_{\vec d}$ is supported within the set 
\begin{equation}
\label{eq:suppdefXndab}
X_{n,\vec d}^{ab} = \mathcal N_{nR}\bigl(\operatorname{supp}(O^a_{\vec0})\bigr) \cup \operatorname{supp}(O^b_{\vec d}),
\end{equation}
independently of $N$, at fixed $n,\vec d,a,b$. The coefficients of $A_{n,\vec d}^{ab}(\vec\lambda)$ in an operator expansion are polynomials in $\vec\lambda$. We now turn to the task of controlling the dependence of Eq.~\eqref{eq:mnd_appendix} on $\vec\lambda$.

\subsection{High-temperature expansion and feature map}

\noindent We perform a high-temperature expansion on the DSF spectral moments, writing
\begin{equation}\label{eq:smallbetaexpansion}
m^{ab}_{n,\vec d}(\beta,\vec\lambda) = \sum_i \m^{ab}_{n,\vec d,i}(\vec\lambda)\beta^i.
\end{equation}
We refer to the $\m^{ab}_{n,\vec d,i} = \frac{1}{i!}\pdv[i]{\beta}m_{n,\vec d}^{ab}|_{\beta=0}$ as high-temperature coefficients (or simply coefficients). We will show that these coefficients are analytically computable polynomials of degree $n+i$ and that their expansion converges for $\beta\in[0,\beta_{\mathsf{c}})$ independent of system size. To this end, we first define the \textit{dual interaction graph} $\mathfrak{G}=(V_{\mathfrak{G}},E_{\mathfrak{G}})$~\cite{HaahKothariTangNatPhys}, whose vertices $V_{\mathfrak{G}}$ are pairs $(\alpha,\vec r)$ labeling interaction terms $h_{\vec r}^{\alpha}$. $\mathfrak{G}$ has an edge between two distinct vertices iff their supports intersect:
\begin{equation}
    \{(\alpha,\vec r), (\gamma,\vec r')\} \in E_{\mathfrak{G}} \quad \longleftrightarrow\quad \mathrm{supp}(h_{\vec r}^\alpha)\cap \mathrm{supp}(h_{\vec r'}^\gamma) \neq \emptyset. 
\end{equation}
Let $\mathfrak{d}$ be the maximum degree of $\mathfrak{G}$. Under our locality assumptions, $\mathfrak{d}$ is a constant independent of $N$. For example, in their standard forms, both the one-dimensional transverse-field Ising model and the Kitaev honeycomb model have $\mathfrak d=4$. 
\par 
Following Ref.~\cite{HaahKothariTangNatPhys}, we study the convergence of an operator expectation value on a fixed finite set of sites $X\subset\Lambda$. To carry this out, we adjoin one vertex corresponding to $X$ to $V_\mathfrak{G}$ and to $E_\mathfrak{G}$ we adjoin an edge connecting $X$ to each interaction term that intersects $X$. We define 
\begin{equation}
\begin{aligned}
    \nu(X) &= \#\left\{ (\alpha,\vec r) \mid \mathrm{supp}(h_{\vec r}^\alpha) \cap X \neq \emptyset \right\},\\
\mathfrak{d}_X &= \max\{2,\mathfrak{d}+1,\nu(X)\}.
\end{aligned}
\end{equation}
The augmented graph has a maximum degree at most $\mathfrak{d}_X$. For fixed $X$, $ \mathfrak{d}_X$ is again independent of $N$. We then define a corresponding inverse temperature~\cite{HaahKothariTangNatPhys}
\begin{equation}
    \beta_X = \frac{1}{2e^2h(\mathfrak{d}_X+1)^2}
\end{equation}
which provides a rigorous bound such that for $\beta < \beta_X$, the high-temperature expansion of $\langle A\rangle_\beta$ converges for any operator $A$ with $\mathrm{supp}(A)\subset X$.\par 

Recall that $A_{n,\vec d}^{ab}(\vec\lambda)$ is supported within $X_{n,\vec d}^{ab}$, defined in Eq.~\eqref{eq:suppdefXndab}. For any finite collection of spectral moments, we define the smallest of the $\beta_{X_{n,\vec d}^{ab}}$ as the convergence threshold $\beta_{\mathsf{c}}$. This defines a rigorous convergence threshold for the high-temperature expansion used in this section, though $\beta_{\mathsf{c}}$ may be below the true radius of convergence. \par 

For any polynomial $f(\vec\lambda) = \sum_{\vec k} f_{\vec k}\vec\lambda^{\vec k}$ (where $\vec k = (k_1,\ldots, k_p)$ and $\vec\lambda^{\vec k} = \lambda_1^{k_1}\cdots \lambda_p^{k_p}$), we define a \textit{coefficient norm} by 
\begin{equation}
    \|f\|_{\mathrm{coef}} \coloneqq \sum_{\vec k} |f_{\vec k}|. 
\end{equation}
We note that $|f(\vec \lambda)|\leq \|f\|_{\mathrm{coef}}$ for all $\vec \lambda\in [-1,1]^p$. We then have the following:

\begin{proposition}[Convergent high-temperature expansion for spectral moments]
Fix $n,\vec d,a,b$ and let $X = \mathcal{N}_{nR}(\mathrm{supp} \, O_{\vec 0}^a)\cup \mathrm{supp}\, O_{\vec d}^b$. Then each nonzero coefficient $\m^{ab}_{n,\vec d,i}(\vec\lambda)$ is a homogeneous polynomial of degree $n+i$ in $\vec \lambda$ which can be computed using the terms of $H$ contained in $\mathcal{N}_{iR}(X)$ alone. Moreover, there is a constant $C^{ab}_{n,\vec d}$ independent of $i$ and $N$ such that
\begin{equation}\label{eq:propc1bound}
    \|\m_{n,\vec d,i}^{ab}\|_{\mathrm{coef}} \leq C_{n,\vec d}^{ab}(i+1)\beta_X^{-i}.
\end{equation}
Hence the high-temperature series Eq.~\eqref{eq:smallbetaexpansion} converges absolutely for $\beta < \beta_X$.
\end{proposition}
\begin{proof}
Let us write $A= \mathrm{ad}_H^n( O_{\vec 0}^a) O_{\vec d}^b$ and $m_{n,\vec d}^{ab}=2\pi \langle A\rangle_\beta$. The nested commutators introduce $n$ factors of $H$ and the $i$'th derivative with respect to $\beta$ introduces $i$ factors of $H$. Since $H$ is linear in $\vec \lambda$, each nonzero $\m_{n,\vec d,i}^{ab}$ is a homogeneous polynomial of degree $n+i$ in $\vec \lambda$. By locality, it is clear that this polynomial can be computed from the terms of $H$ contained in $\mathcal{N}_{iR}(X)$ alone, using the fact that surviving contributions are connected clusters, with disconnected contributions cancelling against the partition function in the denominator.\par 
To establish Eq.~\eqref{eq:propc1bound}, we expand $A$ in a fixed orthonormal Hermitian operator basis on $X$, with its expansion coefficients being degree-$n$ polynomials in $\vec\lambda$. We may apply Theorem 3.1 of Ref.~\cite{HaahKothariTangNatPhys} to each normalized basis operator, adjoining it to $H$ with an auxiliary coupling set to zero. The bounds of Ref.~\cite{HaahKothariTangNatPhys} then imply a bound on the coefficient norm of $A$ proportional to $(i+1)\beta_X^{-i}$, with the augmented interaction graph used in defining $\beta_X$ as above. Multiplying by the polynomial coefficients of $A$ and summing over the local operator basis changes only a prefactor, because the coefficient norm is submultiplicative. Absorbing these factors and $2\pi$ into $C^{ab}_{n,\vec d}$ gives 
\begin{equation}
    \|\m_{n,\vec d,i}^{ab}\|_{\mathrm{coef}} \leq C_{n,\vec d}^{ab}(i+1)\beta_X^{-i},
\end{equation}
with $C_{n,\vec d}^{ab}$ independent of $i$ and $N$. Finally, for $\vec\lambda \in [-1,1]^p$ and $\beta < \beta_X$, we have
\begin{equation}
    \sum_{i=0}^{\infty}|\beta|^i \left| \m_{n,\vec d,i}^{ab}(\vec \lambda)\right| \leq C_{n,\vec d}^{ab}\sum_{i=0}^{\infty} (i+1)\left(\frac{\beta}{\beta_X}\right)^i = \frac{C_{n,\vec d}^{ab}}{(1-\beta/\beta_X)^2}<\infty,
\end{equation}
which proves absolute convergence. 
\end{proof}
In conclusion, we have established that each high-temperature coefficient $\m_{n,\vec d, i}^{ab}(\vec \lambda)$ ($n,i \in \mathbb{Z}_{\geq 0}$, $\vec d\in \Lambda$) is a homogeneous polynomial in $\vec \lambda$ of degree $n+i$. Furthermore, using the absolute convergence of the high-temperature series, we can bound the error of any degree-$R$ truncation as follows (writing $q=\beta/\beta_X$):
\begin{equation}
\label{eq:HTtruncerror}
\begin{aligned}
\left|m_{n,\vec d}^{ab}(\beta)-\sum_{i=0}^{R}\beta^i\mathsf m_{n,\vec d,i}^{ab}\right|
&\le C_{n,\vec d}^{ab}\sum_{i=R+1}^{\infty}(i+1)q^i=C_{n,\vec d}^{ab}q^{R+1}\frac{R+2-(R+1)q}{(1-q)^2}.
\end{aligned}
\end{equation}
Eq.~\eqref{eq:HTtruncerror} holds uniformly in system size. For fixed $R$, the error is $O((\beta/\beta_X)^{R+1})$ as $\beta/\beta_X\to 0$.\par 

We will denote a finite collection of $n_f$ high-temperature coefficients by $\F(\vec\lambda)$:
\begin{equation}
    \F(\vec\lambda) := \{ \m_{n,\vec d,i}^{ab}(\vec \lambda) \mid (n,\vec d,i,a,b) \in \mathcal{I}\},
\end{equation}
where $\mathcal{I}$ is a finite set of multi-indices. We call $\F(\vec\lambda)$ the \textit{feature map}, with $|\F|=n_f$ features. We will mostly be interested in the first few moments $n$, lowest harmonics $\vec d$, and lowest $i$. We will distinguish \textit{full DSF access} ($a,b \in \{x,y,z\}$) and \textit{diagonal DSF access} ($a=b \in\{x,y,z\}$).\par

\subsection{Generic local solvability from a fixed feature map}
\label{sec:genericidentifiability}

\noindent Consider a fixed feature map
$\F(\vec\lambda)$ comprising polynomials in the couplings $\vec\lambda\in[-1,1]^p$.
In this section we study the generic solvability of the inverse problem 
\begin{equation}
    \F(\vec\lambda)=\FF
\end{equation}
where $\FF\in \mathbb{C}^{n_f}$ denotes measured DSF high-temperature coefficients, computed from DSF of an unknown Hamiltonian, and $\F(\vec\lambda)$ denotes exact, symbolic high-temperature coefficients. We define and establish conditions under which the couplings are locally and / or globally solvable. In the following, we rely on several results from Ref.~\cite{ChenCotlerHuangQPT}.
\par 

As a minor bookkeeping point, when a coefficient $\m_{n,\vec d,i}^{ab}(\vec\lambda) \in \F(\vec\lambda)$ is complex-valued, we replace it with its real and imaginary parts, regarding each as a real polynomial. We continue to denote the resulting map by $\F$ and its size by $n_f$. We then formally extend these real polynomials to the complex plane to obtain a map $\F: \mathbb{C}^p\to \mathbb{C}^{n_f}$. Call the solution set 
\begin{equation}
    F^{-1}_{\mathrm{th}}(\FF) = \left\{ \vec \lambda \in \mathbb{C}^p \mid \F(\vec \lambda) = \FF \right\}
\end{equation}
the \textit{fiber} of $\F$ over $\FF$. For a given $\FF$, the physical solution set is $F^{-1}_{\mathrm{th}}(\FF) \cap [-1,1]^p$. \par

Let $J\!\F(\vec \lambda)$ denote the $n_f\times p$ Jacobian of the features with respect to the couplings: 
\begin{equation}
    \label{eq:jacobiansupp}
    [J\!\F]_{i,j} = \frac{\partial [\F]_i}{\partial \lambda_j}.
\end{equation}
We assume there are at least as many features as unknown couplings: $n_f\geq p$. We say the couplings are \textit{locally solvable}
at $\vec\lambda$ if $\vec\lambda$ is the only physical solution of $\F(\vec\lambda')=\F(\vec\lambda)$ in a sufficiently small neighborhood of $\vec\lambda$. A sufficient condition is that $J\!\F$ has full column rank at $\vec\lambda$: $\mathrm{rank}\,J\!\F(\vec \lambda)=p$. When this holds, some $p$ components of $\F$ have an invertible Jacobian and therefore admit a local inverse by the inverse function theorem. We say the couplings are \textit{globally solvable} at $\vec\lambda$ if no other solution exists.

\par

\begin{figure}
\includegraphics[width=0.6\linewidth]{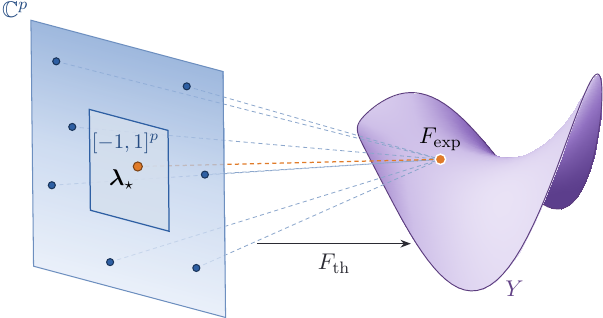}
\caption{\textbf{Polynomial feature map.} The map $\F$ sends coupling vectors to $\FF\in Y$. Several complex coupling vectors may produce the same data $\FF$, forming its fiber. A physical solution is $\vec\lambda_\star \in [-1,1]^p$. This schematic illustrates one physical solution and several nonphysical solutions.}
\label{fig:polynomial}
\end{figure}

We now introduce standard algebraic definitions~\cite{Fulton1969}. An \textit{algebraic set} is a set  given by the intersection of polynomial zero sets.  We will say a statement holds for almost all $\vec\lambda \in [-1,1]^p$, or everywhere except a measure zero set, if the statement holds for all $\vec\lambda\in[-1,1]^p$ except possibly on a proper algebraic subset of $\mathbb{R}^p$. We define the \textit{image variety}
\begin{equation}
    Y = \overline{\F(\mathbb C^p)}^{\,\mathrm{Zar}}\subset \mathbb{C}^{n_f}
\end{equation}
to be the Zariski closure of $\F(\mathbb C^p)$, or equivalently the intersection of all algebraic sets that contain $\F(\mathbb C^p)$ [Fig.~\ref{fig:polynomial}]. A subset $U\subseteq Y$ is \textit{Zariski-open} if its complement in $Y$ is an algebraic set. Having defined local solvability below Eq.~\eqref{eq:jacobiansupp}, we now seek to define \textit{generic local solvability}. We establish the following Proposition, with proof adapted from Ref.~\cite{ChenCotlerHuangQPT}. We will call $\mathsf{g}$ the \textit{generic degree} of the feature map.

\begin{proposition}[Generic local solvability] Suppose that
\begin{equation}
\label{eq:rankassumptionsupp}
    \mathrm{rank}\,J\!\F(\vec \lambda_0) = p
\end{equation}
at one coupling vector $\vec\lambda_0\in[-1,1]^p$. Then, $\F$ is locally solvable for almost all $\vec\lambda \in [-1,1]^p$. Moreover, $\dim Y=p$, and there exist a nonempty Zariski-open subset $U\subseteq Y$ and integer $\mathsf g\geq1$ such that every fiber over $U$ contains exactly $\mathsf g$ complex solutions, at each of which $\mathrm{rank}\,J\!\F=p$.
\label{thm:Generic_identifiability}
\end{proposition}
\begin{proof}
Eq.~\eqref{eq:rankassumptionsupp} implies that some $p\times p$ submatrix of $J\!\F$ has determinant $D(\vec\lambda)$ satisfying $D(\vec\lambda_0)\neq0$. Since $D(\vec\lambda)$ is a nonzero polynomial, its zero set is of measure zero in $[-1,1]^p$. Wherever $D(\vec\lambda)\neq0$, the corresponding $p$ features have an invertible Jacobian. The inverse function theorem therefore gives a local inverse for these features, establishing local solvability.\par 

The remaining statements are standard consequences of generic finiteness of polynomial maps, and follow directly from Lemma E.1 and Theorem E.2 of Ref.~\cite{ChenCotlerHuangQPT}. In particular, full Jacobian rank at one point implies $\dim Y=p$; the generic fiber is zero dimensional and therefore finite; and after restricting to a nonempty Zariski-open subset of $Y$, the number of nonsingular points in the fiber is constant. We denote this number by $\mathsf g$.
\end{proof}

\noindent Consequently, for almost all physical coupling vectors $\vec\lambda$, the fiber over $\F(\vec\lambda)$ contains exactly $\mathsf g$ distinct complex solutions. Prop.~\ref{thm:Generic_identifiability} gives a route to certifying \textit{generic local solvability} with a finite number $\mathsf{g}$ of solutions at generic points. In particular, the existence of a \textit{single} coupling $\vec\lambda_0$ at which the Jacobian has rank $p$ serves as a certificate for the \textit{generic} local solvability across the entire coupling domain. A practical strategy for establishing generic local solvability for any feature map $\F$ is to simply evaluate $\mathrm{rank}\, J\!\F$ at some trial couplings. 

\par 

The generic degree $\mathsf{g}$ counts all complex solutions, whereas physical solutions must lie in $[-1,1]^p$. It follows that $\mathsf g=1$ guarantees uniqueness for generic data, while $\mathsf g>1$ may imply that there are spurious complex solutions outside $[-1,1]^p$---which we omit on physical grounds--- or that the feature map has multiple \textit{physical} solutions.
\par 
To understand the second scenario, we consider a toy example, the transverse-field Ising chain 
\begin{equation}
    H(\vec \lambda) = -J\sum_{i} \sigma^z_i\sigma^z_{i+1} -g \sum_i \sigma^x_i
\end{equation}
with $\vec\lambda = (J,g)$. Suppose we restrict our input data to the diagonal of the DSF, $S^{aa}(\vec q,\omega)$. Let $U= \prod_i \sigma^z_i$. It is straightforward to check that $U H(J,g) U^\dagger = H(J,-g)$ and that $S^{aa}(\vec q,\omega)$ is identical for both $H(J,g)$ and $H(J,-g)$. Therefore any high-temperature coefficients derived from $S^{aa}(\vec q,\omega)$ will suffer from a $g\to -g$ ambiguity; for instance, consider
\begin{equation}
F_{\mathrm{th}}^{\mathrm{TFIM}}(\vec\lambda) \coloneqq \frac{1}{2\pi} (\m_{0,1,1}^{zz},\m_{2,0,0}^{zz}) = (J,4g^2).
\end{equation}
The Jacobian determinant is $8g$ but the fiber contains both $(J,g)$ and $(J,-g)$. In this case, for $g\neq 0$, $F_{\mathrm{th}}^{\mathrm{TFIM}}$ is locally but not globally solvable.  \par 

Since this depends on the measured data comprising the feature map, we dub this ambiguity \textit{measurement equivalence}. We say that physical coupling vectors $\vec\lambda,\vec\lambda'\in[-1,1]^p$ are measurement-equivalent, and denote this by $\vec\lambda'\sim\vec\lambda$, if
\begin{equation}
\F(\vec\lambda')=\F(\vec\lambda).
\end{equation}
A symmetry of the Hamiltonian family $H(\vec\lambda)$ may account for measurement equivalence. For instance, let $G$ be a finite group acting on $\mathbb C^p$ by polynomial automorphisms such that
\begin{equation}
\F(\gamma\cdot\vec\lambda)
= \F(\vec\lambda)
\end{equation}
for every $\gamma\in G$. We emphasize that these transformations need not preserve an individual Hamiltonian belonging to the family. In the preceding example, $G=\mathbb{Z}_2$ and $\gamma: (J,g) \to (J,-g)$, and the measurement equivalence is entirely explained by $G$.

\par 
 We say that the couplings are \textit{uniquely determined up to $G$} if every measurement-equivalent physical coupling vector belongs to $G\cdot\vec\lambda$, i.e. 
\begin{equation}
\vec\lambda'\sim\vec\lambda
\quad\Longrightarrow\quad
\vec\lambda'\in G\cdot\vec\lambda,
\end{equation}
for $\vec\lambda,\vec\lambda'\in[-1,1]^p$. Supposing that the orbit $\{G\cdot \vec\lambda\}$ contains $|G|$ distinct elements for almost all $\vec\lambda$, since this orbit is contained in the fiber of $\F$ over $\FF$, we have that the generic degree satisfies $\mathsf{g}\geq |G|$. If $\mathsf{g}=|G|$, almost all couplings must be uniquely determined up to $G$.\par 

We now apply these definitions to a concrete setting: one- and two-body spin Hamiltonians. We will establish in this case that $\mathsf{g}=1$ for full access to $S^{ab}(\vec q,\omega)$ while for diagonal access to $S^{aa}(\vec q,\omega)$ coupling vectors are measurement equivalent with group $G \supseteq \mathbb{Z}_2^2$. 

\subsection{Learning one- and two-body spin Hamiltonians}

\noindent We now apply the preceding results to translation-invariant spin-$\frac12$ Hamiltonians containing \textit{arbitrary} one- and two-body Pauli interactions. In this example, many preceding results can be made much more concrete. Consider
\begin{equation}
\label{eq:oneandtwobodypauli}
    H(\vec \lambda) = \sum_{\vec r\in \Lambda}b^a\sigma^a_{\vec r} + \sum_{\vec r\in \Lambda,\vec d\in \mathcal{R} } \lambda_{\vec d}^{ab}\sigma_{\vec r}^a\sigma^b_{\vec r+\vec d},
\end{equation}
where $\mathcal{R}$ is a finite set of nonzero bond vectors in the Bravais lattice (with any undirected bond counted once, $\mathcal{R}\cap (-\mathcal{R}) = \emptyset$). We will write the parameter vector $\vec \lambda$ as the concatenation of all three fields $b^a$ and exchange couplings $\lambda_{\vec d}^{ab}$, and set $\vec\lambda \in [-1,1]^p$ as usual with $p=9|\mathcal R|+3$. Below, diagonal spin labels such as $aa$ are not implicitly summed.\par 

We use the high-temperature coefficients $\m_{n,\vec d,i}^{ab}$ defined earlier. A straightforward calculation yields the remarkably simple identity
\begin{equation}
\label{eq:mab2pilambda}
    -\frac{\m_{0,\vec d,1}^{ab}}{2\pi} = \left\langle \sigma_{\vec 0}^a\sigma_{\vec d}^b H\right\rangle_0 = \lambda_{\vec d}^{ab}. 
\end{equation}
Meanwhile, the fields can be obtained using 
\begin{equation}
    \frac{\m^{ab}_{0,\vec 0,1}}{2\pi} = -\left\langle \sigma_{\vec 0}^a\sigma^b_{\vec 0}H\right\rangle_0 = -\ii\, \epsilon_{abc} \left\langle \sigma_{\vec 0}^c H\right\rangle_0 = -\ii\,\epsilon_{abc} b^c,
\end{equation}
which can be solved as
\begin{equation}
\label{eq:bambc}
    b^a = \frac{\ii}{4\pi} \epsilon^{abc} \m^{bc}_{0,\vec 0,1}.
\end{equation}
Note that access to all components of the DSF $S^{ab}$ was important here and only the \textit{static} structure factor $\int\dd\omega \, S^{ab}(\vec q,\omega)$ was necessary. Thus, with full spin-resolved access, the leading high-temperature coefficients of the static structure factor determine all one- and two-body couplings through linear relations. Here, because this linear map has full rank, $\mathsf{g}=1$. \par 

We next consider restricted access to the DSF through only the diagonal spin channels $S^{aa}$, an idealized restricted measurement model motivated by the comparative experimental ease of \textit{unpolarized} neutron scattering~\cite{squires1996introduction}. When restricted to $a=b$, Eq.~\eqref{eq:mab2pilambda} still determines the diagonal exchange couplings
$-\frac{\m_{0,\vec d,1}^{aa}}{2\pi} = \lambda_{\vec d}^{aa}$, but the off-diagonals and $b^a$ require a different approach. We first consider the second spectral moment at infinite temperature:
\begin{equation}
\frac{\m_{2,\vec d,0}^{aa}}{2\pi} = \left\langle \operatorname{ad}_H^2(\sigma_{\vec0}^{a})\sigma_{\vec d}^{a} \right\rangle_0 = -\left\langle [\sigma_{\vec0}^{a},H][\sigma_{\vec d}^{a},H] \right\rangle_0.
\end{equation}
For $\vec d\in\mathcal R$, evaluating this trace yields the principal minors of the coupling matrix: 
\begin{equation}
\begin{aligned} \frac{\m_{2,\vec d,0}^{xx}}{2\pi} &= -8\left( \lambda_{\vec d}^{yy}\lambda_{\vec d}^{zz} - \lambda_{\vec d}^{yz}\lambda_{\vec d}^{zy} \right),\\ \frac{\m_{2,\vec d,0}^{yy}}{2\pi} &= -8\left( \lambda_{\vec d}^{xx}\lambda_{\vec d}^{zz} - \lambda_{\vec d}^{xz}\lambda_{\vec d}^{zx} \right),\\ \frac{\m_{2,\vec d,0}^{zz}}{2\pi} &= -8\left( \lambda_{\vec d}^{xx}\lambda_{\vec d}^{yy} - \lambda_{\vec d}^{xy}\lambda_{\vec d}^{yx} \right). \end{aligned} 
\label{eq:maa2d0}
\end{equation}
From Eq.~\eqref{eq:maa2d0}, we gain information on off-diagonal couplings but not the Zeeman fields $b^a$. The fields do, however, enter the $\vec d=\vec 0$ second-order moments:
\begin{equation} 
\label{eq:maa200}\frac{\m_{2,\vec0,0}^{aa}}{2\pi} = 4\left[ \sum_{c\neq a} (b^c)^2 + \sum_{\vec d\in\mathcal R} \sum_{\substack{c\neq a\\ b=x,y,z}} \left( (\lambda_{\vec d}^{bc})^2 + (\lambda_{\vec d}^{cb})^2 \right) \right]
\end{equation}
where sums over $b,c$ are explicit. To determine the remaining parameters, we utilize the first-order thermal correction to the second spectral moment, finding
\begin{equation}    \frac{\m^{aa}_{2,\vec d,1}}{2\pi} = \left\langle [\sigma_{\vec 0}^a,H][\sigma_{\vec d}^a, H]H\right\rangle_0. 
\end{equation}
An involved calculation yields, for example, the following for the $xx$ channel:
\begin{equation} 
\begin{aligned} 
\frac{\operatorname{Re}\m_{2,\vec d,1}^{xx}}{2\pi} ={}& -4\Big[ \lambda_{\vec d}^{xx} \big( (\lambda_{\vec d}^{yy})^2 + (\lambda_{\vec d}^{yz})^2 + (\lambda_{\vec d}^{zy})^2 + (\lambda_{\vec d}^{zz})^2 \big) \\ &\qquad - \big( \lambda_{\vec d}^{xy}\lambda_{\vec d}^{yx}\lambda_{\vec d}^{yy} + \lambda_{\vec d}^{xy}\lambda_{\vec d}^{zx}\lambda_{\vec d}^{zy} + \lambda_{\vec d}^{xz}\lambda_{\vec d}^{yx}\lambda_{\vec d}^{yz} + \lambda_{\vec d}^{xz}\lambda_{\vec d}^{zx}\lambda_{\vec d}^{zz} \big) \Big] \\ &+ 4\Big[ \lambda_{\vec d}^{yy}(b^z)^2 + \lambda_{\vec d}^{zz}(b^y)^2 - (\lambda_{\vec d}^{yz}+\lambda_{\vec d}^{zy})b^yb^z \Big] + \mathcal T_{\vec d}^{x}. \end{aligned} 
\label{eq:mxx2d1}
\end{equation}
where $\mathcal T_{\vec d}^{x} = 4\sum_{\vec w}\Big[\lambda_{\vec d}^{yy} \lambda_{\vec w}^{zb}\lambda_{\vec d-\vec w}^{b z} + \lambda_{\vec d}^{zz} \lambda_{\vec w}^{yb}\lambda_{\vec d-\vec w}^{b y} - \lambda_{\vec d}^{yz} \lambda_{\vec w}^{zb}\lambda_{\vec d-\vec w}^{b y} - \lambda_{\vec d}^{zy} \lambda_{\vec w}^{yb}\lambda_{\vec d-\vec w}^{b z} \Big]$. Here, we have extended the coupling notation by
$\lambda_{-\vec d}^{ab}=\lambda_{\vec d}^{ba}$, setting it to zero for displacements outside $\mathcal R\cup(-\mathcal R)$, including $\vec0$. Thus, only common neighbors $\vec w$ of bonds contribute. We note that $\mathcal T_{\vec d}^x=0$ when the interaction graph has no triangles. Taken together, the collection
\begin{equation}  
\label{eq:FdiagI}
F_{\mathrm{th}}^{\mathrm{diag}}(\vec \lambda) = \left\{\m_{0,\vec d,1}^{aa}, \m^{aa}_{2,\vec d,0}, \Re \m^{aa}_{2,\vec d,1}\right\}_{a=x,y,z; \vec d\in \mathcal{R}} \cup \left\{ \m^{aa}_{2,\vec 0,0}\right\}_{a=x,y,z}
\end{equation}
constitutes $9|\mathcal{R}|+3$ real polynomial equations in as many unknown parameters. We next establish that $\F^{\mathrm{diag}}(\vec \lambda)$ is generically locally solvable using the method of Prop.~\ref{thm:Generic_identifiability}. 
\begin{theorem}
[Generic local solvability of one- and two-body Pauli Hamiltonians] 
\label{thm:HT2body}
Consider $H(\vec\lambda)$ in Eq.~\eqref{eq:oneandtwobodypauli} with unknown parameters $\vec \lambda \in [-1,1]^{9|\mathcal{R}|+3}$. We have the following:
\begin{itemize}
    \item If the entire DSF $S^{ab}$ is accessible, the first-order high-temperature coefficients of the static structure factor determine all couplings linearly, and $\mathsf{g}=1$. 
    \item If only the diagonal of the DSF $S^{aa}$ is accessible, the couplings are generically locally solvable from $F_{\mathrm{th}}^{\mathrm{diag}}$ [Eq.~\eqref{eq:FdiagI}]. This involves solving a polynomial system of at most cubic degree. 
\end{itemize}
\end{theorem}
\begin{proof}
The full access case follows from Eqs.~\eqref{eq:mab2pilambda}-\eqref{eq:bambc}, which recover the couplings linearly. For diagonal access, Prop.~\ref{thm:Generic_identifiability} reduces the proof to exhibiting one coupling vector at which the Jacobian of $F_{\mathrm{th}}^{\mathrm{diag}}$ has full rank. It is useful first to summarize the dependence of the four families of high-temperature coefficients in Eq.~\eqref{eq:FdiagI}. For $(a,\mu,\nu)\in\{(x,y,z),(y,z,x),(z,x,y)\}$, Eqs.~\eqref{eq:mab2pilambda},\eqref{eq:maa2d0}, \eqref{eq:maa200}, and \eqref{eq:mxx2d1} have the form
\begin{equation}
\begin{aligned}
\frac{\m_{0,\vec d,1}^{aa}}{2\pi}
&=-\lambda_{\vec d}^{aa},\\
\frac{\m_{2,\vec d,0}^{aa}}{2\pi}
&=-8\left(
\lambda_{\vec d}^{\mu\mu}\lambda_{\vec d}^{\nu\nu}
-\lambda_{\vec d}^{\mu\nu}\lambda_{\vec d}^{\nu\mu}
\right),\\
\frac{\operatorname{Re}\m_{2,\vec d,1}^{aa}}{2\pi}
&=
4\left[
\lambda_{\vec d}^{\mu\mu}(b^\nu)^2+
\lambda_{\vec d}^{\nu\nu}(b^\mu)^2
-
(\lambda_{\vec d}^{\mu\nu}+\lambda_{\vec d}^{\nu\mu})
b^\mu b^\nu
\right]
+\text{terms cubic in }\lambda,\\
\frac{\m_{2,\vec0,0}^{aa}}{2\pi}
&=
4\left[(b^\mu)^2+(b^\nu)^2
+\text{terms quadratic in }\lambda\right],
\end{aligned}
\end{equation}
with no implicit summation over $a,\mu,\nu$. We now choose
\begin{equation}
b^x=b^y=b^z=\frac12,
\qquad
\left(\lambda_{\vec d}^{ab}\right)_{ab}
=
\epsilon
\begin{pmatrix}
0&1&-1\\
-1&0&1\\
1&-1&0
\end{pmatrix}_{ab}
\end{equation}
for every $\vec d\in\mathcal R$, with $0<\epsilon\ll1$. For each off-diagonal pair we make the invertible change of variables $(\lambda_{\vec d}^{\mu\nu},\lambda_{\vec d}^{\nu\mu})
\longmapsto
\left(
\lambda_{\vec d}^{\mu\nu}-\lambda_{\vec d}^{\nu\mu},
\lambda_{\vec d}^{\mu\nu}+\lambda_{\vec d}^{\nu\mu}
\right).$ We order the coupling coordinates as the diagonal exchanges, antisymmetric off-diagonal exchanges, symmetric off-diagonal exchanges, and finally $(b^x,b^y,b^z)$. Ordering the rows according to the four families above, direct evaluation gives
\begin{equation}
J\!F^{\rm diag}_{\rm th}
=2\pi
\begin{pmatrix}
-I_{3|\mathcal R|} & 0 & 0 & 0\\[1mm]
0 & -8\epsilon I_{3|\mathcal R|} & 0 & 0\\[1mm]
\mathsf B+O(\epsilon^2) & O(\epsilon^2) &
-I_{3|\mathcal R|}+O(\epsilon^2) & 0\\[1mm]
0 & O(\epsilon) & O(\epsilon) &
4
\mathsf{B}_0
\end{pmatrix}
\end{equation}
where $\mathsf B
=
I_{|\mathcal R|}
\otimes \mathsf{B}_0$ and $\mathsf{B}_0=
\begin{pmatrix}
0&1&1\\
1&0&1\\
1&1&0
\end{pmatrix}$. The matrix is block lower triangular and its four diagonal blocks are invertible for sufficiently small nonzero $\epsilon$. Hence $J\!F^{\rm diag}_{\rm th}$ has full rank at this coupling vector, and Prop.~\ref{thm:Generic_identifiability} implies generic local solvability.\footnote{We credit OpenAI's GPT 5.6 for the diagonal access part of this proof. We verified and refined its output.}
\end{proof}
While the above establishes local solvability, it does not rule out measurement-equivalence for couplings under $F^{\mathrm{diag}}_{\rm th}$. Consider the group of $\pi$ rotations about the three coordinate axes, $\mathbb{Z}_2\times \mathbb{Z}_2 = \{\pm 1\}^3 /\{\pm 1\}$, under which the DSF transforms as $S^{ab}\mapsto s_as_b S^{ab}$, and hence the diagonal DSF $S^{aa}\mapsto S^{aa}$ is invariant ($s_a=\pm 1$). Under this group
\begin{equation}
b^a\mapsto s_a b^a,\qquad
\lambda_{\vec d}^{ab}\mapsto s_as_b\lambda_{\vec d}^{ab}.
\end{equation}
This implies an unavoidable $\mathbb Z_2^2$ measurement equivalence in the case of only diagonal DSF access. In addition, $F_{\mathrm{th}}^{\mathrm{diag}}$ is invariant under $b^a\mapsto-b^a,
\lambda_{\vec d}^{ab}\mapsto\lambda_{\vec d}^{ab}$. Hence, with diagonal access, the inverse problem is measurement equivalent up to a group $G\supset \mathbb{Z}_2^2$ (or, under $F_{\mathrm{th}}^{\mathrm{diag}}$, a group $G\supset \mathbb{Z}_2^3$).

\subsection{Learning higher-body spin Hamiltonians: worked example }
\label{sec:supp_tfim_end2end}

\noindent In this section we explicitly demonstrate the HT protocol with higher-body interactions. We focus on a one-dimensional translation-invariant spin-$\frac12$ chain. This establishes the HT protocol is not limited to two-body Hamiltonians and illuminates its implementation when $p$ is large.
\par
We consider a Hamiltonian comprising contiguous
Pauli operators of length at most three:
\begin{equation}
H(\vec\lambda)=\sum_{w\in\mathcal{W}_{\le 3}}\lambda^w\,H^w,\qquad
H^w=\sum_{r\in\mathbb{Z}} P_{w,r},
\end{equation}
with $\mathcal{W}_{\le 3}$ the set of length $\leq 3$ words comprising $\{\sigma^x,\sigma^y,\sigma^z\}$ and $P_{w,r}$ the
Pauli string corresponding to $w\in \mathcal{W}_{\le 3}$. $H(\vec\lambda)$ has
$|\mathcal{W}_{\le3}|=39$ couplings. To ease numerics, we also impose global $\mathbb{Z}_2$
symmetry generated by $U=\prod_j \sigma^x_j$. A word $w$ is invariant if and only if it contains an even number of $\sigma^y$ and $\sigma^z$ factors. This restriction yields $p=19$ couplings, with
$1$ word of length $1$, $5$ of length $2$, and $13$ of length $3$. The transverse-field Ising and ZXZ models belong to this family. 
\par
We consider local single-site Pauli observables $O_1,O_2\in \{\sigma^x,\sigma^y,\sigma^z\}$. For each displacement $d\in\mathbb{Z}$ and moment order $n\ge 0$, we define the spectral moment harmonics as before,
\begin{equation}
m^{ab}_{n,d}
= 2\pi \big\langle \mathrm{ad}_{H(\vec\lambda)}^{\,n}\!\big(O^a_0\big)\,O^b_d\big\rangle_\beta=\sum_{i\ge 0} \m^{ab}_{n,d,i}(\vec\lambda)\,\beta^i,
\end{equation}
We have $\m^{ab}_{n,d,0}= 2\pi\big\langle \mathrm{ad}_H^{\,n}(\sigma^a_0)\,\sigma^b_d\big\rangle_0$ and $\m^{ab}_{n,d,1}(\vec\lambda)
=-2\pi\big\langle \mathrm{ad}_H^{\,n}(\sigma^a_0)\,\sigma^b_d\,H(\vec\lambda)\big\rangle_0.$ To obtain a concrete finite polynomial system we truncate to a small feature index set:
\begin{equation}
\mathcal I=
\left\{(a,b,n,d,i):
a,b\in\{x,y,z\},\ |d|\leq2,\ 
(n,i)\in\{(0,1),(1,0),(1,1)\}\right\}.
\end{equation}
$\mathcal I$ contains 135 complex features. Splitting real and imaginary parts and removing zero components and conjugate copies leaves 21 independent real features. To evaluate $\F$ systematically, we represent Paulis using the standard
stabilizer encoding $(x|z)\in\mathbb{F}_2^{2N}$ and evaluate commutators symbolically. We compute the coefficients $\m^{ab}_{n,d,i}$ as multivariate
polynomials using \texttt{DynamicPolynomials.jl}. After constructing $\F$, we compute $J\!\F$ symbolically and evaluate it at a random coupling vector $\vec\lambda$. We find that $J\!\F$ has full rank 19. By the previous results, this certifies that $\F$ is generically locally solvable: for generic data, the inverse problem has finitely many complex solutions, and in favorable instances a unique real solution. \par 

We next solve $\F(\vec\lambda)=\FF$. We generated synthetic data by sampling a
random $\vec\lambda_*\in[-1,1]^{19}$ and evaluating the feature vector
$\FF=\{\m^{ab}_{n,d,i}(\vec\lambda_\star)\}_{(a,b,n,d,i)\in\mathcal I}$ explicitly. We then select a
square subsystem of $p=19$ equations from $\F$ and solve numerically over
$\com^{19}$ using \texttt{HomotopyContinuation.jl}, a polynomial solver which uses homotopy methods to identify all roots. Explicitly, we solved
\begin{tiny}
\begin{equation}
\begin{aligned}
c_{1} &= \lambda^{xx} \\
c_{2} &= \lambda^{yy} \\
c_{3} &= \lambda^{zy} \\
c_{4} &= \lambda^{yz} \\
c_{5} &= \lambda^{zz} \\
c_{6} &= 4 \lambda^{zxy} \lambda^{yxz} - 4 \lambda^{yxy} \lambda^{zxz} \\
c_{7} &= 4 \lambda^{xzy} \lambda^{xyz} - 4 \lambda^{xyy} \lambda^{xzz} + 4 \lambda^{zyx} \lambda^{yzx} - 4 \lambda^{yyx} \lambda^{zzx} + 4 \lambda^{zy} \lambda^{yz} - 4 \lambda^{yy} \lambda^{zz} \\
c_{8} &= 4 (\lambda^{xzz})^{2} + 4 (\lambda^{xyz})^{2} + 4 (\lambda^{zxz})^{2} + 4 (\lambda^{yxz})^{2} + 4 (\lambda^{xzy})^{2} + 4 (\lambda^{xyy})^{2} + 4 (\lambda^{zxy})^{2}+ 4 (\lambda^{yxy})^{2} \\&
+ 4 (\lambda^{zzx})^{2} + 4 (\lambda^{yzx})^{2} + 4 (\lambda^{zyx})^{2} + 4 (\lambda^{yyx})^{2} + 4 (\lambda^{zz})^{2} + 4 (\lambda^{yz})^{2} + 4 (\lambda^{zy})^{2} + 4 (\lambda^{yy})^{2} \\
c_{9} &= 4 \lambda^{zzx} \lambda^{xzz} + 4 \lambda^{zyx} \lambda^{xyz} - 4 \lambda^{xxx} \lambda^{zxz} \\
c_{10} &= 4 \lambda^{zxz} \lambda^{xzz} + 4 \lambda^{zxy} \lambda^{xzy} + 4 \lambda^{zzx} \lambda^{zxz} + 4 \lambda^{yzx} \lambda^{yxz} - 4 \lambda^{xxx} \lambda^{xzz} - 4 \lambda^{xxx} \lambda^{zzx} - 4 \lambda^{xx} \lambda^{zz} \\
c_{11} &= 6 (\lambda^{xzz})^{2} + 4 (\lambda^{xyz})^{2} + 6 (\lambda^{zxz})^{2} + 4 (\lambda^{yxz})^{2} + 4 (\lambda^{xzy})^{2} + 2 (\lambda^{xyy})^{2} + 4 (\lambda^{zxy})^{2} + 2 (\lambda^{yxy})^{2} + 6 (\lambda^{zzx})^{2} \\&+ 4 (\lambda^{yzx})^{2} + 4 (\lambda^{zyx})^{2} + 2 (\lambda^{yyx})^{2} + 6 (\lambda^{xxx})^{2} + 4 (\lambda^{zz})^{2} + 2 (\lambda^{yz})^{2} + 2 (\lambda^{zy})^{2} + 4 (\lambda^{xx})^{2} + 2 (\lambda^{x})^{2}\\
c_{12} &= - 4 \lambda^{yzx} \lambda^{xzz} - 4 \lambda^{yyx} \lambda^{xyz} + 4 \lambda^{xxx} \lambda^{yxz} \\
c_{13} &= - 4 \lambda^{yxz} \lambda^{xzz} - 4 \lambda^{yxy} \lambda^{xzy} - 4 \lambda^{zyx} \lambda^{zxz} - 4 \lambda^{yyx} \lambda^{yxz} + 4 \lambda^{xxx} \lambda^{xyz} + 4 \lambda^{xxx} \lambda^{yzx} + 4 \lambda^{xx} \lambda^{yz} \\
c_{14} &= - 2 \lambda^{xyz} \lambda^{xzz} - 2 \lambda^{yxz} \lambda^{zxz} - 2 \lambda^{xzy} \lambda^{xzz} - 2 \lambda^{xyy} \lambda^{xyz} - 2 \lambda^{xyy} \lambda^{xzy} - 2 \lambda^{zxy} \lambda^{zxz} - 2 \lambda^{yxy} \lambda^{yxz} \\&
- 2 \lambda^{yxy} \lambda^{zxy} - 2 \lambda^{yzx} \lambda^{zzx} - 2 \lambda^{zyx} \lambda^{zzx} - 2 \lambda^{yyx} \lambda^{yzx} - 2 \lambda^{yyx} \lambda^{zyx} + 2 \lambda^{zz} \lambda^{xzz} + 2 \lambda^{zz} \lambda^{zzx} \\ 
&+ 2 \lambda^{yz} \lambda^{xyz} + 2 \lambda^{yz} \lambda^{yzx} - 2 \lambda^{yz} \lambda^{zz} + 2 \lambda^{zy} \lambda^{xzy} + 2 \lambda^{zy} \lambda^{zyx} - 2 \lambda^{zy} \lambda^{zz} + 2 \lambda^{yy} \lambda^{xyy} + 2 \lambda^{yy} \lambda^{yyx}\\
&- 2 \lambda^{yy} \lambda^{yz} - 2 \lambda^{yy} \lambda^{zy} + 4 \lambda^{xx} \lambda^{xxx} + 4 \lambda^{x} \lambda^{xx} \\
c_{15} &= - 4 \lambda^{zxz} \lambda^{xyz} - 4 \lambda^{zxy} \lambda^{xyy} - 4 \lambda^{zzx} \lambda^{zxy} - 4 \lambda^{yzx} \lambda^{yxy} + 4 \lambda^{xxx} \lambda^{xzy} + 4 \lambda^{xxx} \lambda^{zyx} + 4 \lambda^{xx} \lambda^{zy} \\
c_{16} &= - 4 \lambda^{zzx} \lambda^{xzy} - 4 \lambda^{zyx} \lambda^{xyy} + 4 \lambda^{xxx} \lambda^{zxy} \\
c_{17} &= 4 \lambda^{yzx} \lambda^{xzy} + 4 \lambda^{yyx} \lambda^{xyy} - 4 \lambda^{xxx} \lambda^{yxy} \\
c_{18} &= 4 \lambda^{yxz} \lambda^{xyz} + 4 \lambda^{yxy} \lambda^{xyy} + 4 \lambda^{zyx} \lambda^{zxy} + 4 \lambda^{yyx} \lambda^{yxy} - 4 \lambda^{xxx} \lambda^{xyy} - 4 \lambda^{xxx} \lambda^{yyx} - 4 \lambda^{xx} \lambda^{yy} \\
c_{19} &= 2 (\lambda^{xzz})^{2} + 4 (\lambda^{xyz})^{2} + 2 (\lambda^{zxz})^{2} + 4 (\lambda^{yxz})^{2} + 4 (\lambda^{xzy})^{2} + 6 (\lambda^{xyy})^{2} + 4 (\lambda^{zxy})^{2} + 6 (\lambda^{yxy})^{2} + 2 (\lambda^{zzx})^{2} \\&
+ 4 (\lambda^{yzx})^{2} + 4 (\lambda^{zyx})^{2} + 6 (\lambda^{yyx})^{2} + 6 (\lambda^{xxx})^{2} + 2 (\lambda^{yz})^{2} + 2 (\lambda^{zy})^{2} + 4 (\lambda^{yy})^{2} + 4 (\lambda^{xx})^{2} + 2 (\lambda^{x})^{2} 
\end{aligned}
\end{equation}
\end{tiny}
where 
\begin{multline}
2\pi\vec c =
\bigl(
-\mathsf m_{0,1,1}^{xx},
-\mathsf m_{0,1,1}^{yy},
-\mathsf m_{0,1,1}^{zy},
-\mathsf m_{0,1,1}^{yz},
-\mathsf m_{0,1,1}^{zz},\\
\operatorname{Re}\mathsf m_{1,2,1}^{xx},
\operatorname{Re}\mathsf m_{1,1,1}^{xx},
\operatorname{Re}\mathsf m_{1,0,1}^{xx},
\operatorname{Re}\mathsf m_{1,2,1}^{yy},
\operatorname{Re}\mathsf m_{1,1,1}^{yy},
\operatorname{Re}\mathsf m_{1,0,1}^{yy},\\
\operatorname{Re}\mathsf m_{1,2,1}^{zy},
\operatorname{Re}\mathsf m_{1,1,1}^{zy},
\operatorname{Re}\mathsf m_{1,0,1}^{yz}
+\operatorname{Im}\mathsf m_{1,0,1}^{yz},\\
\operatorname{Re}\mathsf m_{1,1,1}^{yz},
\operatorname{Re}\mathsf m_{1,2,1}^{yz},
\operatorname{Re}\mathsf m_{1,2,1}^{zz},
\operatorname{Re}\mathsf m_{1,1,1}^{zz},
\operatorname{Re}\mathsf m_{1,0,1}^{zz}
\bigr).
\label{eq:ht19features}
\end{multline}
In random instances for $\vec c$, we found 2,328
complex solutions, with $\sim 12$ real solutions including $\vec\lambda_\star$ to machine precision. By Prop.~\ref{thm:Generic_identifiability}, the generic degree is $\mathsf{g}= 2,328$.

\subsection{Numerics}
\label{sec:htsuppnumerics}

\begin{figure}
    \centering
    \includegraphics[width=0.63\linewidth]{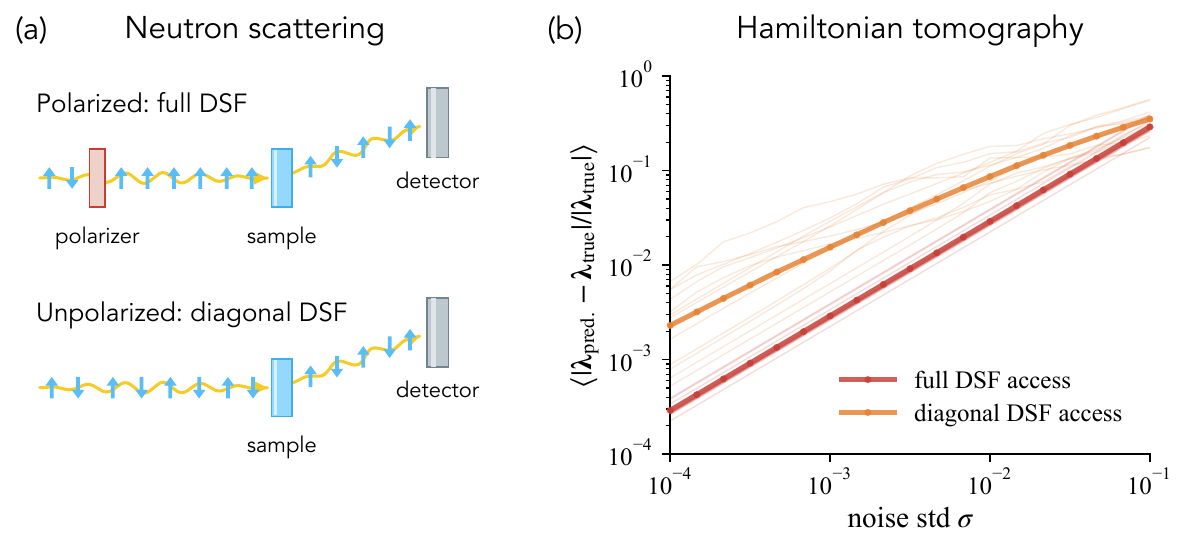}
    \caption{\textbf{High-temperature (HT) Hamiltonian tomography.} (a) Polarized and unpolarized neutron scattering schematics. (b) Recovery of the three field components and nine exchange couplings of
$H=\sum_i\left[\boldsymbol b\cdot\boldsymbol\sigma_i+
\lambda^{ab}\sigma_i^a\sigma_{i+1}^b\right]$
on a periodic chain ($N=12$). We sample 1,000 Hamiltonians with couplings drawn independently and uniformly from $[-0.6,0.6]$; Gaussian noise of standard deviation $\sigma$ was added to HT coefficients. The protocol outlined in Thm.~\ref{thm:HT2body} was performed. Curves show mean relative coupling error; faint lines show example instances.}
    \label{fig:highTtomography}
\end{figure}

\noindent In this section, we numerically test the robustness of the HT protocol to measurement noise. We adopt a simple noise model: additive Gaussian noise on the DSF spectral moments. \par 
First, in Fig.~\ref{fig:highTtomography}, we compare full and diagonal DSF access for a general one- and two-body Hamiltonian as detailed in Thm.~\ref{thm:HT2body}, quantifying performance by the relative coupling error $\frac{|\boldsymbol{\lambda}_{\mathrm{pred}}-\boldsymbol{\lambda}_{\mathrm{true}}|}
{|\boldsymbol{\lambda}_{\mathrm{true}}|}$. In particular, we consider a periodic $N=12$ chain with $H=\sum_i\left[
b^a\sigma_i^a+
\lambda^{ab}\sigma_i^a\sigma_{i+1}^b
\right]$ ($\vec\lambda = \{b^a,\lambda^{ab}\}$). We sample 1,000 Hamiltonians, with i.i.d. $\vec\lambda$ uniform on $[-0.6,0.6]$. \par 

With full DSF access, we learn all twelve parameters using Eqs.~\eqref{eq:mab2pilambda}–\eqref{eq:bambc}. With only diagonal DSF access, we solve the cubic polynomial system in Eq.~\eqref{eq:FdiagI}; we initialize at the noise-free solution, so this experiment tests local noise stability. Before solving, we compute HT coefficients from the DSF and add i.i.d. noise of standard deviation $\sigma$. As expected for a linear system, full DSF access results in an error linear in $\sigma$, while the diagonal DSF access shows ($\sim 10\times$) larger error with greater instance-dependence and nonlinearity. This reflects distinct but appreciable robustness of the cubic polynomial solution to additive noise. \par

\begin{figure}
    \centering
    \includegraphics[width=\linewidth]{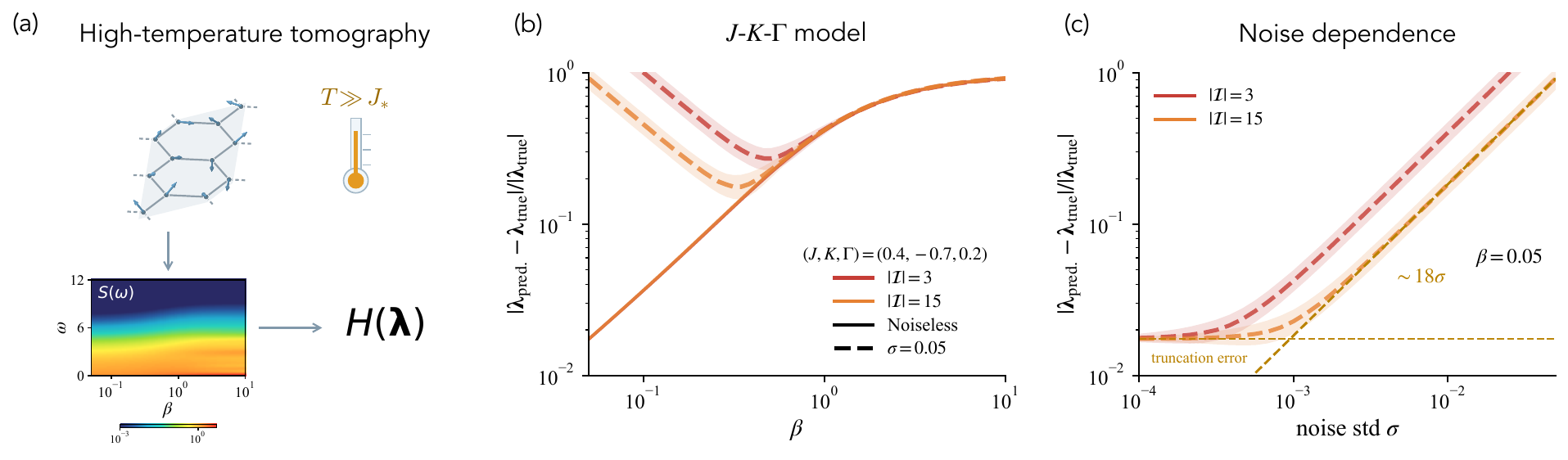}
    \caption{\textbf{$J-K-\Gamma$ model.} (a) Schematic of HT Hamiltonian tomography on $J-K-\Gamma$ model. (b) Performance of HT protocol for $J-K-\Gamma$ model with $(J,K,\Gamma)=(0.4,-0.7,0.2)$, using $n_f=|\mathcal{I}|=3,15$ features. Noise-free (solid) and noisy (dashed, $\sigma=0.05$ Gaussian noise) curves shown. (c) Dependence of (b) on $\sigma$ ($\beta=0.05$). Curves are 5000 noise instance medians with 37.5th--62.5th percentile ribbons.}
    \label{fig:HTJKGamma}
\end{figure}

Next, in Fig.~\ref{fig:HTJKGamma}, we illustrate HT tomography for the $J-K-\Gamma$ model on a honeycomb lattice:
\begin{equation}
\label{eq:JKGammaHTsupp}
H=\sum_{\langle ij\rangle_\gamma}
\left[J(\sigma^x_i\sigma^x_j+\sigma^y_i\sigma^y_j+\sigma^z_i\sigma^z_j)
+K\sigma_i^\gamma\sigma_j^\gamma
+\Gamma(\sigma_i^a\sigma_j^b+\sigma_i^b\sigma_j^a)\right].
\end{equation}
Each bond type has a label $\gamma=x,y,z$, with $a,b$ the complementary axes~\cite{matsuda2025kitaev}. We choose $\vec\lambda_{\mathrm{true}} = (J,K,\Gamma) = (0.4,-0.7,0.2)$. Letting $\vec d_\gamma$ denote the honeycomb bond corresponding to $\langle ij\rangle_\gamma$, we consider two feature maps, with three and fifteen HT coefficients, respectively:
\begin{equation}
\begin{aligned}
    F_{\mathrm{th},3} &= 
\left\{
\mathsf m_{0,\vec d_z,1}^{xx},
\mathsf m_{0,\vec d_z,1}^{zz},
\mathsf m_{0,\vec d_z,1}^{xy}
\right\},\\
F_{\mathrm{th},{15}}
&=
\left\{
\mathsf m_{0,\vec d_\gamma,1}^{aa}
:\gamma,a=x,y,z
\right\}\cup
\left\{
\mathsf m_{0,\vec d_x,1}^{yz},
\mathsf m_{0,\vec d_x,1}^{zy},
\mathsf m_{0,\vec d_y,1}^{xz},
\mathsf m_{0,\vec d_y,1}^{zx},
\mathsf m_{0,\vec d_z,1}^{xy},
\mathsf m_{0,\vec d_z,1}^{yx}
\right\}.
\end{aligned}
\end{equation}
These depend on $J,K,\Gamma$ linearly, with forms listed in Table~\ref{tab:JKG-HT-coefficients}. We choose both to study the effect of redundant features for Hamiltonian learning in the presence of noise: $F_{\mathrm{th},3}$ is necessary and sufficient, but we see in Fig.~\ref{fig:HTJKGamma} that $F_{\mathrm{th},15}$ improves recovery in the presence of noise.  \par 

In both cases, we solve $\vec c
=-\beta\mathsf M\vec{\lambda}$ for a coefficient matrix $\mathsf M$ defined using Table~\ref{tab:JKG-HT-coefficients}. We add i.i.d. noise to $\vec c$ and reconstruct $\vec\lambda$ by ordinary least squares,
$\vec{\lambda}_{\mathrm{pred}}
=-\frac{\mathsf M^{+}\widetilde{\vec c}}{\beta}$, where $\mathsf M^{+}$ is the Moore-Penrose inverse. In  Fig.~\ref{fig:HTJKGamma}b, we vary $\beta$ with $\sigma=0.05$. Without noise, decreasing $\beta$ improves recovery by reducing the $O(\beta)$ truncation error. With noise, however, dividing by $\beta$ amplifies measurement errors. The competition between these two effects produces an intermediate-temperature optimum. Finally, in Fig.~\ref{fig:HTJKGamma}c we vary $\sigma$ at fixed $\beta=0.05$. At small $\sigma$, both curves approach a floor of approximately $1.7\%$; when noise dominates, errors grow linearly with $\sigma$.

\begin{table}
\centering
\begin{tabular}{c|ccc|l}
\hline
bond $\gamma$ & $ab=xx$ & $yy$ & $zz$ & Off-diag. \\
\hline
$x$ & $J+K$ & $J$ & $J$ & $yz,zy:\ \Gamma$ \\
$y$ & $J$ & $J+K$ & $J$ & $xz,zx:\ \Gamma$ \\
$z$ & $J$ & $J$ & $J+K$ & $xy,yx:\ \Gamma$ \\
\hline
\end{tabular}
\caption{Coefficients $-\mathsf m_{0,\vec d_\gamma,1}^{ab}/(2\pi)$
for Eq.~\eqref{eq:JKGammaHTsupp} ($a,b=x,y,z$). Unlisted components vanish.}
\label{tab:JKG-HT-coefficients}
\end{table}

\section{Low-temperature obstruction to learning}
\label{sec:supphardness}

\noindent In this section, we establish that, in the worst case, it is exponentially hard in the inverse temperature $\beta$ to learn the Hamiltonian from DSF data at low temperatures. We prove this assertion through a simple example. In particular, consider a Hamiltonian defined on two qubits,
\begin{equation}
\label{eq:hardnessH}
    H = \frac h2(\sigma^z_1+\sigma^z_2)+\frac U4(1+\sigma^z_1)(1+\sigma^z_2) + h
\end{equation}
where $h,U>0$. We'll denote the eigenstates of $\sigma^z_i$ by $\ket{\uparrow}$ and $\ket{\downarrow}$, with eigenvalues $+1$ and $-1$, respectively. The additive constant is chosen to set the ground-state energy to zero and does not affect physical statements. This Hamiltonian---which is entirely classical---has energies
\begin{equation}
   E_{\downarrow\downarrow}=0,\qquad
E_{\uparrow\downarrow}=E_{\downarrow\uparrow}=h,\qquad
    E_{\uparrow\uparrow}=2h+U.
\end{equation}
We assume $h>0$ is known and let the unknown coupling $U$ vary. Since the system hosts only two sites, we drop the $\vec q$ label and simply work in real space. The DSF takes the real-space form 
\begin{equation}
S^{ab}(\omega)
=2\pi
\sum_{m,n}p_m
\langle m|O^a|n\rangle
\langle n|O^b|m\rangle
\delta\!\left(\omega-(E_n-E_m)\right),
\end{equation}
where $O^a,O^b\in \{\sigma^x_1,\sigma^x_2,\sigma^y_1,\sigma^y_2,\sigma^z_1,\sigma^z_2\}$ are Pauli operators and $p_m=\frac{e^{-\beta E_m}}{\mathcal{Z}}$, with $\mathcal Z=1+2e^{-\beta h}+e^{-\beta(2h+U)}$ the partition function.

We remark that while this example is defined for a two-site system for maximal simplicity, one could easily generalize (e.g., by tiling a lattice) to infinitely many sites; our hardness-of-learning results extend straightforwardly to the latter setting.  
\par 
At zero temperature, only $\ket{\downarrow\downarrow}$ contributes to the DSF, yielding 
\begin{equation}
 S^{ab}(\omega) =2\pi \sum_{n} \langle \downarrow\downarrow|O^a|n\rangle\langle n|O^b|\downarrow\downarrow\rangle \, \delta(\omega - E_n). 
    \label{eq:S_0}
\end{equation}
For each single-site operator $O^a\in \{\sigma^x_1,\sigma^x_2,\sigma^y_1,\sigma^y_2,\sigma^z_1,\sigma^z_2\}$ we clearly have $\braket{\downarrow\downarrow|O^a}{\uparrow\uparrow}= 0$. Therefore $\ket{\uparrow\uparrow}$ does not contribute in Eq.~\eqref{eq:S_0}, and $S^{ab}(\omega)$ is strictly $U$-independent at $T=0$:
\begin{equation}
    \pdv{U} S^{ab}(\omega) = 0. 
\end{equation}
Hence even perfect knowledge of the DSF cannot help us learn $U$. This simple result establishes proof of principle that Hamiltonian learning from the DSF is, in general, not possible at $T=0$.\par 

At finite temperature, in this simple model we can compute the DSF explicitly. For example, the on-site transverse
DSF is
\begin{equation}
\label{eq:hardness_exact_dsf}
\begin{aligned}
S_{11}^{xx}(\omega)
=
\frac{2\pi}{\mathcal Z}\Big[
&\delta(\omega-h)
+e^{-\beta h}\delta(\omega+h)+e^{-\beta h}\delta(\omega-h-U)
+e^{-\beta(2h+U)}\delta(\omega+h+U)
\Big],
\end{aligned}
\end{equation}
with $S_{22}^{xx}=S_{11}^{xx}$ and
$S_{ii}^{yy}=S_{ii}^{xx}$. Moreover,
\begin{equation}
S_{11}^{zz}(\omega)=S_{22}^{zz}(\omega)=2\pi\delta(\omega),
\qquad
S_{12}^{zz}(\omega)
=
\frac{1-2e^{-\beta h}+e^{-\beta(2h+U)}}{\mathcal Z}2\pi
\delta(\omega).
\end{equation}
We observe that the dependence on $U$ always occurs in terms with a thermal suppression factor $e^{-\beta h}$ (with a smaller dependence
on $U$ through the denominator). In other words, denoting by $\widetilde S^{ab}(\omega)$ the DSF with each $\delta$-function broadened to a finite-width Gaussian, we have
\begin{equation}
\label{eq:partialStildesmall}
    \left|\pdv{U}\widetilde S^{ab}(\omega) \right|\leq C e^{-\beta h}
\end{equation}
as $\beta \to \infty$ for some constant $C$. (Owing to subtleties of the $\mathcal{Z}^{-1}$ and $\delta$-functions, proving Eq.~\eqref{eq:partialStildesmall} requires some care, and we leave the derivation to the end of this section). The intuition for this result is that the only way for $S^{ab}$ to ``know" about $U$ is through the states $\ket{\downarrow\uparrow}$, $\ket{\uparrow\downarrow}$, and $\ket{\uparrow\uparrow}$, since there are no other eigenstates with single-Pauli matrix elements with $\ket{\uparrow\uparrow}$, and only the energy of $\ket{\uparrow\uparrow}$ depends on $U$. The contributions of these states, at low temperatures, are all thermally suppressed by $e^{-\beta h}$. \par 

\begin{figure}[t]\centering
\includegraphics[width=0.8\columnwidth]{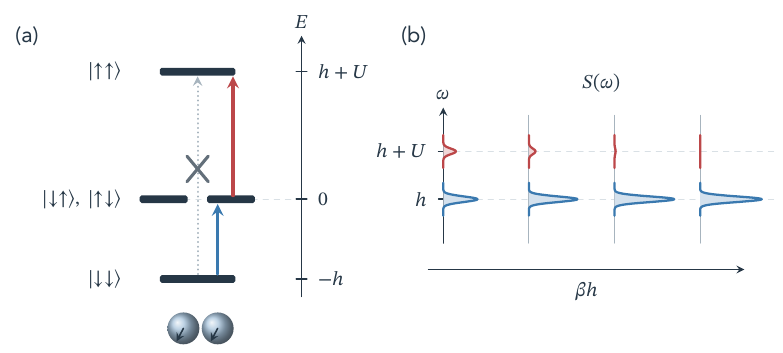}
\caption{\textbf{Hardness at low temperatures.} (a) Energy levels of Eq.~\eqref{eq:hardnessH},
with $h,U>0$. 
A single spin operator connects levels separated by $h$ (blue)
and $h+U$ (red), but cannot connect $\ket{\downarrow\downarrow}$ directly to
$\ket{\uparrow\uparrow}$ (crossed arrow). (b) Schematic dynamical structure factor $S(\omega)$ for increasing
$\beta h$, where $\beta$ is the inverse temperature.
The peak at $h+U$ relies on thermal occupation,
and its weight relative to the peak at $h$ is suppressed by
$e^{-\beta h}$. At $\beta=\infty$ the DSF is independent of $U$.}
\label{fig:hardness}
\end{figure}

Therefore, as sketched in Fig.~\ref{fig:hardness}, $S^{ab}(\omega)$ depends \textit{exponentially weakly} on $U$ at low temperatures, and hence learning $U$ is exponentially difficult in the presence of experimental noise. Alternatively, assume we have access to independent noisy measurements of the smeared DSF integrals in Eq.~\eqref{eq:nonzeroT}, with additive Gaussian noise of standard deviation $\sigma$. Averaging $M$ observations reduces the noise to $\sigma/\sqrt{M}$ by central limit scaling. In order to resolve $U$ to a precision $\delta$, we need $O(Ce^{-\beta h}\delta ) \gtrsim O( \sigma/\sqrt{M})$ and hence $M = \Omega\left(\frac{\sigma^2e^{2\beta h}}{C^2\delta^2}\right)$, exponential in $\beta h$. \par

The example above is inspired by Ref.~\cite{haah2022optimal}, which considered two Hamiltonians: 
\begin{equation}
    H_0 = \begin{pmatrix}
        -2 &&&\\
        &0&&\\
        &&1&\\
        &&&1
    \end{pmatrix},\quad  H_1 = \begin{pmatrix}
        -2 &&&\\
        &0&&\\
        &&1+2\varepsilon&\\
        &&&1-2\varepsilon
    \end{pmatrix},
\end{equation}
and noted that any protocol for learning Hamiltonians from Gibbs state copies within $\varepsilon$ must be able to distinguish these two. Using the KL divergence between the Gibbs states of $H_0,H_1$, Ref.~\cite{haah2022optimal} proved $\Omega(e^{2\beta}/\beta^2\varepsilon^2)$ Gibbs state copies are necessary.

We now derive Eq.~\eqref{eq:partialStildesmall}. In fact, we derive a slightly more general statement, 
\begin{equation}\label{eq:nonzeroT}
    \left| \pdv{U} \int \frac{\mathrm{d}\omega}{2\pi} \, \varphi(\omega) S^{ab}(\omega) \right| \leq C e^{-\beta h}
\end{equation}
as $\beta \to \infty$, for some $\beta$-independent constant $C$, where  $\varphi$ is a smooth function. For ease of notation, let $\ket{d}=\ket{\uparrow\uparrow}$
and $\delta_{m,d}=1$ if $\ket{m}=\ket{d}$,
and zero otherwise. Since only the energy of $\ket{d}$
depends on $U$,
$E_m(U)=E_m(0)+U\delta_{m,d}$.
We also define $ W^{ab}_{mn} = \langle m|O^a|n\rangle\langle n|O^b|m\rangle.$ We consider
\begin{equation}
    S^{ab}_\varphi = \int \frac{\mathrm{d}\omega}{2\pi} \, \varphi(\omega) S^{ab}(\omega)= \sum_{mn} p_m(U) W_{mn}^{ab}\varphi(E_n(U)-E_m(U))
\end{equation}
where $p_m(U) = e^{-\beta E_m(U)}/\mathcal{Z}$. Differentiating gives 
\begin{equation}
    \partial_U S_\varphi^{ab} = \sum_{mn} W^{ab}_{mn}\left[ \partial_U p_m(U) \varphi(E_{nm}) + p_m(U) (\delta_{n,d}-\delta_{m,d})\varphi'(E_{nm})  \right]
\end{equation}
where $E_{nm} = E_n-E_m$. We bound the two terms separately. We have $\partial_U e^{-\beta E_m(U)} = -\beta \delta_{m,d} e^{-\beta E_m(U)}$ and $\partial_U \mathcal{Z} = -\beta \mathcal{Z} p_d(U)$ and hence \begin{equation}
    \partial_U p_m(U) = \frac{-\beta \delta_{m,d} e^{-\beta E_m}\mathcal{Z} - e^{-\beta E_m} \partial_U \mathcal{Z}}{\mathcal{Z}^2} = -\beta p_m(U)( \delta_{m,d}  -p_d).
\end{equation}
Considering $m=d$ and $m\neq d$ separately gives $|\partial_U p_m(U)|\leq\beta p_d(U)
\leq\beta e^{-\beta(2h+U)}$. Since we assumed $h,U\geq 0$, using $xe^{-x} \leq \frac1e$ for $x\geq 0$, we have $\beta e^{-2\beta h -\beta U} \leq \beta e^{-2\beta h} \leq \frac{1}{eh} e^{-\beta h}. 
$ Hence 
\begin{equation}
    |\partial_U p_m(U)| \leq \frac{1}{eh}e^{-\beta h}. 
\end{equation}
For the second term, note that $\delta_{n,d}-\delta_{m,d}$ is nonzero only if exactly one of the states $\ket{n},\ket{m}$ is equal to $\ket{\uparrow\uparrow}$. A one-body Pauli has zero matrix element between $\ket{\downarrow\downarrow}$ and $\ket{\uparrow\uparrow}$. Hence, when $W^{ab}_{mn} (\delta_{n,d}-\delta_{m,d}) \neq 0,$ the state not equal to $\ket{\uparrow\uparrow}$ must be one of $\ket{\downarrow\uparrow}$ or $\ket{\uparrow\downarrow}$. Therefore 
\begin{equation}
    p_m(U) \leq \max(p_d(U), e^{-\beta h}/\mathcal{Z}) \leq e^{-\beta h}. 
\end{equation}
Combining the two bounds, we obtain
\begin{equation}
    \left| \partial_U S_\varphi^{ab}\right| \leq C_\varphi e^{-\beta h}
\end{equation}
as was to be shown, where one may take, for instance, $C_\varphi = 16\left(
\max_{m,n}|\varphi'(E_{mn})|
+\frac{1}{eh}\max_{m,n}|\varphi(E_{mn})|
\right).$
This may depend on $h,U$ but is independent of $\beta$. \par

\section{Strong-field Hamiltonian tomography}
\label{sec:suppprobefield}

\noindent In this section, we give an in-depth treatment of Hamiltonian learning from the DSF using a strong applied Zeeman field. Throughout this Appendix, Hamiltonians are assumed to be of the form 
\begin{equation}
    H_{\b}^{(\hat n)}(\vec \lambda) = -\b\sum_{\vec r} \hat n\cdot \vec \sigma_{\vec r} + H(\vec\lambda)
\end{equation}
where $\vec r\in\Lambda$, $\vec\sigma_{\vec r}$ labels spins on which $H(\vec\lambda)$ acts, and $\hat n$ is a unit vector defining the field direction. In \textsection\ref{sec:yarotsky}-\ref{sec:sfpnumerics}, we assume $\hat n$ points along discrete directions, and in \textsection\ref{sec:angular_tomography}, we generalize this to continuously varying directions for the protocol of \textit{angular tomography}. \par 

\subsection{Motivation and examples}

\noindent Our primary motivation is the following. Recall that the Hamiltonian Eq.~\eqref{eq:hardnessH} is exponentially hard to learn at low temperatures $T =1/\beta$, presenting a no-go for general Hamiltonian learning at low $T$. It is natural to investigate whether such hardness persists under perturbations~\cite{ChenCotlerHuangQPT}. In strong-field Hamiltonian tomography, we focus on \textit{strong} perturbations due to the analytical control they afford. Perhaps the simplest, experimentally accessible example is a global Zeeman field. Indeed, in quantum magnetism, polarization by a strong Zeeman field is an established technique for extracting Hamiltonian parameters~\cite{ColdeaTennantHabicht2002PRL}, and in quantum simulators single-site controls are typically easiest. As a word of caution, we note that for many quantum materials the strong-field limit may not be available because the exchange couplings may be too large (e.g. tens of meVs). 

\par 

Complementing and substantially expanding upon Ref.~\cite{ColdeaTennantHabicht2002PRL}, we will provide a general, rigorous treatment of Hamiltonian learning from a strong Zeeman field which is in principle unconstrained by the weight of interaction terms or number of couplings. Denoting the Hamiltonian in Eq.~\eqref{eq:hardnessH} of App.~\ref{sec:supphardness} by $H$, we illustrate the basic idea by orienting a Zeeman field along $\hat x$ and studying
\begin{equation}\label{eq:HB2qubit}
    H_\b^{(\hat x)} = -\b(\sigma^x_1+\sigma^x_2) + H,
\end{equation}
treating $|h|,|U| \ll \b$ as perturbations. The unperturbed eigenstates are
\begin{equation}
    \ket{g} = \ket{++},\,\, \ket{s} = \frac{\ket{-+}+\ket{+-}}{\sqrt{2}},\,\, \ket{a} = \frac{\ket{-+}-\ket{+-}}{\sqrt{2}},\,\, \ket{e} = \ket{--}
\end{equation}
in the $\sigma^x$ basis, with $E_g^{(0)} = -2\b, E_s^{(0)}=E_a^{(0)} = 0, E_e^{(0)} = 2\b$. Two transition energies are 
\begin{align}
    \Delta_s &\coloneqq E_s -E_g = 2\b + \frac{U}{4} + \frac{1}{\b} \left[\frac{(2h+U)^2+(U/2)^2}{16} \right] + O(\b^{-2})\\
    \Delta_a &\coloneqq E_a-E_g = 2\b - \frac{U}{4}+ \frac{1}{\b} \left[\frac{(2h+U)^2+(U/2)^2}{16} \right] + O(\b^{-2}).
\end{align}
Since there are only two sites, we work in real space and write $S_{ij}$ for spin correlations between sites $i,j=1,2$. We recall that at $T=0$, $U$ was previously unlearnable. In contrast, here we compute $S_{ij}^{zz}(\omega)/2\pi = \sum_{m\neq g}\langle g|\sigma^z_i|m\rangle\langle m|\sigma^z_j|g\rangle \delta(\omega-E_{mg})$ at $T=0$ and find, to second order,
\begin{equation}
\label{eq:S11-two-site}
\begin{aligned}
\frac{S_{11}^{zz}(\omega)}{2\pi}
=&
\left[
\frac12-\frac{U}{16\b}
-\frac{(2h+U)^2}{32\b^2}
\right]
\delta(\omega-\Delta_s)
+
\left[
\frac12+\frac{U}{16\b}
-\frac{(2h+U)^2}{32\b^2}
\right]
\delta(\omega-\Delta_a)
+
O\!\left(\b^{-3}\right),\\
\frac{S_{12}^{zz}(\omega)}{2\pi}
=&
\left[
\frac12-\frac{U}{16\b}
-\frac{(2h+U)^2}{32\b^2}
\right]
\delta(\omega-\Delta_s)
-
\left[
\frac12+\frac{U}{16\b}
-\frac{(2h+U)^2}{32\b^2}
\right]
\delta(\omega-\Delta_a)
+
O\!\left(\b^{-3}\right),
\end{aligned}
\end{equation}
dropping an unimportant elastic term $\propto \delta(\omega)$. We define spectral moments $m_{ij,n}^{zz} = \int \dd \omega \, \omega^n\, S_{ij}^{zz}(\omega)$ and find by a perturbative calculation
\begin{equation}
\begin{aligned}
\frac{m_{11,n}^{zz}}{2\pi}
&=
(2\b)^n
\left[
1
+
\frac{n-2}{\b^2}
\left(
\frac{(2h+U)^2}{32}
+
\frac{nU^2}{128}
\right)
+
O\!\left(\b^{-3}\right)
\right],\\
\frac{m_{12,n}^{zz}}{2\pi}
&=
(2\b)^n
\left[
\frac{(n-1)U}{8\b}
+
O\!\left(\b^{-3}\right)
\right].
\end{aligned}
\end{equation}
$m_{12,0}^{zz}$ and $m_{11,0}^{zz}$ provide two independent equations in two unknowns and thus $\b \hat x$ renders $h,U$ learnable at $T=0$ from the DSF, illustrating how a field can overcome this obstruction to learning.

Before treating more general cases, we work through another example, this time in the thermodynamic limit. Consider the translation-invariant spin-$\frac12$ chain 
\begin{equation}\label{eq:HXYZBfield}
H_\b^{(\hat x)}=
-\b\sum_r\sigma^x_r+ 
\sum_r
\left(
J_x\sigma^x_r\sigma^x_{r+1}+J_y\sigma^y_r\sigma^y_{r+1}+J_z\sigma^z_r\sigma^z_{r+1}
\right)\coloneqq -\b\sum_r\sigma^x_r+H,
\end{equation}
with $J \coloneqq |J_x|+|J_y|+|J_z|\ll \b$. The unperturbed ground state is $\ket{0} = \ket{+}^{\otimes N}$ (we take periodic boundary conditions and later $N\to \infty$) with energy $E_g^{(0)} = -\b N$. Define single-spin-flip states
\begin{equation}
    \ket{r} = \ket{+}\cdots \ket{-}_r \cdots \ket{+}. 
\end{equation}
These have unperturbed energy $ E_1^{(0)} = E_g^{(0)} + 2\b.$ Define single-flip momentum states and Fourier-space Paulis $\ket{q} = \frac{1}{\sqrt{N}} \sum_r e^{-\ii qr} \ket{r}, \sigma^z_q = \frac{1}{\sqrt{N}} \sum_r e^{-\ii qr}\sigma^z_r.$ We focus on the $T=0$ DSF
\begin{equation}
\label{eq:dsfxyzzeemanfield}
\begin{aligned}
\frac{S^{zz}(q,\omega)}{2\pi} &= \sum_{m\neq g} |\langle m|\sigma^z_q|g\rangle|^2 \delta(\omega- E_{mg})
\end{aligned}
\end{equation}
where $\ket{g}$ is the exact ground state of $H_\b^{(\hat x)}$ and $m$ runs over exact eigenstates.\par 

To leading order in $J/\b$, the DSF is dominated by the one-spin-flip sector $\mathcal{H}_1\coloneqq\mathrm{span}\{\ket r\}$. Let $P_1$ project onto $\mathcal{H}_1$; then $P_1 HP_1 = J_x (N-4) P_1 +(J_y+J_z)\sum_r (\ket{r}\bra{r+1}+\ket{r+1}\bra{r}).$ This is diagonal in the $\ket{q}$ basis with energies $E_q^{(1)} = J_x(N-4) + 2(J_y+J_z)\cos q.$ Next, let $\ket{r,r'}$ be the state with $\sigma^x$-basis spin flips at sites $r,r'$:
\begin{equation}
    \ket{r,r'} = \ket{+}\cdots \ket{-}_r \cdots\ket{-}_{r'} \ket{+}. 
\end{equation}
First-order perturbation theory yields $\ket{g} = \ket{0} - \frac{J_z-J_y}{4\b} \sum_r \ket{r,r+1} + O(J^2/\b^2).$ Next, we use $P_1\sigma^z_q \ket{r,r+1} = N^{-1/2} (e^{-\ii qj}\ket{r+1} + e^{-\ii q(r+1)}\ket{r})$ and hence $\langle q|\sigma^z_q |r,r+1\rangle = N^{-1} (e^{\ii q}+e^{-\ii q}) = 2\cos q/N$. Summing over $r$, we obtain $|\langle q|\sigma^z_q|g\rangle|^2 = 1-\frac{(J_z-J_y)\cos q}{\b} + O(J^2/\b^2)$. Define the transition energy $\Delta(q) \coloneqq 2\b - 4J_x +2(J_y+J_z)\cos q + O(J^2/\b)$. Combining the above, we arrive at 
\begin{equation}
    \begin{aligned} 
\frac{S^{zz}(q,\omega)}{2\pi}
&=
\left[
1-\frac{(J_z-J_y)\cos q}{\b}
\right]
\delta\!\left(
\omega-\Delta(q)
\right)
+
O\!\left(\frac{J^2}{\b^2}\right).
\end{aligned}
\end{equation}
The frequency moments $m_n^{zz}(q)
=
\int \dd\omega\,\omega^n S^{zz}(q,\omega)$
follow as
\begin{equation}
\begin{aligned}
\frac{m_n^{zz}(q)}{2\pi}
&=
(2\b)^n
\bigg[
1
+
\frac{1}{\b}
\bigg(
\frac{n}{2}
\left[
-4J_x+2(J_y+J_z)\cos q
\right]
-
(J_z-J_y)\cos q
\bigg)
+
O\!\left(\frac{J^2}{\b^2}\right)
\bigg].
\end{aligned}
\end{equation}
\par 

The zeroth and first spectral moments suffice to
reconstruct all three couplings. Taking their spatial
harmonics, we obtain
\begin{equation}
\begin{aligned}
\frac{\b\,m_{0,1}^{zz}}{2\pi}
&=\frac{J_y-J_z}{2}+O(\b^{-1}),\\
\frac{m_{1,0}^{zz}-2\b m_{0,0}^{zz}}{2\pi}
&=-4J_x+O(\b^{-1}),\\
\frac{m_{1,1}^{zz}-2\b m_{0,1}^{zz}}{2\pi}
&=J_y+J_z+O(\b^{-1}).
\end{aligned}
\end{equation}
These three independent linear relations determine
$J_x,J_y,J_z$, demonstrating that the XYZ chain
can be reconstructed using a single field direction
and diagonal DSF access.

\subsection{Controlled strong-field expansion in the thermodynamic limit}
\label{sec:yarotsky}
\par

\noindent It is not \textit{a priori} clear that a perturbative expansion of the spectral moments in $1/\b$ converges in the thermodynamic limit. In analogy to App.~\ref{sec:hightemp}, where we invoked the convergence of the high-temperature expansion for local lattice models below a sufficiently small $\beta_{\mathsf{c}}$, we seek rigorous guarantees on the convergence of the $1/\b$ expansion. We establish this now. \par 
To set the stage, consider a translation-invariant finite-range
Hamiltonian on $\mathbb Z^d$ of the form
\begin{equation}
H_\b(\vec\lambda)
=
-\b\sum_{\vec r}\sigma^x_{\vec r}
+
H(\vec\lambda),
\qquad
H(\vec\lambda)
=
\sum_{\vec r'}h_{\vec r'}(\vec\lambda),
\label{eq:strong-field-general}
\end{equation}
where $h_{\vec r'}$ is a bounded operator with
support contained in a fixed-radius neighborhood of $\vec r'$. We
assume $H$ depends analytically on
couplings $\vec\lambda\in [-1,1]^p$ and write $J_\star
=
\sup_{\vec\lambda}
\|h_{\vec 0}(\vec\lambda)\|$ ($J_\star/\b\ll 1$). Up to an irrelevant additive constant, we can rescale the Hamiltonian to $\b^{-1}H_\b(\vec\lambda) = \sum_{\vec r}(1-\sigma^x_{\vec r})+ \b^{-1}H(\vec\lambda).$ At $\b=\infty$, the first term is a sum of commuting on-site
terms, with unique product ground state
$\ket{+}^{\otimes\infty}$ and gap $2$. The second term is a finite-range perturbation with bounded local strength $J_\star/\b$. Weakly perturbed classical Hamiltonians of precisely this form were studied by Refs.~\cite{yarotsky2006ground,yarotsky2005uniqueness}, where a $\b_{\mathsf{c}}$ independent of system size was established such that
for $\b>\b_{\mathsf{c}}$ the ground state is unique and the spectral gap is bounded below by a constant independent of system size. In this regime, Refs.~\cite{yarotsky2006ground,yarotsky2005uniqueness} prove connected
correlations decay exponentially and expectation values of local
observables depend \textit{analytically} on $1/\b$ and $\vec\lambda$. \par 

We apply these analyticity results to the DSF spectral moments of $H_\b(\vec\lambda)$. The $n$'th spectral moment harmonic of operators separated by $\vec d$ is a ground-state expectation value of a local operator. For $n=0$, $m_{0,\boldsymbol d}^{AA'}/2\pi
=
\langle A_{\boldsymbol 0}A'_{\boldsymbol d}\rangle$ and for $n\geq 1$, 
\begin{equation}
m_{n,\boldsymbol d}^{AA'}
=
2\pi\left\langle
\operatorname{ad}_{H_\b}^{n}(A_{\boldsymbol0})
A'_{\boldsymbol d}
\right\rangle,
\label{eq:strong-field-moment-commutator}
\end{equation}
where $\operatorname{ad}_H(O)\coloneqq[O,H]$. Because $H_\b$ is finite range, the nested commutator in Eq.~\eqref{eq:strong-field-moment-commutator} has finite support, independent of system size. Moreover, $\b^{-n}m_{n,\boldsymbol d}^{AA'}
=
2\pi\left\langle
\operatorname{ad}_{H_\b/\b}^{n}(A_{\boldsymbol0})
A'_{\boldsymbol d}
\right\rangle$. Therefore, for sufficiently large
$\b$, each normalized moment is analytic in $1/\b$ and
$\vec\lambda$. Thus
\begin{equation}
\b^{-n}m_{n,\boldsymbol d}^{AA'}
=
\sum_{k=0}^{\infty}\b^{-k}
\,\m_{n,\boldsymbol d,k}^{AA'}(\vec\lambda),
\label{eq:strong-field-moment-expansion}
\end{equation}
with a nonzero radius of convergence in $\b^{-1}$ independent of system size. The strong-field coefficients $\m_{n,\boldsymbol d,k}^{AA'}(\vec\lambda)$ can be derived within perturbation theory. We note that these strong-field coefficients should not be confused with high-temperature coefficients, though they are directly analogous, with the small parameter $1/\b$ playing the role of $\beta$.  \par 

We may therefore use strong-field coefficients derived within perturbation theory to establish Hamiltonian learnability at sufficiently large but finite probe field. In the next section, we carry out a general analysis of the strong-field protocol, in a manner paralleling the high-temperature protocol, for one- and two-body Hamiltonians.

\subsection{Learning one- and two-body Hamiltonians from strong-field spectral moments }
\label{sec:strong_field_two_body}

\noindent We now use the strong-field expansion above to establish learnability for arbitrary finite-range one- and two-body Pauli
Hamiltonians. Consider a translation-invariant spin-$\frac12$ Hamiltonian
\begin{equation}
H(\boldsymbol{\lambda})
=
\sum_{\boldsymbol r}
\left[
b^a\sigma_{\boldsymbol r}^a
+
\sum_{\boldsymbol d\in\mathcal R}
\lambda_{\boldsymbol d}^{ab}
\sigma_{\boldsymbol r}^a
\sigma_{\boldsymbol r+\boldsymbol d}^b
\right],
\label{eq:strong-field-general-2body}
\end{equation}
where $\mathcal R$ is a finite set of interaction bonds (which excludes $\vec0$ and includes only one representative of each pair $\{\vec d,-\vec d\}$). The unknown couplings $\boldsymbol{\lambda}\in[-1,1]^p$
comprise the $\lambda_{\boldsymbol d}^{ab}$ and $b^a$,
with $p=9|\mathcal R|+3$.
We apply a Zeeman field of known strength along a coordinate axis $\hat i \in \{\hat x,\hat y,\hat z\}$,
\begin{equation}
H_\b^{(\hat i)}
=
-\b\sum_{\vec r}\sigma_{\vec r}^{i}
+
H(\vec\lambda),
\end{equation}
and denote DSF spectral moments by $m_{n}^{ab}(\vec q)$, keeping dependence on $\hat i$ implicit unless needed. It will be useful to subtract the zeroth moment and define the \textit{centered first moment}
\begin{equation}\label{eq:centeredmoments}
c^{ab}(\vec q)
\coloneqq\frac{
m_{1}^{ab}(\vec q)
-
2\b\,m_{0}^{ab}(\vec q)}{2\pi}.
\end{equation}
The following theorem provides an explicit protocol to learn $\vec\lambda$.

\begin{theorem}[Strong-field protocol for one- and two-body Hamiltonians]
\label{thm:strong-field-2body}
Consider
Eq.~\eqref{eq:strong-field-general-2body}, with couplings
$\vec\lambda\in[-1,1]^{9|\mathcal R|+3}$. Suppose that the full DSF $S^{ab}(\vec q,\omega)$ can be measured at $T=0$ with Zeeman fields along $\hat x$,
$\hat y$, and $\hat z$. Then, for sufficiently large but finite $\b$, all couplings are uniquely determined from the spatial
harmonics of the zeroth and first spectral moments. Moreover, at leading order in $1/\b$, the reconstruction is linear with the following readouts: 
\begin{equation}
\begin{array}{ccl}
\hat n= \hat x &:&
\lambda^{yy}_{\vec d},\,
\lambda^{zz}_{\vec d},\,
\lambda^{yz}_{\vec d},\,
\lambda^{zy}_{\vec d},
\\[1mm]
\hat n= \hat y &:&
\lambda^{zz}_{\vec d},\,
\lambda^{xx}_{\vec d},\,
\lambda^{zx}_{\vec d},\,
\lambda^{xz}_{\vec d},
\\[1mm]
\hat n= \hat z &:&
\lambda^{xx}_{\vec d},\,
\lambda^{yy}_{\vec d},\,
\lambda^{xy}_{\vec d},\,
\lambda^{yx}_{\vec d},
\end{array}
\end{equation}
for every $\vec d\in\mathcal R$, with the $b^a$ determined from the $\vec d=0$ moments and $\lambda_{\vec d}^{ab}$.
\end{theorem}

\begin{proof}
We apply the field along $\hat x$.
For either $a=y,z$ (no sum over $a$), direct computation gives
\begin{equation}
\label{eq:strong-field-cX}
\begin{aligned}
c^{aa}(\boldsymbol q)
=-2b^x
-4\sum_{\boldsymbol d\in\mathcal R}
\lambda_{\boldsymbol d}^{xx}+2\sum_{\boldsymbol d\in\mathcal R}
\Big[
(\lambda_{\boldsymbol d}^{yy}+\lambda_{\boldsymbol d}^{zz})
\cos(\boldsymbol q\cdot\boldsymbol d)+
(\lambda_{\boldsymbol d}^{yz}-\lambda_{\boldsymbol d}^{zy})
\sin(\boldsymbol q\cdot\boldsymbol d)
\Big]
+O(\b^{-1}).
\end{aligned}
\end{equation}
Hence the $\cos(\vec q\cdot\vec d)$ and $\sin(\vec q\cdot\vec d)$ harmonics determine $\lambda_{\vec d}^{yy}+\lambda_{\vec d}^{zz}$ and $\lambda_{\vec d}^{yz}-\lambda_{\vec d}^{zy}$, respectively, for each $\vec d$. Next, the zeroth moments take the form
\begin{equation}
\label{eq:strong-field-m0-diff}
\begin{aligned}
\frac{
m_{0}^{yy}(\vec q)
-
m_{0}^{zz}(\vec q)
}{4\pi}
&=
\frac{1}{\b}
\sum_{\vec d\in\mathcal R}
(\lambda_{\vec d}^{zz}-\lambda_{\vec d}^{yy})
\cos(\vec q\cdot\vec d)
+
O(\b^{-2}),\\
\frac{m_{0}^{yz}(\vec q)}{2\pi}
&=
\ii
-
\frac{1}{\b}
\sum_{\vec d\in\mathcal R}
(\lambda_{\vec d}^{yz}+\lambda_{\vec d}^{zy})
\cos(\vec q\cdot\vec d)
+
O(\b^{-2}).
\end{aligned}
\end{equation}
The cosine harmonics determine $\lambda_{\vec d}^{zz}-\lambda_{\vec d}^{yy},
\lambda_{\vec d}^{yz}+\lambda_{\vec d}^{zy}$ at large $\b$. Altogether, we can learn $\lambda_{\vec d}^{yy},
\lambda_{\vec d}^{zz},
\lambda_{\vec d}^{yz},
\lambda_{\vec d}^{zy}$. Repeating with $\hat y$ and $\hat z$
allows reconstruction of the other couplings. 
\par 

We next turn to the one-body terms. Restoring the Zeeman field-direction label, we have
\begin{equation}
\label{eq:strong-field-zero-harmonic}
\begin{aligned}
c_{\hat i,\boldsymbol0}^{aa}&\coloneqq
\int_{\mathrm{BZ}}
\frac{\mathrm d\boldsymbol q}{\operatorname{vol}(\mathrm{BZ})}
\,c_{\hat i}^{aa}(\boldsymbol q)
= -2b^i
-4\sum_{\boldsymbol d\in\mathcal R}
\lambda_{\boldsymbol d}^{ii}
+O(\b^{-1}),
\end{aligned}
\end{equation}
where $a\neq i$ and no sum over $a$ is taken.
With the $\lambda^{ii}_{\vec d}$ already known, we have
\begin{equation}
\label{eq:strong-field-onsite}
b^i
=
-\frac12 c_{\hat i,\boldsymbol0}^{aa}
-2\sum_{\boldsymbol d\in\mathcal R}
\lambda_{\boldsymbol d}^{ii}
+O(\b^{-1}),
\end{equation}
determining all three one-body couplings, including signs.

\par 

The preceding results indicate that the couplings $\vec\lambda$ can be learned at leading order in $1/\b$, but do not yet establish the Theorem at finite, small $1/\b$ (e.g., higher-order corrections could spoil injectivity). We show this cannot occur for
sufficiently large $\b$ following \textsection~\ref{sec:yarotsky}. 
\par 
For each cyclic choice $(\hat i,\hat j,\hat k) = (x,y,z),(y,z,x),(z,x,y)$
and each $\vec d\in\mathcal R$, we collect the four nonzero spatial harmonics used above:
the cosine and sine harmonics of $c^{jj}$, the cosine
harmonics of $\b(m_0^{jj}-m_0^{kk})/(4\pi)$, and the cosine
harmonics of $\b\,\operatorname{Re}m_0^{jk}/(2\pi)$. We also append
$c_{\hat x,\boldsymbol0}^{yy}$,
$c_{\hat y,\boldsymbol0}^{zz}$, and
$c_{\hat z,\boldsymbol0}^{xx}$.
Denoting this vector as 
$\mathcal M_\b(\boldsymbol{\lambda})$, we write
\begin{equation}
\mathcal M_\b(\vec\lambda)
=
\mathsf{M}\vec\lambda+\vec\rho_\b(\vec\lambda),
\label{eq:strong-field-linear-map}
\end{equation}
where $\mathsf{M}$ is the coefficient matrix giving the leading-order linear relations
derived above and $\boldsymbol{\rho}_\b$ is defined to contain the subleading finite-$\b$ corrections. By \textsection~\ref{sec:yarotsky}, the exact thermodynamic-limit measured quantities are
analytic in $1/\b$ and $\vec\lambda$ and hence we have $\left\|\partial_{\vec\lambda}\vec\rho_\b(\vec\lambda)\right\|
\leq \frac{C}{\b}$  uniformly in
$\vec\lambda$, for some constant $C$ and all sufficiently large $\b$. Since the domain is convex, $\|\vec\rho_\b(\vec\lambda)-\vec\rho_\b(\vec\lambda')\|
\leq
\frac{C}{\b}\|\vec\lambda-\vec\lambda'\|.$ Hence
\begin{align}
\|\mathcal M_\b(\vec\lambda)-\mathcal M_\b(\vec\lambda')\|
&\geq
\|\mathsf{M}(\vec\lambda-\vec\lambda')\|
-
\|\vec\rho_\b(\vec\lambda)-\vec\rho_\b(\vec\lambda')\|\\
&\geq \|\mathsf{M}(\vec\lambda-\vec\lambda')\|- \frac{C}{\b} \|\vec\lambda-\vec\lambda'\|.
\end{align}
Next, let $\sigma_{\min}(\mathsf{M})
\coloneqq
\min_{\|\vec v\|=1}\|\mathsf{M}\vec v\|$. Because $\mathsf{M}$ has full column rank, $\sigma_{\min}(\mathsf{M})>0$. Hence
\begin{equation}
 \|\mathcal M_\b(\vec\lambda)-\mathcal M_\b(\vec\lambda')\|   \geq
\left(
\sigma_{\min}(\mathsf{M})-\frac{C}{\b}
\right)
\|\vec\lambda-\vec\lambda'\|.
\end{equation}
For sufficiently large $\b$,
$\sigma_{\min}(\mathsf{M})-C/\b>0$. Therefore distinct coupling vectors produce
distinct exact spectral moments, proving the Theorem. 
\end{proof}

Theorem~\ref{thm:strong-field-2body} assumed access to the \textit{full,
spin-resolved} DSF $S^{ab}$. For instance, when applying a Zeeman field along $\hat x$, the
off-diagonal spectral moment $m_{0}^{yz}$ is used to determine
$\lambda_{\vec d}^{yz}+\lambda_{\vec d}^{zy}$. In experimental settings, it is often more straightforward to measure only the diagonals $S^{aa}$ of the DSF. We now study the restriction of SF tomography to \textit{diagonal} DSF data. This parallels the diagonal DSF restriction discussion in App.~\ref{sec:hightemp}.\par

We first restrict the probe field to
$\hat i\in\{\hat x,\hat y,\hat z\}$ and assume that, for each $\hat i$, only $S^{aa}(\mathbf q,\omega)$ is accessible. For a field along $\hat x$, our previous results show that harmonics of the diagonal centered first-order moments
determine $\lambda_{\vec d}^{yy}+\lambda_{\vec d}^{zz}$ and
$\lambda_{\vec d}^{yz}-\lambda_{\vec d}^{zy}$.
Moreover, Eq.~\eqref{eq:strong-field-m0-diff}, which also involves only
diagonal channels, determines
$\lambda_{\vec d}^{zz}-\lambda_{\vec d}^{yy}$.
Thus the $\hat x$-field experiment determines
$\lambda_{\vec d}^{yy}$,
$\lambda_{\vec d}^{zz}$, and
$\lambda_{\vec d}^{yz}-\lambda_{\vec d}^{zy}$. Repeating along $\hat y$ and $\hat z$ determines
the remaining diagonal couplings and antisymmetric combinations. Finally, we determine $b^x,b^y,b^z$
through Eq.~\eqref{eq:strong-field-onsite}. Thus, in the diagonal access setting, we have established the following result: 
\begin{proposition}[Strong-field tomography with diagonal DSF access]
\label{prop:coordinate_diagonal_access}
Suppose that a field can be applied \textit{only} along $\hat x$, $\hat y$, and $\hat z$,
and \textit{only} the diagonal DSF is measured. At leading order in
$1/\b$, the zeroth and first spectral moments determine, for every
$\vec d\in\mathcal R$, the three fields $b^x,b^y,b^z$, the diagonal couplings
$\lambda_{\vec d}^{xx}$, $\lambda_{\vec d}^{yy}$,
$\lambda_{\vec d}^{zz}$, and the three
combinations
$\lambda_{\vec d}^{xy}-\lambda_{\vec d}^{yx}$,
$\lambda_{\vec d}^{xz}-\lambda_{\vec d}^{zx}$,
$\lambda_{\vec d}^{yz}-\lambda_{\vec d}^{zy}$.
\end{proposition}
We note, in particular, that the symmetric off-diagonal combinations
$\lambda_{\vec d}^{xy}+\lambda_{\vec d}^{yx}$,
$\lambda_{\vec d}^{xz}+\lambda_{\vec d}^{zx}$, and
$\lambda_{\vec d}^{yz}+\lambda_{\vec d}^{zy}$ do not appear in
the leading order spectral moments. One possibility is to recover these by using higher-order spectral moments. Another possibility, which we pursue, is changing the direction of the Zeeman field; we now show this allows recovery of \textit{all} couplings.\par

To see this, we write $\Lambda_{\vec d}
=
(\lambda_{\vec d}^{ab})_{a,b=x,y,z}$ and for each pair $\mu\nu\in\{xy,xz,yz\}$ define the orthogonal axes $\hat n_{\mu\nu}
=
\frac{\hat e_\mu-\hat e_\nu}{\sqrt2}$ and $\hat u_{\mu\nu}
=
\frac{\hat e_\mu+\hat e_\nu}{\sqrt2}$ where $[\hat e_\mu]_\nu = \delta_{\mu\nu}$. Applying the field along
$\hat n_{\mu\nu}$ and measuring spectral moments
along $\hat u_{\mu\nu}$ gives, by the same calculation as above in a
rotated frame,
\begin{equation}
\label{eq:tilted-diagonal-coupling}
    \lambda_{\vec d}^{u_{\mu\nu}u_{\mu\nu}} = \hat u_{\mu\nu} \cdot \Lambda_{\vec d}\cdot \hat u_{\mu\nu} = \frac12 \left( \lambda_{\vec d}^{\mu\mu} +\lambda_{\vec d}^{\nu\nu}+\lambda_{\vec d}^{\mu\nu} +\lambda_{\vec d}^{\nu\mu}  \right).
\end{equation}
Note the diagonal entries on the right-hand side are known [Prop.~\ref{prop:coordinate_diagonal_access}] and thus $\hat n_{\mu\nu}$-field experiments determine the remaining symmetric off-diagonal combinations $\lambda_{\vec d}^{\mu\nu} + \lambda_{\vec d}^{\nu\mu}$. Together with the combinations in Prop.~\ref{prop:coordinate_diagonal_access}, this determines every entry of $\Lambda_{\vec d}$. In summary, with only diagonal DSF access in these polarization bases, strong-field tomography along the six axes $\hat x,\hat y,\hat z, \hat n_{xy},\hat n_{xz},\hat n_{yz}$ suffices to reconstruct all couplings using leading-order data in $1/\b$. 
\par

The preceding protocols were designed to reconstruct arbitrary one- and two-body Hamiltonians. When the Hamiltonian possesses additional spin symmetries, however, fewer field orientations may suffice. Consider first an SU(2) symmetric Hamiltonian, for which the exchange tensor on each bond is isotropic, $\lambda_{\vec d}^{ab}=\lambda_{\vec d}^{xx}\delta_{ab}$, and all one-body fields vanish. Since global spin rotations leave $H$ invariant, all directions of the applied Zeeman field are equivalent. We find
\begin{equation}
c^{yy}(\vec q) = 4\sum_{\vec d\in\mathcal R}
\lambda_{\vec d}^{xx}
\left[\cos(\vec q\cdot\vec d)-1\right],
\end{equation}
and hence $\lambda_{\vec d}^{xx}$ is learnable from a single field orientation, even with only diagonal DSF access.\par 

It is also instructive to consider a Hamiltonian
U(1) symmetric about $\hat z$. In this case,
$b^x=b^y=0$, $\lambda_{\vec d}^{xx}=\lambda_{\vec d}^{yy}$,
and $\lambda_{\vec d}^{xy}=-\lambda_{\vec d}^{yx}$,
with all other off-diagonal exchanges vanishing.
Unlike the SU(2) case, not all field directions are
equivalent; for a field along $\hat z$, perturbation
theory gives
\begin{equation}
\begin{aligned}
c^{yy}(\vec q)
=-2b^z-4\sum_{\vec d\in\mathcal R}
\lambda_{\vec d}^{zz}+4\sum_{\vec d\in\mathcal R}
\left[
\lambda_{\vec d}^{xx}\cos(\vec q\cdot\vec d)
+\lambda_{\vec d}^{xy}\sin(\vec q\cdot\vec d)
\right].
\end{aligned}
\end{equation}
Thus, $\lambda_{\vec d}^{xx}$ and $\lambda_{\vec d}^{xy}$ can be learned from the cosine and sine harmonics, respectively, whereas $\lambda_{\vec d}^{zz}$ enters only through
$b^z+2\sum_{\vec d}\lambda_{\vec d}^{zz}$. We can resolve this ambiguity by using a field along
$\hat n=(\hat x+\hat z)/\sqrt{2}$ and measuring only $S^{yy}$, in which case perturbation theory yields
\begin{equation}
\begin{aligned}
c^{yy}(\vec q)
&=-\sqrt{2}b^z
-2\sum_{\vec d\in\mathcal R}
(\lambda_{\vec d}^{xx}+\lambda_{\vec d}^{zz})+\sum_{\vec d\in\mathcal R}\Big[
(3\lambda_{\vec d}^{xx}+\lambda_{\vec d}^{zz})
\cos(\vec q\cdot\vec d)
+2\sqrt{2}\lambda_{\vec d}^{xy}
\sin(\vec q\cdot\vec d)\Big]
+O(\b^{-1}),\\
\frac{m_0^{yy}(\vec q)}{2\pi}
&=
1+\frac{1}{2\b}\sum_{\vec d\in\mathcal R}
(\lambda_{\vec d}^{zz}-\lambda_{\vec d}^{xx})
\cos(\vec q\cdot\vec d)
+O(\b^{-2}).\nonumber
\end{aligned}
\end{equation}
Thus, cosine harmonics determine
$3\lambda_{\vec d}^{xx}+\lambda_{\vec d}^{zz}$
and $\lambda_{\vec d}^{zz}-\lambda_{\vec d}^{xx}$,
while sine harmonics determine
$\lambda_{\vec d}^{xy}$. These measurements
recover all exchange couplings, after which $b^z$
follows.
We conclude that, for U(1)-symmetric one- and two-body
Hamiltonians, a single suitably tilted field direction
suffices.

\subsection{Numerics}
\label{sec:sfpnumerics}
\begin{figure}
    \centering
    \includegraphics[width=\linewidth]{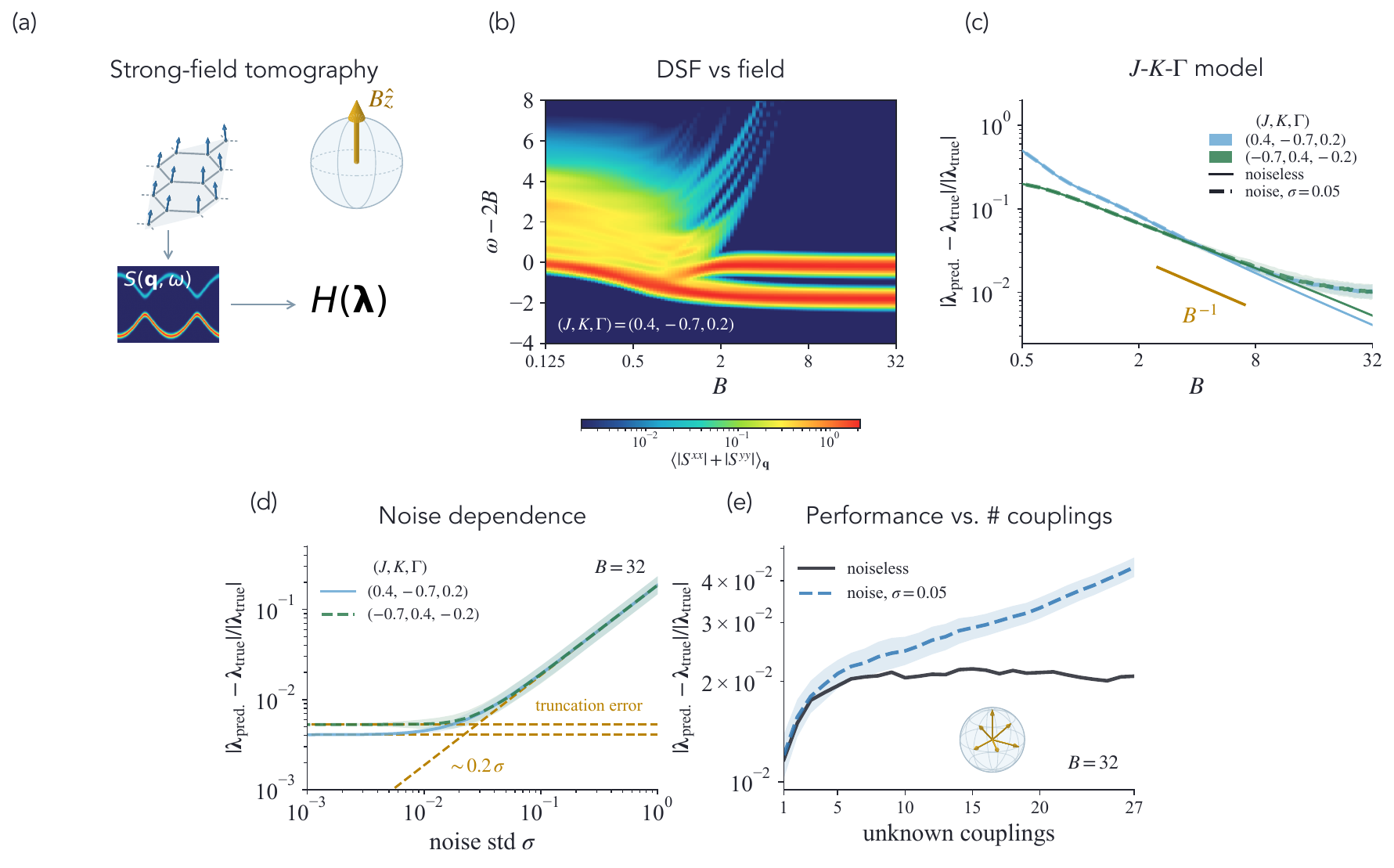}
\caption{\textbf{Strong-field (SF) Hamiltonian tomography.} (a) Polarizing a quantum system with a Zeeman field $\b$ simplifies the dynamical structure factor (DSF) $S(\vec q,\omega)$, enabling learning the Hamiltonian $H(\vec \lambda)$ even at low temperature, contrasting with and complementing high-temperature Hamiltonian tomography. Here, calculations use a periodic $N=12$-spin honeycomb cluster at $T=0$. (b) Momentum-averaged transverse DSF for the first Hamiltonian in (c), showing two isolated bands at large $\b$ at $\omega \approx 2\b$. For display, spectra are broadened by a Gaussian of standard deviation $0.15$, using frequency spacing $0.05$. (c) Performance of SF tomography for two $J$-$K$-$\Gamma$ Hamiltonians $H_\b^{(\hat n)}$ [Eq.~\eqref{eq:HB-honeycomb-JKG}] using fields along $\hat z$ and $(\hat x-\hat y)/\sqrt{2}$ and full spin-resolved DSF access. Solid curves use noiseless centered moments, defined in Eq.~\eqref{eq:centeredmoments}, and dashed curves include Gaussian noise of standard deviation $\sigma = 0.05$. The noise-free curves approach expected $\b^{-1}$ decay. (d) Dependence of results in (c) on $\sigma$ at $\b=32$. (e) Performance of SF tomography (with fields along the six directions shown in the inset) at $\b=32$ for a general $\leq 27$-unknown-coupling two-body exchange Hamiltonian $H(\vec \lambda)$. Statistical curves show medians, and ribbons contain the central $25\%$ of errors. Errors in (c--e) are relative Euclidean coupling errors $|\vec \lambda_{\mathrm{pred}}-\vec\lambda_{\mathrm{true}}|/|\vec\lambda_{\mathrm{true}}|$.}   \label{fig:sftomography}
\end{figure}

\noindent Next, we detail numerical tests of SF tomography. We investigate the dependence of the SF protocol on $\b$ and noise, within a simple noise model. We simulate DSF data for Hamiltonians $H(\vec\lambda)$ using exact diagonalization, add random Gaussian noise of standard deviation $\sigma$, and extract $\vec\lambda$ using the SF protocol outlined previously. Figure~\ref{fig:sftomography} presents main results. 
\par 
We first consider a honeycomb $J$-$K$-$\Gamma$ (Kitaev-Heisenberg-Gamma) model~\cite{rau2014generic} in $\b$ field, 
\begin{equation}
\begin{aligned}
H_\b^{(\hat n)}(\boldsymbol{\lambda})
&= -\b\sum_{\vec r}\hat n\cdot\boldsymbol{\sigma}_{\vec r}+\sum_{\langle \vec r\vec r'\rangle_\gamma}
\left[
J\,\boldsymbol{\sigma}_{\vec r}\cdot\boldsymbol{\sigma}_{\vec r'}
+K\sigma_{\vec r}^\gamma\sigma_{\vec r'}^\gamma
+\Gamma\left(\sigma_{\vec r}^\alpha\sigma_{\vec r'}^\beta
+\sigma_{\vec r}^\beta\sigma_{\vec r'}^\alpha\right)
\right]\\
&\coloneqq -\b\sum_{\vec r}\hat n\cdot\boldsymbol{\sigma}_{\vec r} + H(\vec\lambda)
\end{aligned}
\label{eq:HB-honeycomb-JKG}
\end{equation}
where $\boldsymbol{\lambda}=(J,K,\Gamma)$ and $\alpha,\beta$ specify two axes other than $\gamma$. We assemble centered first moments [Eq.~\eqref{eq:centeredmoments}] into a vector $\vec c$ for a $N=12$ honeycomb cluster [Fig.~\ref{fig:sftomography}a] at $T=0$. In Fig.~\ref{fig:sftomography}b, for $(J,K,\Gamma)=(0.4,-0.7,0.2)$ and field along $\hat z$, we plot $\frac{1}{2\pi N} \sum_{\vec q} \left[|S^{xx}(\vec q,\omega)|+|S^{yy}(\vec q,\omega)|\right]$ to illustrate broad features of the DSF. At large $\b$, we observe the emergence of two bands at $\omega \approx 2\b$.\par 

Sec.~\ref{sec:yarotsky} establishes the leading expansion
$\vec c=\mathsf{M}\boldsymbol{\lambda}+O(\b^{-1})$, with coefficient matrix $\mathsf{M}$ independent of $\vec \lambda$ and computable from perturbation theory around $\ket{\Omega_{\hat n}} \equiv \ket{+_{\hat n}}^{\otimes N}$ (the $+1$ eigenstate of each $\hat n\cdot\vec\sigma_{\vec r}$). An explicit form is as follows: parameterizing DSF channels by $p_a^*p_bS^{ab}(\vec q,\omega)$ by polarization vectors $\vec p$, for fixed $\vec p$ and $\vec q$, let
$O_{\vec q,\vec p}=N^{-1/2}\sum_{\vec r} e^{\mathrm{i}\vec q\cdot\vec r}
\vec p\cdot\boldsymbol{\sigma}_{\vec r}$. Then
\begin{equation}
c_{\vec q,\vec p}
= \frac{1}{2\pi}\int_{0^+}^{\infty} \dd\omega\,(\omega-2\b)\, p_a^* p_bS^{ab}(\vec q,\omega)
=\langle v_{\vec q,\vec p}|(H_\b^{(\hat n)}-E_0-2\b)|v_{\vec q,\vec p}\rangle,
\label{eq:SF-numerical-moment}
\end{equation}
where $|v_{\vec q,\vec p}\rangle=(1-P_0)O_{\vec q,\vec p}|g\rangle $, $|g\rangle$ is the ground state with energy $E_0$, and $P_0$ projects onto the $|g\rangle$. Writing $V_j=\partial H_\b^{(\hat n)}/\partial\lambda_j$ and
$|v_{\vec q,\vec p}^{(0)}\rangle=(1-|\Omega_{\hat n}\rangle\langle \Omega_{\hat n}|)O_{\vec q,\vec p}|\Omega_{\hat n}\rangle$, we have
\begin{equation}
\mathsf{M}_{(\vec q,\vec p), j}
=\langle v_{\vec q,\vec p}^{(0)}|V_j|v_{\vec q,\vec p}^{(0)}\rangle
-\langle \Omega_{\hat n}|V_j|\Omega_{\hat n}\rangle
 \langle v_{\vec q,\vec p}^{(0)}|v_{\vec q,\vec p}^{(0)}\rangle.
\label{eq:SF-numerical-inverse}
\end{equation}
Since $H_\b^{(\hat n)}$ is linear in $\lambda_j$, $V_j$ is independent of couplings. Hence Eq.~\eqref{eq:SF-numerical-inverse} specifies $\mathsf{M}$ entirely as a matrix of $O(1)$ coefficients (independent of $\vec\lambda$) which can be computed straightforwardly. \par 

We now construct an estimator for the solution to $\vec c= \mathsf{M}\vec\lambda +O(\b^{-1})$. Our estimator will be  
\begin{equation}
\label{eq:SFestimator}    \boldsymbol{\lambda}_{\mathrm{pred}}=\mathsf{M}^+\vec c
\end{equation}
where $\mathsf{M}^+$ is the Moore--Penrose pseudoinverse of the (generally rectangular) matrix $\mathsf{M}$. Equivalently, $\vec\lambda_{\mathrm{pred}}$ is the least squares solution to the linear system with $O(\b^{-1})$ terms discarded. By previous results, $\boldsymbol{\lambda}_{\mathrm{pred}}$ is guaranteed to be exact when $\b$ is infinite and noise is absent.\par 

In Fig.~\ref{fig:sftomography}c, we study the performance of $\boldsymbol{\lambda}_{\mathrm{pred}}$ for two choices of $\vec \lambda = (J,K,\Gamma)$, denoted $\vec\lambda_{\mathrm{true}}$. We construct $\mathsf{M}$ as in Eq.~\eqref{eq:SF-numerical-inverse} using fields along $\hat n = \hat z$ and $\hat n_{xy} \equiv (\hat x-\hat y)/\sqrt{2}$. For each direction we use 18 $\vec q$'s and the full DSF. We find $\mathsf{M}$ has full column rank equal to three. Next, we compute $\widetilde c_{\vec q,\vec p}=c_{\vec q,\vec p} +\sigma \mu$ with $\mu$ standard Gaussian noise and $\sigma=0.05$. We plot the median and $25\%$ error bars of $|\vec\lambda_{\mathrm{pred}}-\vec\lambda_{\mathrm{true}}|/|\vec\lambda_{\mathrm{true}}|$ to gauge the performance of Eq.~\eqref{eq:SFestimator}. We observe steady $\sim \b^{-1}$ decrease for the noiseless curves as expected from the truncation error being $O(\b^{-1})$ in the SF protocol and a leveling off of error in the presence of noise.\par 

In Fig.~\ref{fig:sftomography}d, we plot similar results with varying $\sigma$ at $\b=32$. We point out two clear regimes in the performance as a function of $\sigma$: a plateau at small $\sigma$ and a linear-in-$\sigma$ regime at large $\sigma$. The plateau in the low-noise regime can be understood as the $O(\b^{-1})$ truncation error coming from the finite field $\b=32$. Meanwhile, adding noise to $\vec c$ changes $\vec\lambda_{\mathrm{pred}}$ by $\sigma\mathsf{M}^+\vec z$, which is linear in $\sigma$. The slope, which is specific to $\mathsf{M}$, is found to be $\sim 0.2$ for both $\vec\lambda_{\mathrm{true}}$ values. 
\par 

Finally, Fig.~\ref{fig:sftomography}e illustrates the performance of the SF protocol as a function of the number of unknown Hamiltonian parameters.  For this diagnostic, we consider translation-invariant nearest-neighbor
exchange on the honeycomb lattice, with nine couplings for each of the
three bond orientations. We sample 64 Hamiltonians with independent
Gaussian couplings and normalize each 27-component coupling vector. For each $H$, we randomly order
couplings and treat the first $p$ as unknown, supplying the remaining
$27-p$ exactly. We learn the unknown couplings using the SF protocol and plot performance as a function of the number of unknown couplings in both the noiseless and noisy cases. We observe increase of error in the noisy case with the number of unknown couplings, while performance plateaus in the noiseless case.

\subsection{Angular tomography}
\label{sec:angular_tomography}

\noindent So far, we have focused on learning $H$ using fields pointed along a discrete set of axes. In this section, we present a protocol, which we term \textit{angular tomography}, that takes advantage of continuously varying field axes $\hat n$. The basic insight is that varying the field over continuous spherical angles $(\theta,\phi)$ and recording DSF data as a function of angle yields incredibly rich data. Our protocol is relevant to neutron scattering experiments (with vector magnets or varying sample mountings) and quantum simulation experiments with access to angle-tunable global fields. \par 

While angular tomography can be useful for general Hamiltonian learning, we focus on a specific application for concreteness and to avoid overlap with prior sections: determining the \textit{bodyness} of the Hamiltonian, defined roughly to be the minimum $K$ such that all interactions are $k$-body with $k\leq K$. We will show that a nonzero angular harmonic of order $\ell$
certifies the presence of interactions involving at least
$\ell$ spins.

\par 
We define the \textit{bodyness} of a Hamiltonian more precisely as follows. Take $H$ to be of the form  
\begin{equation} H(\vec\lambda) = \sum_{j=1}^p \lambda^jH^j,\qquad H^j = \sum_{\vec r\in \Lambda} h_{\vec r}^j
\end{equation}
with $\vec \lambda \in [-1,1]^p$ and with $p$ a fixed, possibly large, integer. Take $\Lambda$ to be a periodic Bravais lattice with a fixed finite-dimensional spin Hilbert space at each site and let $h_{\vec r}^j$ be translated versions of a fixed local interaction. This is the standard form assumed in the main text. The \textit{bodyness} of $H$ is the largest number of sites acted upon by any nonzero interaction term.
\par 
In Hamiltonians of real quantum magnets or analog simulators, small nonzero interactions with large support may generally occur---yet be irrelevant to observations. Bodyness as defined above can correspondingly be fragile (i.e., the value can change upon adding exponentially small terms) and potentially difficult to measure in practice depending on the chosen model. Thus, it is more practical to consider an approximate version of bodyness which accounts for small high-weight terms or measurement errors. To do this unambiguously, we define
\begin{equation}
    H(\vec\lambda) = \sum_X h_X(\vec\lambda)
\end{equation}
where $h_X(\vec\lambda)$ is the sum of \textit{all} terms with exactly the same support $X\subset \Lambda$. We then define the \textit{higher-body interaction strength} as 
\begin{equation}
    \delta_K(H) \coloneqq \sup_i \sum_{\substack{X\ni i\\ |X|>K}} \|h_X\|_{\mathrm{op}}
\end{equation}
where $\|\cdot \|_{\mathrm{op}}$ denotes the operator norm. Finally, we define the $\epsilon$\textit{-bodyness} of $H(\vec \lambda)$ as
\begin{equation}
    K_{\epsilon}(H) = \min \{ K \mid \delta_K(H) \leq \epsilon\}.
\end{equation}
When $\epsilon=0$, this reduces to the bodyness. Here, $\epsilon$ sets a tolerance scale for the couplings $\vec\lambda$ below which measurements cannot probe and / or interaction terms are discarded.
\par 
Let us now consider the ground state of 
\begin{equation}
    H_\b^{(\hat n)}(\vec\lambda) = -\b\sum_{\vec r}\hat n\cdot \vec\sigma_{\vec r} + H(\vec\lambda).
\end{equation}
We assume $H(\vec\lambda)$ is finite-range and translation-invariant and $\hat n\in S^2$. Though $\hat n$ may vary, we assume $H(\vec\lambda)$ is fixed. Our main result is that the presence of $k$'th order angular harmonics (with a suitably defined strength $> \varepsilon$) in the DSF indicates that the $\varepsilon$-bodyness of $H$ is at least $k$.
\par 
We consider centered first spectral moment harmonics, the Fourier transform of Eq.~\eqref{eq:centeredmoments}. For $\vec d\neq \vec 0$, define $c^{aa}_{\hat{n}}(\vec d) \coloneqq (m^{aa}_{1}(\vec d) - 2\b m^{aa}_{0}(\vec d))/2\pi$, with dependence on $\b$ implicit. It is useful to define the following:
\begin{equation}
    \mathcal{C}_{\vec d}(\hat n) \coloneqq \lim_{\b\to\infty} \left[ \b+\frac12\sum_{a=x,y,z}c_{\hat n}^{aa}(\vec d) \right].
\end{equation}
We note the following useful identity, whose derivation we leave to the end of this section:
\begin{equation}
\label{eq:cvecdhatnidentity}
\mathcal C_{\vec d}(\hat n)
=
\bra{\vec 0;\hat n}H\ket{\vec d;\hat n}
\end{equation}
for $\vec d\neq\vec0$. Next, we decompose $\mathcal C_{\vec d}(\hat n)$ into spherical harmonics. Let $Y_{\ell m}(\hat n)$ denote the standard spherical harmonics
$\ell=0,1,\ldots$ and $m=-\ell,\ldots,\ell$. Define the angular
coefficients
\begin{equation}
\mathcal C_{\vec d;\ell m}
\coloneqq
\int_{S^2}\mathrm \sin\,\theta\,\mathrm \dd\theta\,\mathrm \dd\varphi\, 
Y_{\ell m}^*(\hat n)\,
\mathcal C_{\vec d}(\hat n).
\label{eq:angular_coefficients}
\end{equation}
To quantify the strength of each, we define
\begin{equation}
\label{eq:angulartomographyAell}
A_\ell\coloneqq
\left[\frac{1}{4\pi}\sum_{m=-\ell}^{\ell}
|\mathcal C_{\vec d;\ell m}|^2\right]^{1/2}.
\end{equation}
We then have the following Proposition:
\begin{proposition}[Bodyness from angular harmonics]\label{prop:angular_body_order}
For $\vec d \neq \vec 0$ and every $K\geq 0$, we have 
\begin{equation}
\label{eq:Aellinequality}
    \sum_{\ell > K} A_{\ell}^2 \leq \delta_K(H)^2.
\end{equation}
Consequently, the presence of an amplitude $A_\ell > \epsilon$ implies that $H$ has $\epsilon$-bodyness $K_\epsilon(H) \geq \ell$. 
\end{proposition}
\begin{proof}
First, consider a general $k$-body term
$H=J^{a_1\cdots a_k}
\sigma_{\mathbf r_1}^{a_1}\cdots
\sigma_{\mathbf r_k}^{a_k}$
acting on $k$ distinct sites.
By Eq.~\eqref{eq:cvecdhatnidentity}, the contribution of $H$ vanishes unless its support contains both $\vec 0$ and $\vec d$, so let us assume this is the case. We label these sites as
$\vec r_1=\vec 0$ and $\vec r_2=\vec d$. By Eq.~\eqref{eq:cvecdhatnidentity}, other sites contribute factors of $\langle+_{\hat n}|\sigma^{a_u}|+_{\hat n}\rangle=n_{a_u}$ while $\vec r_{1,2}$ contribute
$\langle-_{\hat n}|\sigma^{a_1}|+_{\hat n}\rangle\langle+_{\hat n}|\sigma^{a_2}|-_{\hat n}\rangle$, which can be simplified as
$\delta_{a_1a_2}-n_{a_1}n_{a_2}-\ii\epsilon_{a_1a_2a_3}n_{a_3}$. Hence, for this choice of $H$, we have
\begin{equation}
\mathcal C_{\vec d}(\hat n)
=
J^{a_1\cdots a_k}
\left(
\delta_{a_1a_2}
-
n_{a_1}n_{a_2}
-
\ii\epsilon_{a_1a_2c}n_c
\right)
\prod_{u=3}^{k}n_{a_u}
\label{eq:angular_k_body_kernel}
\end{equation}
for $k>2$ (with $\prod_{u=3}^{k}n_{a_u}$ absent when $k=2$). Eq.~\eqref{eq:angular_k_body_kernel} is a polynomial in
$(n_x,n_y,n_z)$ of degree at most $k$, so its restriction to $S^2$ contains no harmonics above degree $k$. Thus harmonics of $\mathcal C_{\vec d}(\hat n)$ of degree above $k$ come entirely from interactions with support sizes greater than $k$. \par 

Next, define $f(\hat n)=
\sum_{\substack{X\supset\{\vec0,\vec d\}\\|X|>K}}
\langle\vec0;\hat n|h_X|\vec d;\hat n\rangle.$ Since each term has magnitude at most $\|h_X\|_{\mathrm{op}}$, we have $|f(\hat n)|
\leq
\sum_{\substack{X\supset\{\vec0,\vec d\}\\|X|>K}}
\|h_X\|_{\mathrm{op}}
\leq\delta_K(H),
$ with the final inequality coming from the definition of $\delta_K(H)$. We also expand $f(\hat n)$ in spherical harmonics: $f(\hat n) = \sum_{\ell,m}f_{\ell m}Y_{\ell m}(\hat n),
$ noting that $f_{\ell m}=\mathcal C_{\vec d;\ell m}$ when $\ell>K$. We then have 
\begin{equation}
    \begin{aligned}
        \sum_{\ell > K} A_\ell^2 &= \frac{1}{4\pi} \sum_{\ell>K,m}|\mathcal{C}_{\vec d;\ell m}|^2\\
        &= \frac{1}{4\pi} \sum_{\ell>K,m} |f_{\ell m}|^2\\
        &\leq \frac{1}{4\pi} \sum_{\ell,m} |f_{\ell,m}|^2\\
        &= \frac{1}{4\pi} \int_{S^2}|f(\hat n)|^2 \dd \Omega \\
        &\leq \delta_K(H)^2,
    \end{aligned}
\end{equation}
which is Eq.~\eqref{eq:Aellinequality}. Now, suppose $A_\ell>\epsilon$ for some $\ell$. Then for every $0\leq j < \ell$ we have
\begin{equation}
    \delta_j(H) \geq \delta_{\ell-1}(H) \geq \left(\sum_{\ell'\geq \ell}A_{\ell'}^2\right)^{1/2}\geq A_{\ell} > \epsilon.
\end{equation}
Hence $K_\epsilon(H)=\min\{j:\delta_j(H)\leq\epsilon\}\geq\ell.$
\end{proof}

\begin{figure}
    \centering
\includegraphics[width=\linewidth]{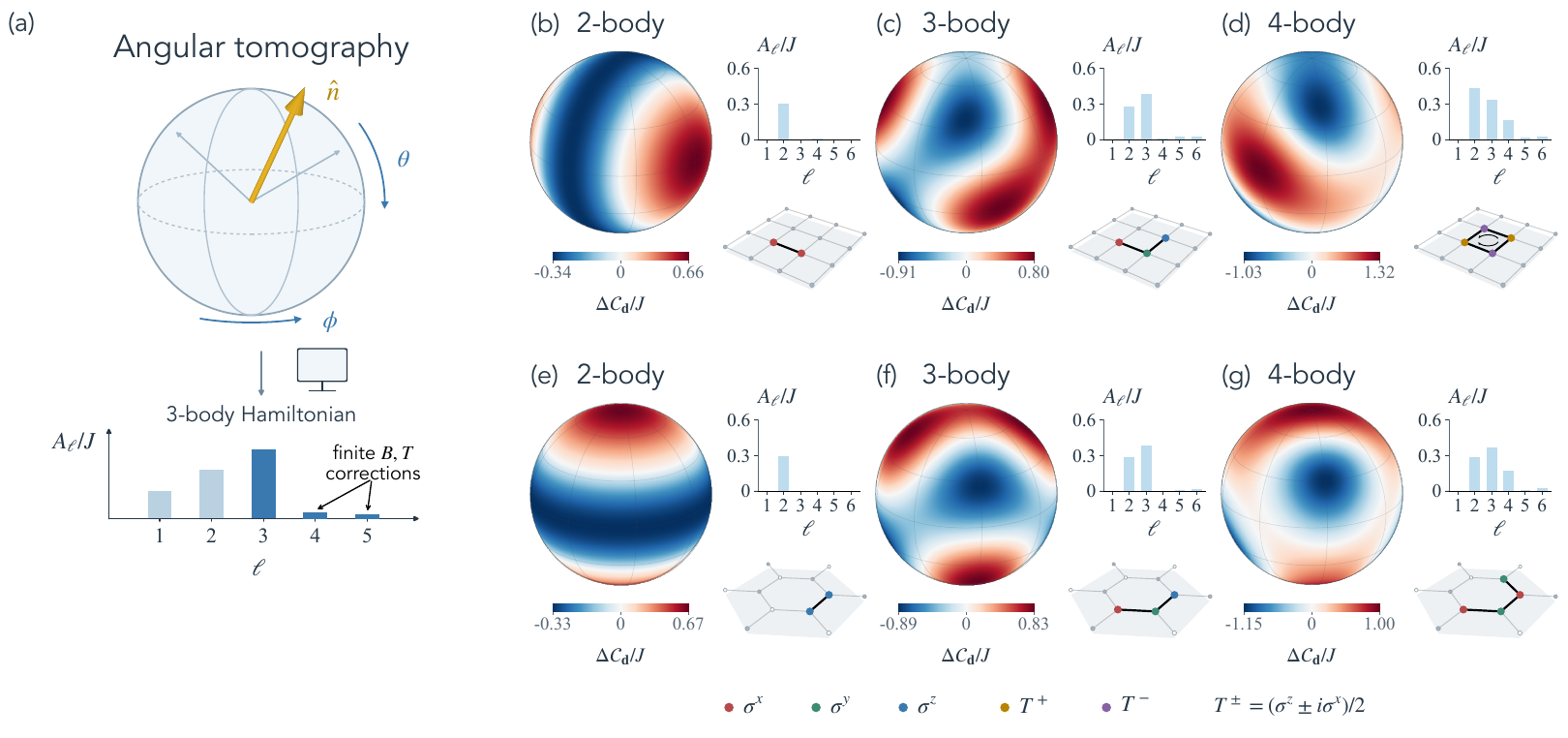}
\caption{\textbf{Angular tomography with a strong field.}
(a) In angular tomography, we apply a strong field $\b \hat n$ and continuously vary $\hat n$, shedding light on the bodyness (or minimum $k$ such that all interactions of $H$ are at most $k$-body) of the Hamiltonian through a spherical harmonic decomposition of data.
(b)--(g) Angular tomography signals at finite field and
temperature for $H_\b^{(\hat n)}=-\b\sum_i\hat n\cdot\boldsymbol\sigma_i+H$, with $H$ of the following forms.
(b)
$H_{s,2}=-J\sum_{\vec r}(\sigma^x_{\mathbf r}\sigma^x_{\vec r+\hat e_x}
+\sigma^z_{\vec r}\sigma^z_{\vec r+\hat e_y})$.
(c)
$H_{s,3}=H_{s,2}-J\sum_{\vec r}\sum_{s,t=\pm1}
\sigma^x_{\vec r+s\hat e_x}\sigma^y_{\vec r}\sigma^z_{\vec r+t\hat e_y}$.
(d)
$H_{s,4}=H_{s,3}-8J\sum_{\square=(1234)}
(T_1^+T_2^-T_3^+T_4^-+\mathrm{h.c.})$,
where $T_i^\pm=(\sigma^z_i\pm i\sigma^x_i)/2$ and each plaquette is counted once with
$(1,2,3,4)=(\vec r,\vec r+\hat e_x,
\vec r+\hat e_x+\hat e_y,\vec r+\hat e_y)$.
(e) Kitaev model,
$H_{h,2}=-J(\sum_{\langle ij\rangle_x}\sigma^x_i\sigma^x_j
+\sum_{\langle ij\rangle_y}\sigma^y_i\sigma^y_j
+\sum_{\langle ij\rangle_z}\sigma^z_i\sigma^z_j)$.
(f) With $j_\alpha$ the neighbor of $j$ along $\alpha$,
$H_{h,3}=H_{h,2}-J\sum_j(\sigma^x_{j_x}\sigma^z_j\sigma^y_{j_y}
+\sigma^x_{j_x}\sigma^y_j\sigma^z_{j_z}+\sigma^y_{j_y}\sigma^x_j\sigma^z_{j_z})$.
(g)
$H_{h,4}=H_{h,3}+J\sum_{\langle
ijkl\rangle_{\alpha\beta\gamma}}^{\prime}
\sigma_i^\alpha\sigma_j^\gamma\sigma_k^\alpha\sigma_l^\gamma$,
where $\sum'$ counts each path once. ($N=12$, $\b=32J$, $T/\b=0.1$). Spheres show $\Delta\mathcal C_{\vec d}/J$, where $\Delta\mathcal C_{\vec d}$ denotes the angular signal with its mean subtracted and $\vec d$ is a nearest-neighbor bond.
Histograms show $A_\ell$ [Eq.~\eqref{eq:angulartomographyAell}] for $\ell=1,\ldots,6$. The small tail of harmonics comes predominantly from finite-$\b$ corrections. In the limit $\b\to\infty$ at $T=0$, the highest nonzero harmonic gives a lower bound on the bodyness of the Hamiltonian [Prop.~\ref{prop:angular_body_order}].}
\label{fig:angulartomography}
\end{figure}

As a simple, concrete example, consider $H=g\sigma_1^z\sigma_2^z\sigma_3^z$. For $\vec d = \vec r_2-\vec r_1$, we have
\begin{equation}
\mathcal C_{\vec d}(\hat n) = g\,n_z(1-n_z^2) = \frac{2g}{5}\left[P_1(n_z)-P_3(n_z)\right],
\label{eq:angular_three_body_example}
\end{equation}
where $P_\ell$ is the $\ell$'th Legendre polynomial. The content of Prop.~\ref{prop:angular_body_order} is that the existence of this nonzero $\ell=3$ harmonic
indicates $H$ has a bodyness at least three.\par

Angular tomography is a protocol for determining $\varepsilon$-bodyness of $H$. We illustrate it in Fig.~\ref{fig:angulartomography} for several Hamiltonians. In Figs.~\ref{fig:angulartomography}b-d, we plot $\mathcal{C}_{\vec d}(\hat n)$ and $A_\ell$ for a $2-,3-$ and $4$-body square lattice Hamiltonian. Angular tomography simply consists of decomposing $\mathcal{C}_{\vec d}(\hat n)$ into spherical harmonics, computing $A_\ell$, and applying Prop.~\ref{prop:angular_body_order}. As established in Prop.~\ref{prop:angular_body_order}, a $k$-body Hamiltonian can produce nonzero $\ell=k$ body harmonic components. In Figs.~\ref{fig:angulartomography}e-g we repeat a similar analysis for Kitaev-type models on a honeycomb lattice. Again in these examples, the appearance of $A_\ell$ is a direct witness for the presence of $\ell$-body interactions. We propose that angular tomography can be a useful tool for lower bounding $\varepsilon$-bodyness in quantum magnets or simulators. 
\par 

We close with a few qualifications about practical, experimental aspects of angular tomography. First, we note that the $\epsilon$-bodyness and angular tomography protocol are defined relative to a fixed set of spin degrees of freedom with a fixed onsite Hilbert space. This set of spins must be those that are polarized by the Zeeman term in $H_\b^{(\hat n)}$; if other spins exist in the physical system which are not polarized by the Zeeman field, these do not figure into the definition or measurement of $\epsilon$-bodyness. Such interactions can arise when high-energy degrees of
freedom are integrated out to obtain an effective spin
Hamiltonian. As a rule of thumb, strong-field tomography is sensitive to couplings $J_*$ when the Zeeman field $\b$ and the high-energy degrees of freedom $\Delta$ satisfy $J_* \ll \b \ll \Delta$. Second, in general, Prop.~\ref{prop:angular_body_order} lacks a strict converse. In particular, the existence of $k$-body interactions does not imply the existence of nonzero $k$'th spherical harmonics in $\mathcal{C}_{\vec d}(\hat n)$. For example, $H=\chi\,\epsilon_{abc}\sigma_1^a\sigma_2^b\sigma_3^c$ gives $\mathcal C_{\vec d}=-2\ii\chi$, which has only an $\ell=0$ component. Finally, the idealized limits of zero temperature and infinite $\b$ are not essential provided $\b\gg J_*$ and $T\ll \b$, as evidenced by Fig.~\ref{fig:angulartomography}. Thermal contributions are suppressed by $e^{-2\b/T}$ and finite-$\b$ contributions by powers of $J_*/\b$. In practice, extrapolations to $\b\to\infty$ and $T\to0$ could help eliminate finite-field and finite-temperature effects.
\par

We now derive Eq.~\eqref{eq:cvecdhatnidentity}. First, let $\ket{\Omega_{\hat{ n}}} = \ket{+_{\hat{ n}}}^{\otimes N}$, the $+1$ eigenstate in the $\hat{ n}$-basis. Let $\ket{\vec r;\hat{ n}}$ denote $\ket{\Omega_{\hat{n}}}$ with one spin flipped at $\vec r$. Let $\ket{m_\b}$ denote the $m$'th eigenstate of $H_\b^{\hat n}$, with $m=0$ the ground state, and let $E_m$ be the corresponding energy. The added $\b$ cancels the leading elastic contribution. In this eigenbasis,
\begin{equation}
\begin{aligned}
    \mathcal{C}_{\vec d}(\hat{ n}) &= \lim_{\b\to\infty} \frac12 \sum_{i=x,y,z}\sum_{m\neq 0} (E_m-E_0-2\b)\langle 0_\b|\delta \sigma^i_{\vec 0}|m_\b\rangle\langle m_\b|\delta \sigma^i_{\vec d}|0_\b\rangle \\
&= \lim_{\b\to\infty}  \frac12\sum_{i=x,y,z} \left\langle 0_\b\left| \delta\sigma_{\vec 0}^i (H_\b^{(\hat n)}-E_0-2\b) \delta\sigma_{\vec d}^i \right|0_\b\right\rangle \end{aligned} 
\end{equation}
where
$\delta\sigma_{\vec r}^i
=\sigma_{\vec r}^i-\langle 0_\b|\sigma_{\mathbf r}^i|0_\b\rangle$. At large $\b$, we have $\delta\sigma_{\vec r}^i\ket{0_\b}
=
(\sigma_{\vec r}^i-n_i)\ket{\Omega_{\hat n}}
+O(\b^{-1})
=
\langle-_{\hat n}|\sigma^i|+_{\hat n}\rangle\ket{\vec r;\hat n}
+O(\b^{-1}).$ Similarly, $H_\b^{(\hat n)}-E_0-2\b= -\b\sum_{\vec r}\hat n\cdot\vec\sigma_{\vec r}+\b N-2\b
+
H-\langle\Omega_{\hat n}|H|\Omega_{\hat n}\rangle
+
O(\b^{-1})$. Since every one-spin-flip state has unperturbed energy
$-\b N+2\b$, we have $(-\b\sum_{\vec r}\hat n\cdot\vec\sigma_{\vec r}+\b N-2\b)\ket{\vec r;\hat n}=0.$ Substituting therefore yields 
\begin{align}
\mathcal C_{\vec d}(\hat n)
&=\lim_{\b\to \infty}\left(
\frac12\sum_i
|\langle-_{\hat n}|\sigma^i|+_{\hat n}\rangle|^2
\left[
\bra{\vec 0;\hat n}H\ket{\vec d;\hat n}
-
\langle\Omega_{\hat n}|H|\Omega_{\hat n}\rangle
\langle\vec 0;\hat n|\vec d;\hat n\rangle
\right]
+
O(\b^{-1})\right)\\
&=
\frac12\sum_i
|\langle-_{\hat n}|\sigma^i|+_{\hat n}\rangle|^2
\bra{\vec 0;\hat n}H\ket{\vec d;\hat n}
\label{eq:angular_projection_before_trace}
\end{align}
for $\vec d\neq\vec 0$. Finally, we may use $|\langle-_{\hat n}|\sigma^i|+_{\hat n}\rangle|^2 
=
\langle+_{\hat n}|
\sigma^i
\left(1-|+_{\hat n}\rangle\langle+_{\hat n}|\right)
\sigma^i
|+_{\hat n}\rangle=
1-n_i^2$, and hence $\frac12\sum_{i=x,y,z}|\langle-_{\hat n}|\sigma^i|+_{\hat n}\rangle|^2
=
\frac12\left(3-\sum_a n_a^2\right)
=
1$, to finally obtain Eq.~\eqref{eq:cvecdhatnidentity}.

\section{Fluctuation-dissipation Hamiltonian tomography }
\label{sec:suppfdt}
\noindent In this section, we present in detail the fluctuation-dissipation (FD) protocol for Hamiltonian learning from the DSF. We start with the fluctuation-dissipation theorem (FDT), relating the DSF to the dissipative part of the susceptibility, $\chi''(\vec q,\omega)$. The FDT motivates us to take a closer look at particular linear combinations of the DSF spectral moments and uncovers useful relations among moments at different orders. In principle, the FD protocol allows Hamiltonian learning from the DSF at arbitrary temperatures, without the need for a strong field or further experimental control. This and its extreme simplicity are the main advantages of the FD protocol over the HT and SF protocols. One potential disadvantage is that higher-body Hamiltonian interactions appear to be less tractable within the FD protocol. \par 

We consider $H(\vec\lambda)=\sum_{j=1}^{p}\lambda^j H^j,$
with $H^j=\sum_{\vec r\in\Lambda}h^j_{\vec r}$ as usual, where interaction terms are fixed and $\vec\lambda\in[-1,1]^p$ are the unknown couplings to be learned. Let $\langle \cdot \rangle_\beta$ denote the expectation value in the Gibbs state at inverse temperature $\beta$. We use the same definitions of the DSF, spectral moments, and spatial harmonics as in App.~\ref{sec:hightemp}: 
\begin{equation}
\begin{aligned}
    S^{ab}(\vec q,\omega)
&=
\frac1N\sum_{\vec r,\vec r'\in\Lambda}
\int\mathrm dt\,
e^{-\ii\vec q\cdot(\vec r-\vec r')}e^{\ii\omega t}
\langle O^a_{\vec r}(t)O^b_{\vec r'}(0)\rangle_\beta,\\
m_n^{ab}(\vec q)
&=
\int_{-\infty}^{\infty}\mathrm d\omega\,
\omega^n S^{ab}(\vec q,\omega),\\
m_{n,\vec d}^{ab}
&=\frac1N \sum_{\vec q\in \Lambda^\star} 
e^{-\ii\vec q\cdot\vec d}m_n^{ab}(\vec q),
\end{aligned}
\end{equation}
with $O^a_{\vec r}=\sigma^a_{\vec r}$ for spin-$\frac12$ and with the last sum replaced by
$\int_{\mathrm{BZ}}\mathrm d\mathbf q/\operatorname{vol}(\mathrm{BZ})$
in the thermodynamic limit. In particular, $m_{n,\vec d}^{ab}/2\pi = \left\langle \mathrm{ad}_H^{\,n}\left(O_{\vec 0}^a\right)O_{\vec d}^b\right\rangle_\beta$ where $\mathrm{ad}_H(O) = [O,H]$. \par 

Let us define the retarded spin susceptibility in real time and frequency space:
\begin{equation}
    \chi^{ab}(\vec q,t) = \ii\Theta(t) \left\langle [O^a_{\vec q}(t),O_{-\vec q}^b]\right\rangle_\beta,\qquad \chi^{ab}(\vec q,\omega) = \int_{-\infty}^{\infty} \dd t \, e^{\ii\omega t} \chi^{ab}(\vec q,t).
\end{equation}
We define the dissipative part as $\chi^{\prime \prime\,ab}(\vec q,\omega) = \frac{\chi^{ab}(\vec q,\omega)-\chi^{ba}(\vec q,\omega)^*}{2\ii}$. In thermal equilibrium, the FDT relates the dissipative part of the retarded spin susceptibility to the DSF. In particular, we have from detailed balance that $S^{ba}(-\vec q,-\omega) = e^{-\beta \omega}S^{ab}(\vec q,\omega)$ and hence 
\begin{equation}\label{eq:FDTrelation}
    \begin{aligned}
        2\chi''^{ab}(\vec q,\omega) &= \int_{-\infty}^{\infty} \dd t\, e^{i\omega t} \left\langle [O_{\vec q}^a(t),O_{-\vec q}^b]\right\rangle_\beta \\
        &= S^{ab}(\vec q,\omega) - S^{ba}(-\vec q,-\omega)\\
        &= (1-e^{-\beta \omega})S^{ab}(\vec q,\omega).
    \end{aligned}
\end{equation}
which is one form of the FDT~\cite{coleman2015introduction}.
\par 
Next, we take spectral moments of Eq.~\eqref{eq:FDTrelation}. Integrating against $\omega$ yields
\begin{equation}
\label{eq:intSversusomegaFDT}
\begin{aligned}
2\int\dd\omega\,\omega\chi''^{ab}(\vec q,\omega)
&=\int\dd\omega\,\omega
\left[S^{ab}(\vec q,\omega)-S^{ba}(-\vec q,-\omega)\right]\\
&=m_1^{ab}(\vec q)+m_1^{ba}(-\vec q).
\end{aligned}
\end{equation}
We discuss higher powers of $\omega$ in \textsection\ref{sec:fdthigherbody}. Next, taking spatial harmonics of Eq.~\eqref{eq:intSversusomegaFDT} yields
\begin{equation}
  \frac{  m_{1,\vec d}^{ab} + m_{1,-\vec d}^{ba}}{2\pi} = \left\langle [O_{\vec 0}^a,H]O_{\vec d}^b\right\rangle_\beta + \left\langle [O_{\vec d}^b,H]O_{\vec 0}^a\right\rangle_\beta.
\end{equation}
Stationarity, in the form $[\rho_\beta,H]=0$ for the Gibbs state $\rho_\beta$, implies that $0 = \left\langle [O_{\vec d}^bO_{\vec 0}^a,H]\right\rangle_\beta = \left\langle [O_{\vec d}^b,H]O_{\vec 0}^a\right\rangle_\beta + \left\langle O_{\vec d}^b[O_{\vec 0}^a,H]\right\rangle_\beta$. Consequently, 
\begin{equation}
\label{eq:mab1mba1}
\begin{aligned}
     \frac{  m_{1,\vec d}^{ab} + m_{1,-\vec d}^{ba}}{2\pi} &= \left\langle [O_{\vec 0}^a,H]O_{\vec d}^b - O_{\vec d}^b[O_{\vec 0}^a,H]\right\rangle_\beta \\
     &= \left\langle [O_{\vec d}^b,[H,O_{\vec 0}^a]]\right\rangle_\beta.
\end{aligned}
\end{equation}
Crucially, Eq.~\eqref{eq:mab1mba1} holds when expectation values are with respect to any stationary state, not just the Gibbs state $\rho_\beta$. The FD protocol will thus apply to nonthermal
stationary states as well, with the same learning guarantees. Next, we substitute the Hamiltonian to obtain
\begin{equation}
\label{eq:mplusmequalA}
   \frac{ m_{1,\vec d}^{ab} + m_{1,-\vec d}^{ba} }{2\pi}= \sum_{j=1}^p \mathsf{A}_{\vec d,j}^{ab} \lambda^j,\qquad \mathsf{A}_{\vec d,j}^{ab} \coloneqq \left\langle [O_{\vec d}^b,[H^j,O_{\vec 0}^a]]\right\rangle_\beta,
\end{equation}
which is the central formula of the FD protocol. \par 

The left-hand side of Eq.~\eqref{eq:mplusmequalA} is obtained from the DSF data. However, it is not \textit{a priori} clear if the right-hand side can be obtained purely from the DSF (i.e., expressible as linear combinations of DSF spectral moments). While Eq.~\eqref{eq:mplusmequalA} becomes an extremely useful linear system for the couplings $\lambda^j$ if $\mathsf{A}_{\vec d,j}^{ab}$ can also be obtained directly from the DSF data, in general this need not be the case. \par

When is $\mathsf{A}_{\vec d,j}^{ab}$ obtainable from the DSF alone? One instance, as we prove in the following section, is when $H$ comprises only one- and two-body Pauli terms. As an illustrative example, consider the XYZ model $H = \sum_{\langle \vec r\vec r'\rangle } (J_x \sigma^x_{\vec r}\sigma^x_{\vec r'} + J_y \sigma^y_{\vec r}\sigma^y_{\vec r'} + J_z \sigma^z_{\vec r}\sigma^z_{\vec r'})$. Furthermore,  consider Eq.~\eqref{eq:mplusmequalA} with $a=b=y$ and $\vec d$ a nearest-neighbor vector. In this case, we find
\begin{equation}
\label{eq:XYZFDTexample}
\frac{1}{2\pi}(m_{1,\vec d}^{yy}+m_{1,-\vec d}^{yy})
=
4 J_x\langle \sigma^z_{\vec0}\sigma^z_{\vec d}\rangle_\beta
+
4J_z\langle \sigma^x_{\vec0}\sigma^x_{\vec d}\rangle_\beta.
\end{equation}
Notably, both $\langle \sigma^z_{\vec0}\sigma^z_{\vec d}\rangle_\beta$ and $\langle \sigma^x_{\vec0}\sigma^x_{\vec d}\rangle_\beta$ are zeroth-order DSF spectral moments: $\langle \sigma^z_{\vec0}\sigma^z_{\vec d}\rangle_\beta=m_{0,\vec d}^{zz}/2\pi$ and $\langle \sigma^x_{\vec0}\sigma^x_{\vec d}\rangle_\beta=m_{0,\vec d}^{xx}/2\pi$. Hence Eq.~\eqref{eq:XYZFDTexample} can be written
\begin{equation}
\label{eq:XYZFDTexample2}
m_{1,\vec d}^{yy}+m_{1,-\vec d}^{yy}
=4J_xm_{0,\vec d}^{zz} + 4J_zm_{0,\vec d}^{xx}. \end{equation}
Eq.~\eqref{eq:XYZFDTexample2} holds at \textit{arbitrary} temperature. Repeating the analysis for $a=b=x$ and $a=b=z$, we obtain the linear system
\begin{equation}
\begin{pmatrix}
m_{1,\vec d}^{xx}+m_{1,-\vec d}^{xx}\\
m_{1,\vec d}^{yy}+m_{1,-\vec d}^{yy}\\
m_{1,\vec d}^{zz}+m_{1,-\vec d}^{zz}
\end{pmatrix}
=
4
\begin{pmatrix}
0&m_{0,\vec d}^{zz}&m_{0,\vec d}^{yy}\\
m_{0,\vec d}^{zz}&0&m_{0,\vec d}^{xx}\\
m_{0,\vec d}^{yy}&m_{0,\vec d}^{xx}&0
\end{pmatrix}
\begin{pmatrix}
J_x\\J_y\\J_z
\end{pmatrix}.
\end{equation}
We note $a\neq b$ could yield more equations, but the above generically suffices for learning $J_x,J_y$, and $J_z$. The matrix on the right-hand side is $\mathsf{A}^{aa}_{\vec d,j}$, indexed by $(a,j)$; for brevity we simply refer to it as $\mathsf{A}$. If $\mathsf{A}$ is nonsingular, solving the matrix equation allows us to learn the couplings. While invertibility of $\mathsf{A}$ can have detailed dependence on the physics of the model, we will prove for this model that $\mathsf{A}$ is generically invertible in the thermodynamic limit at $T=0$.\par 

In fact, a similar construction applies to all one- and two-body Hamiltonians. For such Hamiltonians, each entry of $\mathsf{A}$ contains only one- and two-body operators, whose expectations are available from the zeroth-order spectral moments, yielding a matrix equation for $\vec \lambda$ which is exact at every temperature. The ability to invert this linear system depends on $\mathsf A$ having full column rank, and its robustness depends on the smallest singular value of $\mathsf{A}$. We dub this approach to solving for $\vec\lambda$ the FD protocol. In the next section, we will give explicit sufficient conditions for the success of the FD protocol. We treat higher than two-body interactions afterwards.

\par 

We note that for the nearest-neighbor isotropic Heisenberg chain, our procedure essentially reduces to applying the classic Hohenberg-Brinkman sum rule~\cite{hohenbergbrinkman}. The FD protocol can thus be regarded as a generalization of this sum rule to more general local Hamiltonians.

\subsection{Learning one- and two-body spin Hamiltonians}

\noindent We now consider a general translation-invariant spin-$\frac12$ Hamiltonian,
\begin{equation}
\label{eq:oneandtwobodypaulifdt}
    H(\vec \lambda) = \sum_{\vec r\in \Lambda} \left(b^a\sigma^a_{\vec r} + \sum_{\vec d\in \mathcal{R} } \lambda_{\vec d}^{ab}\sigma_{\vec r}^a\sigma^b_{\vec r+\vec d}\right),
\end{equation}
where $\mathcal{R}$ is a finite set of nonzero bond vectors excluding $\vec0$, with only one representative of each pair $\{\vec d,-\vec d\}$. The unknown couplings are $\vec \lambda= (b^a,\lambda_{\vec d}^{ab})$, with $\vec\lambda \in [-1,1]^p$ for $p = 9|\mathcal{R}|+3$. The following theorem provides sufficient conditions for learning $\vec\lambda$ using the FD protocol. 

\begin{theorem}[FD protocol for one- and two-body interactions] Consider the Hamiltonian above at a fixed temperature. Assume access to DSF spectral moments $m^{ab}_{n,\vec d}$ for $n=0,1$ and all $\vec d$. If 
\begin{equation}\label{eq:fdprotocoldetmcondition}
\mathrm{det} (m_{0,\vec d}^{ab})_{ab} \neq 0
\qquad \text{for every } \vec d\in\mathcal R,
\end{equation}
then the $\lambda_{\vec d}^{ab}$ can be determined uniquely from these data. Moreover, if $\|\langle \sigma_{\vec 0}^a\rangle_\beta\|_2 \neq 0$, then the $b^a$ can be determined uniquely from these data. 
\label{thm:FD2body}
\end{theorem}
\begin{proof}
    Using Eq.~\eqref{eq:mplusmequalA} and Pauli commutation relations gives
\begin{equation}
\label{eq:epsilonepsilonlambdam}
    \begin{aligned}
        \frac{m_{1,\vec d}^{ab}+m_{1,-\vec d}^{ba}}{2\pi} &= \lambda_{\vec d}^{a'b'} \left\langle [\sigma_{\vec d}^b,[\sigma_{\vec 0}^{a'}\sigma_{\vec d}^{b'},\sigma_{\vec 0}^a]]\right\rangle_\beta \\
        &= 4 \epsilon_{aa'c}\epsilon_{bb'c'}\lambda_{\vec d}^{a'b'}\left\langle \sigma_{\vec 0}^c\sigma_{\vec d}^{c'}\right\rangle_\beta \\
        &= 4 \epsilon_{aa'c}\epsilon_{bb'c'}\lambda_{\vec d}^{a'b'} \frac{m_{0,\vec d}^{cc'}}{2\pi}.
    \end{aligned}
\end{equation}
Thus, for each $\vec d$, we obtain a linear system for the $\lambda_{\vec d}^{ab}$. We now show that Eq.~\eqref{eq:fdprotocoldetmcondition} guarantees a unique solution for $\lambda_{\vec d}^{ab}$. To this end, define 
\begin{equation}
\label{eq:FLambdaCsupp}
\mathsf{F} \coloneqq \left( \frac{m_{1,\vec d}^{ab}+m_{1,-\vec d}^{ba}}{2\pi}\right)_{ab},\quad   \mathsf{\Lambda} \coloneqq \left(4\lambda_{\vec d}^{ab}\right)_{ab},\quad    \mathsf{C} \coloneqq \left(\frac{m_{0,\vec d}^{ab}}{2\pi}\right)_{ab},
\end{equation}
so that Eq.~\eqref{eq:epsilonepsilonlambdam} can be expressed as $\mathsf{F}_{ab} =\epsilon_{aa'c} \epsilon_{bb'c'} \mathsf{\Lambda}_{a'b'}\mathsf{C}_{cc'}$. We solve for $\mathsf{\Lambda}$. If Eq.~\eqref{eq:fdprotocoldetmcondition} holds, we may define $\mathsf{K}=\mathsf{C}^{-1}\mathsf{\Lambda}$ and rewrite this as
\begin{equation}
\label{eq:FepsilonCCK}   
    \mathsf{F}_{ab} = \varepsilon_{aa'a''}\varepsilon_{bb'b''} \mathsf{C}_{a'c}\mathsf{C}_{a''b''} \mathsf{K}_{cb'}.
\end{equation}
Contracting both sides of $\epsilon_{a'''a'a''}\mathsf{C}_{a'''c'}\mathsf{C}_{a'c}\mathsf{C}_{a''b''} = (\det \mathsf{C})\epsilon_{c'cb''}$ with $(\mathsf{C}^{-1})_{c'a}$ yields 
\begin{equation}
    \label{eq:epsilonsandCs}   \epsilon_{aa'a''}\mathsf{C}_{a'c}\mathsf{C}_{a''b''} = (\det \mathsf{C})(\mathsf{C}^{-1})_{c'a}\epsilon_{c'cb''}.
\end{equation}
Substituting Eq.~\eqref{eq:epsilonsandCs} into Eq.~\eqref{eq:FepsilonCCK} and contracting Levi-Civita symbols yields
\begin{equation}
\begin{aligned}
\mathsf{F}_{ab} &= (\det\mathsf{C})(\mathsf{C}^{-1})_{c'a}\epsilon_{c'cb''}\epsilon_{bb'b''}\mathsf{K}_{cb'}\\
&= (\det\mathsf{C})(\mathsf{C}^{-1})_{c'a} (\delta_{c'b}\delta_{cb'}-\delta_{c'b'}\delta_{cb})\mathsf{K}_{cb'}\\
&= (\det \mathsf{C})\left[(\mathsf{C}^{-1})_{ba}\tr \mathsf{K} -(\mathsf{K}\mathsf{C}^{-1})_{ba}\right]
\end{aligned}
\end{equation}
Reorganizing, we find
\begin{equation}
\label{eq:FtransposeCdetCtrKmK}
    \mathsf{F}^\top\mathsf{C} = (\det \mathsf{C})\left[ (\tr\mathsf{K} )\mathsf{I} - \mathsf{K}\right].
\end{equation}
Next, we take the trace to find $\tr(\mathsf{F}^\top \mathsf C) = (\det \mathsf{C})(3-1)(\tr\mathsf{K})$. Rearranging, we have
\begin{equation}
\label{eq:trKformulaFDT}
    \tr\mathsf{K} = \frac{\tr(\mathsf{F}^\top \mathsf{C})}{2\det \mathsf{C}}.
\end{equation}
Multiplying Eq.~\eqref{eq:FtransposeCdetCtrKmK} on the left by $\mathsf{C}$ now gives
\begin{equation}
    \mathsf{C}\mathsf{F}^\top \mathsf{C} = (\det\mathsf{C})[(\tr\mathsf{K})\mathsf{C}-\mathsf{\Lambda}].
\end{equation}
Solving for $\mathsf{\Lambda}$ and using Eq.~\eqref{eq:trKformulaFDT} finally gives us
\begin{equation}
\label{eq:solutionforLambda}
    \mathsf{\Lambda} = \frac{1}{\det \mathsf{C}} \left[\frac12 \tr(\mathsf{F}^\top \mathsf{C}) \mathsf{C} - \mathsf{C}\mathsf{F}^\top \mathsf{C}\right].
\end{equation}
This gives the unique solution whenever $\det \mathsf{C}\neq 0$. \par 

Next, we turn to the Zeeman fields $b^a$. Let us write $M^a=\langle\sigma_{\vec0}^a\rangle_\beta$ and note that
\begin{equation}
    \frac{m_{0,\vec0}^{ab}}{2\pi}
    =\delta_{ab}+i\epsilon_{abc}M^c.
\end{equation}
We return to Eq.~\eqref{eq:mplusmequalA}, now taking
$\vec d=\vec0$. Evaluating the double commutator gives
\begin{equation}
\begin{aligned}
\frac{m_{1,\vec0}^{ab}+m_{1,\vec0}^{ba}}{2\pi}
={}&4(M^a b^b-\delta_{ab}M^c b^c)+\sum_{\vec d\in\mathcal R}
\Big[
(\mathsf C_{\vec d}\mathsf\Lambda_{\vec d}^{\top}
+\mathsf C_{\vec d}^{\top}\mathsf\Lambda_{\vec d})_{ab}
-2\delta_{ab}\tr(\mathsf\Lambda_{\vec d}^{\top}
\mathsf C_{\vec d})
\Big],
\end{aligned}
\end{equation}
where we have added $\vec d$-labels to the matrices
$\mathsf C$ and $\mathsf\Lambda$ defined previously. Using Eq.~\eqref{eq:solutionforLambda}, the entire second term is known. Let $\mathsf G_{ab}$ denote the left-hand
side after subtracting this piece:
\begin{equation}
\label{eq:FDTdefiningG}
    \mathsf G_{ab}
    =4(M^a b^b-\delta_{ab}M^c b^c).
\end{equation}
Taking the trace yields
$\tr\mathsf G=-8M^c b^c$. Substituting this back gives
\begin{equation}
    \mathsf G_{ab}
    -\frac12\delta_{ab}\tr\mathsf G
    =4M^a b^b.
\end{equation}
Since $\vec M\neq0$, there exists at least one component $M^a\neq0$.
Dividing by $4M^a$ for this fixed $a$, the three equations above determine
$b^x,b^y,b^z$ uniquely, completing the proof.
\end{proof}

Under the stated conditions, the preceding theorem establishes global solvability of the couplings $\vec\lambda$ without a high-temperature or perturbative expansion (see Appendix~\ref{sec:strong_field_two_body} for treatment of the latter). We next quantify the accuracy of the solution for $\lambda_{\vec d}^{ab}$ when spectral moments are measured imperfectly. We arrive at the natural conclusion that the value of the determinant in Eq.~\eqref{eq:fdprotocoldetmcondition} determines the robustness of the solution.

\begin{corollary}[Accuracy of FD protocol] For a nonzero $\vec d\in\mathcal{R}$, let $\widehat{\mathsf{F}}$ and $\widehat{\mathsf{C}}$ be measured estimates of the matrices $\mathsf{F}$ and $\mathsf{C}$ [Eq.~\eqref{eq:FLambdaCsupp}], respectively. Suppose
\begin{equation}
\label{eq:FDTcorollaryconditions}
    \max \{\|\widehat{\mathsf{C}}-\mathsf{C}\|_\mathrm{F},\|\widehat{\mathsf{F}}-\mathsf{F}\|_{\mathrm{F}}\} \leq \varepsilon,\quad |\det \widehat{\mathsf{C}}|\geq \gamma > 0,
\end{equation}
where $\|\cdot\|_{\mathrm{F}}$ denotes the Frobenius norm. Let $\widehat{\lambda}^{ab}_{\vec d} = \widehat{\mathsf{\Lambda}}_{ab}/4$, where $\widehat{\mathsf{\Lambda}}$ is the solution obtained by substituting $\widehat{\mathsf{F}}$ and $\widehat{\mathsf{C}}$ into Eq.~\eqref{eq:solutionforLambda}. Then 
\begin{equation}
\label{eq:FDTcorollarylambdasupp}
    \max_{a,b} \left|\widehat{\lambda}^{ab}_{\vec d}-\lambda^{ab}_{\vec d}\right| \leq \frac{25(1+\varepsilon)^2}{4\gamma} \varepsilon
\end{equation}
at any temperature, independent of system size. 
\label{cor:FDaccuracy}
\end{corollary}
\begin{proof}
    We suppress the $\vec d$ label. We have the following relation among $\widehat{\mathsf{\Lambda}}$, $\widehat{\mathsf{F}}$ and $\widehat{\mathsf{C}}$: 
    \begin{equation}
        \widehat{\mathsf{F}}_{ab} = \epsilon_{aa'c}\epsilon_{bb'c'}\widehat{\mathsf\Lambda}_{a'b'}\widehat{\mathsf C}_{cc'}.
    \end{equation}
    Subtracting the exact equation (without $\widehat{\cdot}\,\,$'s) gives
    \begin{equation}
    \label{eq:twoequationsforEabFDTsupp}
        \begin{aligned}        &\epsilon_{aa'c}\epsilon_{bb'c'}\left(\widehat{\mathsf{\Lambda}}- \mathsf{\Lambda}\right)_{a'b'}\widehat{\mathsf{C}}_{cc'} = \mathsf{E}_{ab},\\
            &\mathsf{E}_{ab} = \left(\widehat{\mathsf{F}}-\mathsf{F}\right)_{ab} - \epsilon_{aa'c}\epsilon_{bb'c'}\mathsf{\Lambda}_{a'b'}\left(\widehat{\mathsf{C}}-\mathsf{C}\right)_{cc'}.
        \end{aligned}
    \end{equation}
   It follows that
   \begin{equation}
   \label{eq:EF25epsilonFDT}
       \begin{aligned}
           \|\mathsf{E}\|_{\mathrm{F}} &\leq \|\widehat{\mathsf{F}}-\mathsf{F}\|_{\mathrm{F}} + 2\|\mathsf{\Lambda}\|_{\mathrm{F}} \|\widehat{\mathsf{C}}-\mathsf{C}\|_{\mathrm{F}}\\
           &\leq 25\varepsilon
       \end{aligned}
   \end{equation}
   where in the last line we used $|\lambda_{\vec d}^{ab}|\leq 1$ and hence $\|\mathsf{\Lambda}\|_{\mathrm{F}} = 4\left(\sum_{ab}|\lambda_{\vec d}^{ab}|^2\right)^{1/2} \leq 12$. Next, applying Eq.~\eqref{eq:solutionforLambda} to Eq.~\eqref{eq:twoequationsforEabFDTsupp} gives 
\begin{equation}
\widehat{\mathsf{\Lambda}}-\mathsf{\Lambda} = \frac{\widehat{\mathsf{C}}}{\det\widehat{\mathsf{C}}} \left[ \frac12 \tr(\mathsf{E}^\top \widehat{\mathsf{C}})\mathsf{I}-\mathsf{E}^\top \widehat{\mathsf{C}}\right].
\end{equation}
For any $3\times 3$ matrix $\mathsf{X}$, a direct expansion establishes the inequality 
\begin{equation}
\label{eq:XtrinequalityFDT}
    \left\| \frac12 (\tr \mathsf{X})\mathsf{I}-\mathsf{X}\right\|_{\mathrm{F}}^{2} = \|\mathsf{X}\|_{\mathrm{F}}^2 - \frac14 |\tr \mathsf{X}|^2 \leq \|\mathsf{X}\|_{\mathrm{F}}^2.
\end{equation}
Next, we will need to control multiplication by $\widehat{\mathsf C}$. First consider the exact version, $\mathsf C_{ab} = \langle\sigma_{\vec0}^a\sigma_{\vec d}^b\rangle_\beta.$ For real unit vectors $\vec v,\vec w$,
\begin{equation}
\left|\vec v^{\,\top}\mathsf C\vec w\right|
=
\left|
\left\langle
(\vec v\cdot\vec\sigma_{\vec0})
(\vec w\cdot\vec\sigma_{\vec d})
\right\rangle_\beta
\right|
\leq1,
\end{equation}
because these two spin operators commute and each has eigenvalues $\pm1$. Maximizing over $\vec v$ therefore gives
\begin{equation}
\|\mathsf C\vec w\|_2\leq\|\vec w\|_2.
\end{equation}
The same holds with $\mathsf C^\top$. Applying these inequalities to the columns or rows of any matrix $\mathsf X$ implies
\begin{equation}
\|\mathsf C\mathsf X\|_{\mathrm F}\leq\|\mathsf X\|_{\mathrm F},
\qquad
\|\mathsf X\mathsf C\|_{\mathrm F}\leq\|\mathsf X\|_{\mathrm F}.
\end{equation}
Next, we may use the triangle inequality and submultiplicativity of the Frobenius norm ($\|\mathsf{AB}\|_{\mathrm{F}}\leq \|\mathsf{A}\|_{\mathrm{F}}\|\mathsf{B}\|_{\mathrm{F}}$) to establish
\begin{equation}
\begin{aligned}
\|\widehat{\mathsf C}\mathsf X\|_{\mathrm F}
&\leq
\|\mathsf C\mathsf X\|_{\mathrm F}
+
\|(\widehat{\mathsf C}-\mathsf C)\mathsf X\|_{\mathrm F}\\
&\leq
\|\mathsf X\|_{\mathrm F}
+
\|\widehat{\mathsf C}-\mathsf C\|_{\mathrm F}
\|\mathsf X\|_{\mathrm F}\\
&\leq
(1+\varepsilon)\|\mathsf X\|_{\mathrm F}.
\end{aligned}
\end{equation}
Likewise,
\begin{equation}
\|\mathsf X\widehat{\mathsf C}\|_{\mathrm F}
\leq(1+\varepsilon)\|\mathsf X\|_{\mathrm F}.
\end{equation}
We now apply these bounds twice, together with Eq.~\eqref{eq:XtrinequalityFDT}:
\begin{equation}
\begin{aligned}
\|\widehat{\mathsf\Lambda}-\mathsf\Lambda\|_{\mathrm F}
&\leq
\frac{1+\varepsilon}{|\det\widehat{\mathsf C}|}
\left\|
\frac12\operatorname{tr}(\mathsf E^\top\widehat{\mathsf C})\mathsf I
-\mathsf E^\top\widehat{\mathsf C}
\right\|_{\mathrm F}\\
&\leq
\frac{1+\varepsilon}{|\det\widehat{\mathsf C}|}
\|\mathsf E^\top\widehat{\mathsf C}\|_{\mathrm F}\\
&\leq
\frac{(1+\varepsilon)^2}{|\det\widehat{\mathsf C}|}
\|\mathsf E\|_{\mathrm F}.
\end{aligned}
\end{equation}
Together with $|\det \widehat{\mathsf{C}}|\geq \gamma > 0$ and Eq.~\eqref{eq:EF25epsilonFDT}, we have
\begin{equation}
    \|\widehat{\mathsf\Lambda}-\mathsf\Lambda\|_{\mathrm F} \leq \frac{25(1+\varepsilon)^2}{\gamma}\varepsilon.
\end{equation}
Finally, the Frobenius norm bounds the magnitude of every matrix entry. Thus, using $\mathsf{\Lambda}_{ab} = 4\lambda^{ab}_{\vec d}$, Eq.~\eqref{eq:FDTcorollarylambdasupp} follows. 
\end{proof}

Eq.~\eqref{eq:FDTcorollaryconditions} implies that if $|\det\widehat{\mathsf{C}}|$ is bounded away from zero, then one has an explicit accuracy guarantee for the learning of the couplings $\lambda^{ab}_{\vec d}$.

\subsection{Higher-body interactions and higher spectral moments }
\label{sec:fdthigherbody}

\noindent In the previous section, we established conditions under which the FD protocol can be used at any temperature to learn the couplings of translation-invariant spin-$\frac12$ Hamiltonians with at most two-body interactions. To accomplish this, we need only solve a linear system to find the globally unique solution. In this section, we address possible extensions of the FD protocol to Hamiltonian families with higher-body interactions. 
\par 
Whether the FD protocol can be extended to higher-body interactions requires a nuanced analysis. To appreciate this, we may consider  Eq.~\eqref{eq:mplusmequalA} with $H^j$ comprising three-body interactions. By Pauli commutation algebra, we find that $\mathsf{A}$ consists of expectation values of products of \textit{three} Pauli operators. In general, such an expectation value cannot necessarily be extracted from DSF data. Hence, in general, studying higher-body interactions with Eq.~\eqref{eq:mplusmequalA} can involve solving not only for unknown \textit{couplings} but also unknown \textit{correlation functions}, and the solvability of the inverse problem requires a heavier analysis. \par 

We also observe that there is a natural hierarchy of generalizations of Eq.~\eqref{eq:mplusmequalA}. Recall that $ m^{ab}_{n,\vec d} = 2\pi \left\langle \mathrm{ad}^n_H\left(\sigma_{\vec 0}^a\right)\sigma_{\vec d}^b\right\rangle_\beta.$ Stationarity gives us $\langle [A,H]B\rangle_\beta = -\langle A[B,H]\rangle_\beta$. Applying this identity repeatedly and using translation invariance, we find
\begin{equation}
    \frac{m_{n,-\vec d}^{ba}}{2\pi} = \left\langle \mathrm{ad}_H^n\left(\sigma_{\vec d}^b\right)\sigma_{\vec 0}^a\right\rangle_\beta = (-1)^n \left\langle \sigma_{\vec d}^b\left(\mathrm{ad}_H^n\sigma_{\vec 0}^a\right)\right\rangle_\beta
\end{equation}
and consequently
\begin{equation}
    m_{n,\vec d}^{ab}-(-1)^nm_{n,-\vec d}^{ba} = 2\pi \left\langle \left[ \mathrm{ad}_H^n\left(\sigma_{\vec 0}^a\right),\sigma_{\vec d}^b\right]\right\rangle_\beta. 
\end{equation}
The left-hand side is identified with spectral moment harmonics of $S^{ab}(\vec q,\omega)-(-1)^nS^{ba}(-\vec q,\omega)$. This generalizes Eq.~\eqref{eq:mplusmequalA} to $n>1$. \par 

These additional equations can, again, introduce additional unknown correlation functions into the system of equations. For interactions acting on at most $k$ sites, each $\mathrm{ad}_H$ can add at most $k-1$ sites to a Pauli operator $\sigma_{\vec 0}^a$. The identity above can therefore involve correlation functions of up to $1+n(k-1)$ spins. For example, with three-body interactions, the second-moment equations are quadratic in the couplings but can contain five-spin correlations. Thus higher moments in $n$ do not automatically restore the closure of the system of equations which we observed in the one- and two-body case. A reconstruction based on this hierarchy must additionally establish that the unknown correlation functions which appear can be eliminated or determined jointly with the couplings. We leave a more in-depth analysis to future work.

\subsection{Temperature and field dependence}
\label{sec:temp_and_field_fdt}

\noindent We now return to the case of one- and two-body Hamiltonians and study how the performance of the FD protocol depends on temperature and applied Zeeman field. One of our goals is to compare and contrast the FD protocol with the SF and HT protocols. Theorem~\ref{thm:FD2body} establishes global solvability under conditions on $m_{0,\vec d}^{ab}$ and $\langle \vec\sigma_{\vec 0}\rangle_\beta$, but does not specify when these conditions are met, which ultimately depends on the Hamiltonian family $H(\vec\lambda)$. To analytically better understand the operational regimes of the FD protocol, we first examine the high-temperature and strong-field limits, and then prove for the XYZ model that the conditions of Thm.~\ref{thm:FD2body} are generically met down to zero temperature in the thermodynamic limit. \par 

We first make the elementary observation that at infinite temperature, each entry of $\mathsf{A}$ vanishes because it is the trace of a commutator:
\begin{equation}
    \mathsf{A}^{ab}_{\vec d,j} \left.\right|_{\beta = 0} = \left\langle [\sigma_{\vec d}^b,[H^j,\sigma_{\vec 0}^a]]\right\rangle_{\beta = 0} = 0.
\end{equation}
The HT expansion gives $\mathsf{A}= O(\beta)$ at small $\beta$, so its singular values vanish as $\beta \to 0$. Consequently, even when $\mathsf A$ has full column rank at finite $\beta$, its vanishing singular values increasingly amplify measurement errors. For comparison, increasing the temperature reduces the truncation error in the HT protocol (in the absence of noise, or up to a noise-dependent threshold in the presence of noise) of App.~\ref{sec:hightemp}. Hence the FD and HT protocols are complementary along the temperature axis. \par 

We next consider application of a strong Zeeman field in a fixed direction $\hat n$, as in the SF protocol of App.~\ref{sec:suppprobefield}. When the field is large compared with the scale of the couplings and the temperature, the system approaches a fully polarized state and two-point correlations factorize:
\begin{equation}
    (\mathsf{C}_{\vec d})_{ab} = \langle \sigma_{\vec 0}^a\sigma_{\vec d}^b\rangle \longrightarrow \langle \sigma_{\vec 0}^a\rangle\langle\sigma_{\vec d}^b\rangle.
\end{equation}
This implies $\rank (\mathsf{C}_{\vec d})_{ab}\longrightarrow 1$ in this limit, and hence $\mathsf{C}$ becomes singular, implying by Eq.~\ref{eq:FDTcorollaryconditions} that the FD protocol becomes ill-conditioned. This is in contrast with the SF protocol, which uses a different set of spectral moment identities and becomes most useful at strong fields when perturbation theory is valid. Hence the FD and SF protocols are complementary along the axis of applied Zeeman field. 
\par 
Having identified these limits, it is instructive to ask whether the FD protocol can in principle succeed at low temperatures. To our knowledge, there is no general or even average-case constraint on the invertibility of $\mathsf{A}$ at low temperatures $T\to 0$. For many common Hamiltonians of interest, however, we have found numerically that $\mathsf{A}$ remains well-conditioned down to $T=0$ and hence the FD protocol often succeeds, as we detail in the next section. However, the FD protocol cannot \textit{always} succeed at low temperatures due to the worst-case hardness result in App.~\ref{sec:supphardness}. In particular, App.~\ref{sec:supphardness} provides a Hamiltonian that cannot be learned from the DSF at low temperatures without exponential resources (e.g., exponentially small measurement errors in the moments) because $|\pdv{U} S_{ab}(\omega) | \lesssim e^{-\beta h}\ll 1$, with $h$ and $U$ the energy scales of $H$. Attempting the FD protocol on this Hamiltonian, we would find $\det \mathsf{C} =0$.  \par

Does the FD protocol work in the average case at low temperatures, in the sense of everywhere except a measure-zero set in the space of couplings? It is nontrivial to establish any general functional properties of $\det \mathsf{A}(\vec \lambda)$ at low temperatures in the thermodynamic limit, even for translation-invariant local Hamiltonians, while in the high-temperature limit one could appeal to a high-temperature expansion to establish that $\det \mathsf{A}$ is a low-degree polynomial in $\vec\lambda$. Proving strong structural properties of $\det \mathsf{A}(\vec \lambda)$, such as analyticity, seems essential for showing that it is nonzero except at a measure-zero set. We leave this as an open question for future work.\par

It is enlightening to study the zero-temperature behavior of the FD protocol for a particular model. Moreover, the following analysis may serve as a useful blueprint for other specific models. As in the beginning of this Appendix, let us consider the XYZ model
\begin{equation}
    H(\vec\lambda) = \sum_j (J_x \sigma^x_j\sigma^x_{j+1}+J_y \sigma^y_j\sigma^y_{j+1}+J_z\sigma^z_j\sigma^z_{j+1})
\end{equation}
in one dimension, with $\vec\lambda = (J_x,J_y,J_z)$ unknown. Writing $\vec c= (c_x,c_y,c_z)$ where $c_a = \langle \sigma_0^a\sigma_1^a\rangle$, we have the following linear system:
\begin{equation}
\frac{1}{2\pi}
\begin{pmatrix}
m_{1,1}^{xx}+m_{1,-1}^{xx}\\
m_{1,1}^{yy}+m_{1,-1}^{yy}\\
m_{1,1}^{zz}+m_{1,-1}^{zz}\\
2m_{1,0}^{xx}\\
2m_{1,0}^{yy}\\
2m_{1,0}^{zz}
\end{pmatrix}
=
\mathsf A_{\mathrm{XYZ}}
\begin{pmatrix}
J_x\\J_y\\J_z
\end{pmatrix},
\end{equation}
where
\begin{equation}
\mathsf A_{\mathrm{XYZ}}=4
\begin{pmatrix}
0&c_z&c_y\\
c_z&0&c_x\\
c_y&c_x&0\\
0&-2c_y&-2c_z\\
-2c_x&0&-2c_z\\
-2c_x&-2c_y&0
\end{pmatrix}.
\end{equation}
Direct multiplication gives $\mathsf A_{\mathrm{XYZ}}^\top\mathsf A_{\mathrm{XYZ}}
=16(
\|\vec c\|^2 I
+2\,\operatorname{diag}(c_x^2,c_y^2,c_z^2)
+5\vec c\,\vec c^\top)$. The last two terms are positive semidefinite, implying that the smallest singular value of $\mathsf{A}_{\mathrm{XYZ}}$ satisfies 
\begin{equation}
    \sigma_{\mathrm{min}}(\mathsf{A}_{\mathrm{XYZ}}) \geq 4\|\vec c\|.
\end{equation}
Hence, if $\vec c\neq \vec 0$, the linear system determined by $\mathsf{A}_{\mathrm{XYZ}}$ is solvable. In other words, a sufficient condition for success of FD tomography is $\vec c\neq\vec 0$. We now study when this condition holds. \par 

With a simple variational argument, we may establish that $\vec c\neq \vec 0$, in general, for the XYZ model at zero temperature. Define $J_{\mathrm{max}} = \max\{|J_x|,|J_y|,|J_z|\}$ and choose a product state ansatz ordered along the axis of a coupling attaining this maximum, ferromagnetically if that coupling is negative and antiferromagnetically if it is positive. This ansatz has energy density $-J_{\mathrm{max}}$. Denoting the exact ground-state energy density by $e_0$, we have $e_0 = J_x c_x + J_y c_y + J_z c_z \leq -J_{\mathrm{max}}$. Then, by the Cauchy-Schwarz inequality, we have 
\begin{equation}
    \|\vec c\| \|\vec \lambda\| \geq |e_0| \geq J_{\mathrm{max}}.
\end{equation}
Dividing by $\|\vec \lambda\|$ (assumed nonzero) and using that $J_{\mathrm{max}} / \sqrt{J_x^2+J_y^2+J_z^2} \geq \frac{1}{\sqrt{3}}$, we have 
\begin{equation}
\sigma_{\mathrm{min}}(\mathsf{A}_{\mathrm{XYZ}}) \geq \frac{4}{\sqrt{3}}.
\end{equation}
We remark that this implies the learnability of the XYZ model from the FD protocol at zero temperature for arbitrary nonzero coupling vectors, and we have not relied on the existence of a spectral gap, a nondegenerate ground state, analytic dependence of the ground state on the couplings, or any perturbative small parameter. This example therefore shows that low temperature need not degrade the FD protocol, even though it cannot succeed for every Hamiltonian. We next study the FD protocol's performance numerically across a broad collection of models, with particular attention to temperature, applied field, and measurement noise.

\subsection{Numerics}
\label{sec:fdtnumerics}

\noindent In this section, we numerically test the FD protocol across various Hamiltonian families and temperature and noise regimes. Measurement noise is added to the DSF moments $m_{n,\vec d}^{ab}/(2\pi)$ as independent, zero-mean Gaussian noise of standard deviation $\sigma$. We construct both sides of Eq.~\eqref{eq:mplusmequalA} using noisy DSF spectral moments and solve for the couplings $\boldsymbol{\lambda}_{\mathrm{pred}}$ by least squares. This is compared with the true values $\boldsymbol{\lambda}_{\mathrm{true}}$ by evaluating 
\begin{equation}
\frac{\|\boldsymbol{\lambda}_{\mathrm{pred}}-\boldsymbol{\lambda}_{\mathrm{true}}\|}
{\|\boldsymbol{\lambda}_{\mathrm{true}}\|},
\end{equation}
which serves as our performance metric.
\par 
\begin{figure}
    \centering
    \includegraphics[width=\linewidth]{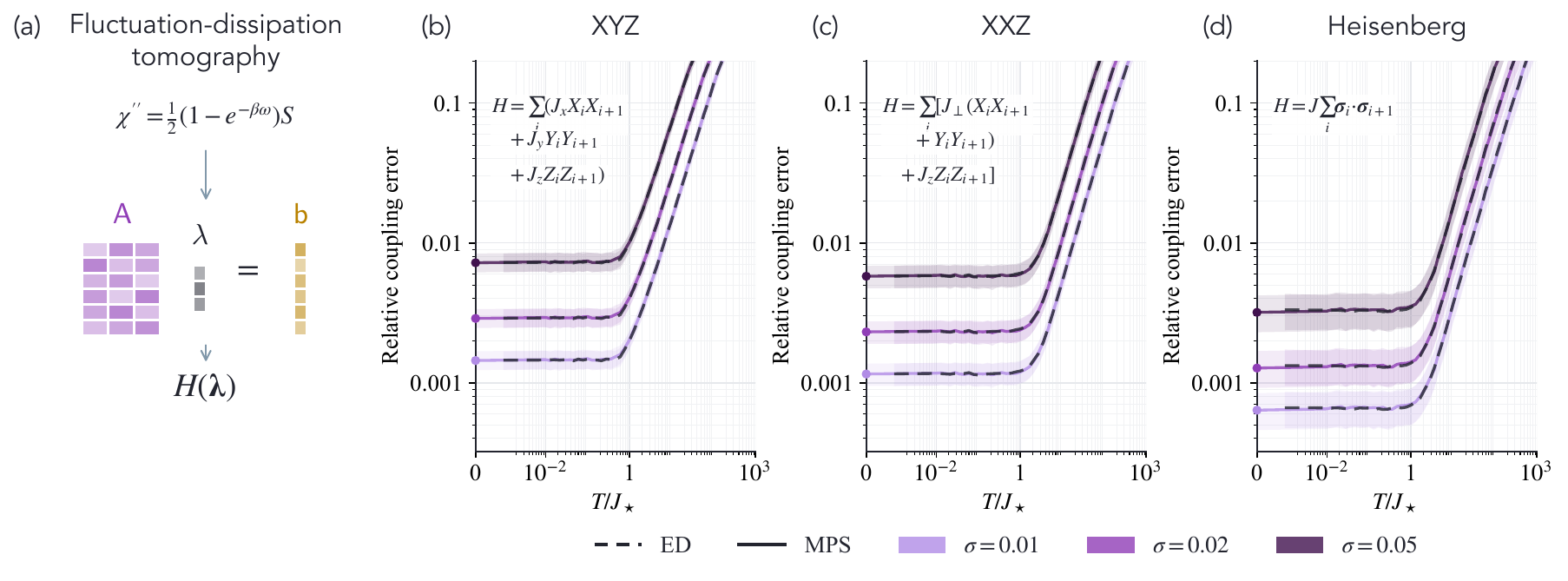}
    \caption{\textbf{Fluctuation-dissipation (FD) tomography.} (a) Fluctuation-dissipation (FD) tomography allows us to extract the Hamiltonian from the DSF at arbitrary temperatures by solving a remarkably simple linear system inspired by the FD theorem. (b-d) We evaluate the practical performance of the FD protocol in the presence of measurement noise $\sigma$, across a wide range of temperatures, for simple 1D spin chains. We plot the relative error between predicted and true couplings for the XYZ, XXZ, and Heisenberg chains with Hamiltonians shown and $(J_x,J_y,J_z)=(0.4,0.2,-1)J_\star$, $(J_\perp,J_z)=(1,1.5)J_\star$, and $J=J_\star$, respectively. DSF data are generated using $N=16$ ED (dashed) or infinite-chain MPS (solid). Lines show medians over 5,000 noise trials, and ribbons have a 25 percentile width. Circles mark MPS ground-state results.}
\label{fig:fig1matrixfdt}
\end{figure}

In Fig.~\ref{fig:fig1matrixfdt}, we consider the FD protocol for the XYZ, XXZ, and Heisenberg chains, varying temperature at three noise strengths, $\sigma=0.01,0.02,0.05$. We observe that, while the FD protocol steadily degrades at larger temperatures, below $T/J_\star \sim 1$ (where $J_\star$ is the coupling scale) it achieves relative coupling errors of order $10^{-2}$. We apply the FD protocol using DSF data derived both from finite-size ED and infinite matrix product states (MPS), which probe the thermodynamic limit, finding excellent agreement between the two. We note that the learnability in these XYZ models using the FD protocol at $T=0$ was established analytically in the previous section, while the loss of accuracy at high temperatures is understood as the result of $\mathsf{A}=O(\beta)$. \par 

\begin{figure}
    \centering
    \includegraphics[width=\linewidth]{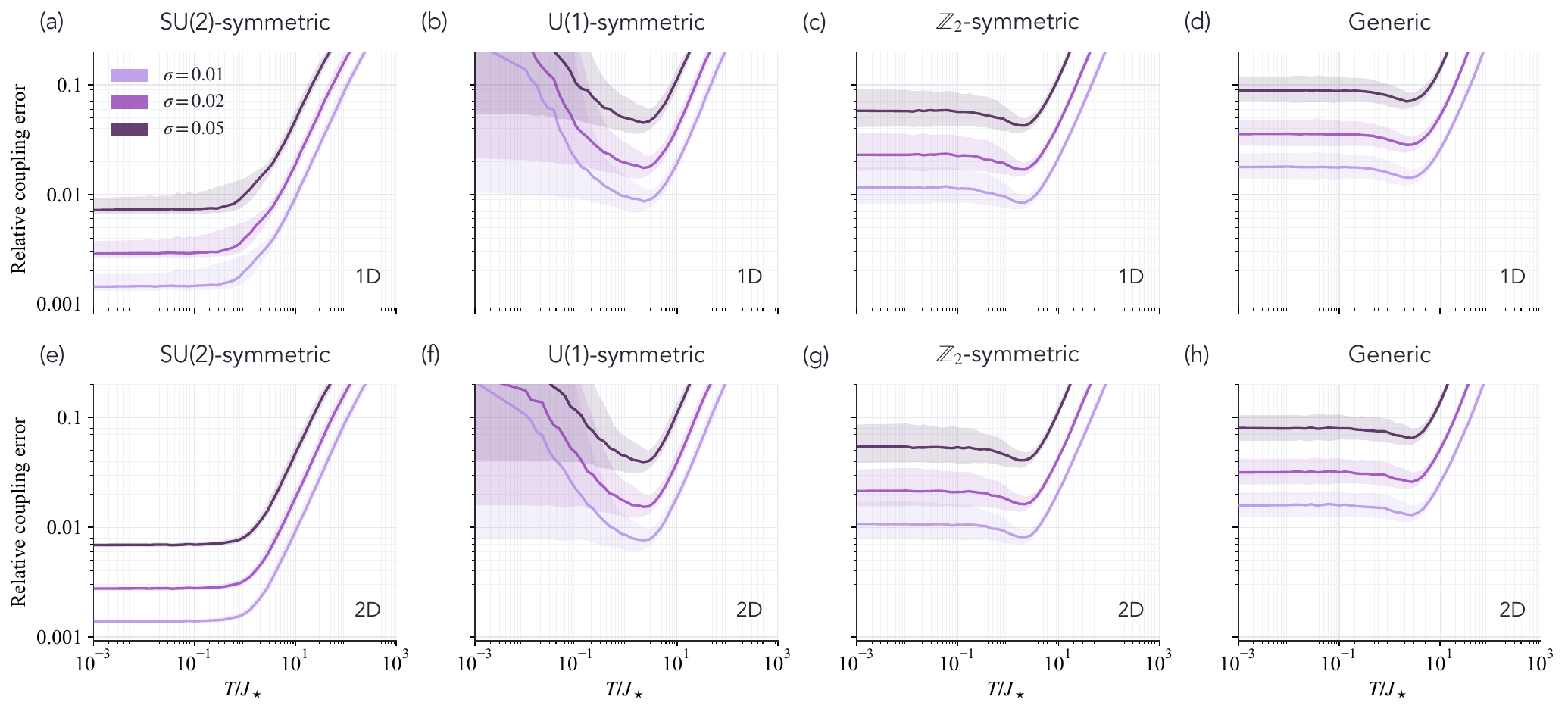}
    \caption{\textbf{Average-case performance of FD tomography.} Performance of FD tomography across broad Hamiltonian families in 1D (top row) and 2D (bottom row) with varying degrees of symmetry. (a--d) Periodic $N=12$ 1D Hamiltonians with nearest- and
    next-nearest-neighbor interactions. (e--h) Periodic $4\times3$ square lattice Hamiltonians with nearest-neighbor interactions. Columns show SU(2), U(1), $\mathbb Z_2$, and generic Hamiltonian families (2, 7, 11, and 21 parameters, respectively). For each family and geometry, we sample 1,000 Hamiltonians with independent symmetry-allowed couplings drawn uniformly from $[-1,1]$. We calculate the DSF spectral moments using ED and apply the FD protocol suited to the given Hamiltonian family. We average $M=100$ noise realizations at each labeled $\sigma$, plot the median relative coupling error, and show 25 percentile width ribbons.}
    \label{fig:fdtsymmetry}
\end{figure}

Having observed the success of the FD protocol at low temperatures for these familiar Hamiltonians, it is natural to ask how generic the success of the FD protocol is across broader Hamiltonian families. While we have pursued an analytical proof of its robustness for XYZ Hamiltonians in the previous section, here we instead numerically study its performance across broad ensembles of Hamiltonians [Fig.~\ref{fig:fdtsymmetry}]. We categorize these ensembles by dimension (1D and 2D) and symmetry (SU(2), U(1), $\mathbb{Z}_2$, or no onsite symmetry).  In particular, we study four translation-invariant subfamilies of
Eq.~\eqref{eq:oneandtwobodypaulifdt}; we take $\mathcal R=\{1,2\}$ in 1D and
$\mathcal R=\{\hat{\boldsymbol{x}},\hat{\boldsymbol{y}}\}$
in 2D
and impose
\begin{itemize}
\item $\mathrm{SU}(2):
b^a=0,
\lambda_{\vec{d}}^{ab}=\lambda_{\vec{d}}\delta_{ab}$;
\item $\mathrm{U}(1):
b^x=b^y=0,
(\lambda_{\vec{d}}^{ab})_{ab}=
\begin{pmatrix}
\lambda_{\vec{d}}^\perp&\lambda_{\vec{d}}^{xy}&0\\
-\lambda_{\vec{d}}^{xy}&\lambda_{\vec{d}}^\perp&0\\
0&0&\lambda_{\vec{d}}^{zz}
\end{pmatrix}$;
\item $
\mathbb Z_2:
b^x=b^y=0,
(\lambda_{\vec{d}}^{ab})_{ab}=
\begin{pmatrix}
\lambda_{\vec{d}}^{xx}&\lambda_{\vec{d}}^{xy}&0\\
\lambda_{\vec{d}}^{yx}&\lambda_{\vec{d}}^{yy}&0\\
0&0&\lambda_{\vec{d}}^{zz}
\end{pmatrix}$
\end{itemize}
with all nonzero couplings above treated as unknowns. Finally, the family labeled ``generic" allows all $b^a$, $\lambda_{\vec d}^{ab}$. These families have 2, 7, 11, and 21 independent parameters,
respectively. We sample each unknown parameter uniformly from $[-1,1]$ and apply the FD protocol to each family.\par 

In Fig.~\ref{fig:fdtsymmetry}, we observe that the FD protocol for the SU(2), $\mathbb Z_2$, and generic ensembles performs reasonably well at low temperatures in both 1D and 2D. The low-temperature error plateau is approximately proportional to $\sigma$, consistent with a coefficient matrix $\mathsf{A}$ whose smallest singular value remains finite down to $T=0$. In contrast, at high temperature, $\mathsf A=O(\beta)$, and we observe the familiar deterioration of the FD protocol. 
\par

To understand the poorer low-temperature performance in
the U(1) case, we note that these Hamiltonians conserve
total $Z$ magnetization and have only one unknown Zeeman
field, $b^z$. Writing $M^z=N^{-1}\sum_{\boldsymbol r}
 \langle\sigma^z_{\boldsymbol r}\rangle_\beta$, Eq.~\eqref{eq:FDTdefiningG} gives $\mathsf G_{xx}=\mathsf G_{yy}=-4M^z b^z$. About three quarters of the failed cases have ground states with $M^z=0$ (i.e., zero-magnetization plateaus), for which $b^z$ cannot be learned from these FD equations. This mechanism is reminiscent of constant magnetization plateaus, arising for instance in the triangular lattice Heisenberg antiferromagnet as a function of Zeeman field. The remaining quarter have fully polarized ground states,
$M^z=\pm1$, for which $b^z$ and the longitudinal exchanges
enter these FD equations only through
$\pm b^z+2\sum_{\vec d}\lambda_{\vec d}^{zz}$
and therefore cannot be learned separately. The distinction between easy-plane and easy-axis Hamiltonians is useful for interpreting the examples which are hard to learn. We find that easy-plane models commonly have partially magnetized ground states and are amenable to the FD protocol. Antiferromagnetic easy-axis models typically have zero-magnetization ground states, while ferromagnetic easy-axis models favor fully polarized ground states. In both cases, Hamiltonian learning with the FD protocol exhibits poor performance.
\par 
In Fig.~\ref{fig:fdt}d of the main text, we plot the performance of the FD protocol across varying temperature and an applied Zeeman field $\b\sum_j \sigma^z_j$, in order to further draw connections to the SF protocol. We sampled 1000 generic $N=12$ spin chains, as in Fig.~\ref{fig:fdtsymmetry}, with a known applied field $\b\sum_j \sigma^z_j$. We add independent Gaussian noise of standard deviation $\sigma=0.02$ to each normalized moment. Accurate reconstruction persists over a broad region of low and moderate temperatures and fields. As discussed in Sec.~\ref{sec:temp_and_field_fdt}, accuracy deteriorates
at high temperature, where $\mathsf A=O(\beta)$, and at strong field, due to polarization into a ferromagnetic state. These trends illustrate the complementary operating regimes of FD, SF, and HT tomography, lending quantitative evidence for the schematic in Fig.~\ref{fig:1}. 

\section{Further experimental considerations}
\label{sec:furtherexperimental}

\subsection{Model mismatch and effective Hamiltonians }

\noindent Thus far, we have assumed that the Hamiltonian family $H(\vec\lambda)$ adopted in our learning protocols contains the true Hamiltonian $H$ governing the
measured DSF. In practice, the true Hamiltonian may include interaction terms of arbitrarily long range or large support, with suitably decaying coefficients consistent with locality, and the assumed Hamiltonian family must inevitably truncate to a subset of such terms. The theoretical problem of Hamiltonian learning without knowing the structure of interaction terms is called \textit{structure learning}, and has been studied in Ref.~\cite{bakshi2024structure} (see also \cite{zhao2025learning,hu2025ansatz}). In quantum materials, this is relevant due to the generic presence of further-neighbor exchange or multispin interactions for which $H(\vec\lambda)$ includes only an effective description, while in quantum simulators, this relates to residual couplings and imperfect controls producing deviations from the intended Hamiltonian. It is therefore important to understand how our protocols perform when the assumed Hamiltonian family is incomplete, and whether the learned couplings (if any) identify the retained microscopic couplings or provide an effective description of the data. \par

To investigate this question, we generate DSF data from Hamiltonians of the form
\begin{equation}
\label{eq:Hexpmismatch}
\begin{aligned}
H_{\mathrm{exp}} &= H(\vec\lambda)+\varepsilon V,\\
H(\vec\lambda) &= \sum_{j=1}^N
\left(b^a\sigma_j^a+\lambda^{ab}\sigma_j^a\sigma_{j+1}^b\right),
\end{aligned}
\end{equation}
where $H(\vec\lambda)$ is a translation-invariant, nearest-neighbor spin chain and $V$ is an interaction omitted
by $H(\vec\lambda)$. We consider four representative omitted interactions: 
\begin{equation}
\label{eq:mismatchVs}
\begin{aligned}
V_{\mathrm{ring}} &=
\frac{1}{\sqrt{21}}\sum_j
\Big[
(\vec\sigma_j\cdot\vec\sigma_{j+1})
(\vec\sigma_{j+2}\cdot\vec\sigma_{j+3})\\&\qquad +
(\vec\sigma_j\cdot\vec\sigma_{j+3})
(\vec\sigma_{j+1}\cdot\vec\sigma_{j+2})-
(\vec\sigma_j\cdot\vec\sigma_{j+2})
(\vec\sigma_{j+1}\cdot\vec\sigma_{j+3})
\Big],\\
V_{\mathrm{NNN}} &=
\frac{1}{\sqrt{3}}\sum_j
\vec\sigma_j\cdot\vec\sigma_{j+2},\\
V_{\mathrm{ZZZ}} &=
\sum_j Z_jZ_{j+1}Z_{j+2},\\
V_{\chi} &=
\frac{1}{\sqrt{6}}\sum_j
\epsilon_{abc}\sigma_j^a\sigma_{j+1}^b\sigma_{j+2}^c,
\end{aligned}
\end{equation}
namely ring-exchange-like terms, next-nearest-neighbor interactions, cubic $Z$ exchange, and scalar spin chirality interactions, each of unit size in the Hilbert-Schmidt norm.  We apply the HT, SF, and FD protocols to DSF data \textit{generated} by $H_{\mathrm{exp}}$ but \textit{assuming} the (incorrect) model $H(\vec\lambda)$. \par

After running our protocols, we assess learning by (i) comparing predicted couplings $\vec\lambda_{\mathrm{pred}}$ with true couplings $\vec\lambda_{\mathrm{true}}$ and (ii) comparing the DSF of $H\left(\vec\lambda_{\mathrm{pred}}\right)$ with that of $H_{\mathrm{exp}}$. We quantify (i) by the relative coupling error $\|\vec\lambda_{\mathrm{pred}}-\vec\lambda_{\mathrm{true}}\|/\|\vec\lambda_{\mathrm{true}}\|$ and (ii) by the normalized DSF residual
\begin{equation}
\label{eq:DSFresidual}
\mathcal E_S(\varepsilon)=
\frac{
\sum_{\boldsymbol q,a,b}\int d\omega\,
\left|
S^{ab}_{H(\boldsymbol\lambda_{\mathrm{pred}})}
(\boldsymbol q,\omega)
-
S^{ab}_{H_{\mathrm{exp}}}
(\boldsymbol q,\omega)
\right|
}{
\sum_{\boldsymbol q,a,b}\int d\omega\,
\left|
S^{ab}_{H_{\mathrm{exp}}}
(\boldsymbol q,\omega)
\right|
}.
\end{equation}
We evaluate the coupling error separately for the field and exchange parameters. We also compare the zeroth and first spectral moments using the normalized `moment residual'
\begin{equation}
\label{eq:momentresidual}
\mathcal E_m(\varepsilon)=
\frac12\sum_{n=0}^{1}
\frac{
\|m_n^{\mathrm{pred}}-m_n^{\mathrm{true}}\|_1
}{
\|m_n^{\mathrm{true}}\|_1
},
\end{equation}
where $m_n^{\mathrm{pred}}$ and $m_n^{\mathrm{true}}$ denote
predicted and true spectral moments, and $\|\cdot\|_1$ is an $L^1$ norm
over momentum and spin indices.
For SF, we project both spin indices transverse to each applied
field and replace $m_1$ by $m_1-2Bm_0$.
Both residuals include sums over the temperatures and field
directions used by each protocol before normalization.
We also evaluate both residuals for the bare Hamiltonian
$H(\vec\lambda)$, obtained by omitting $\varepsilon V$ without
adjusting the retained couplings.

\begin{figure}
    \centering
    \includegraphics[width=\linewidth]{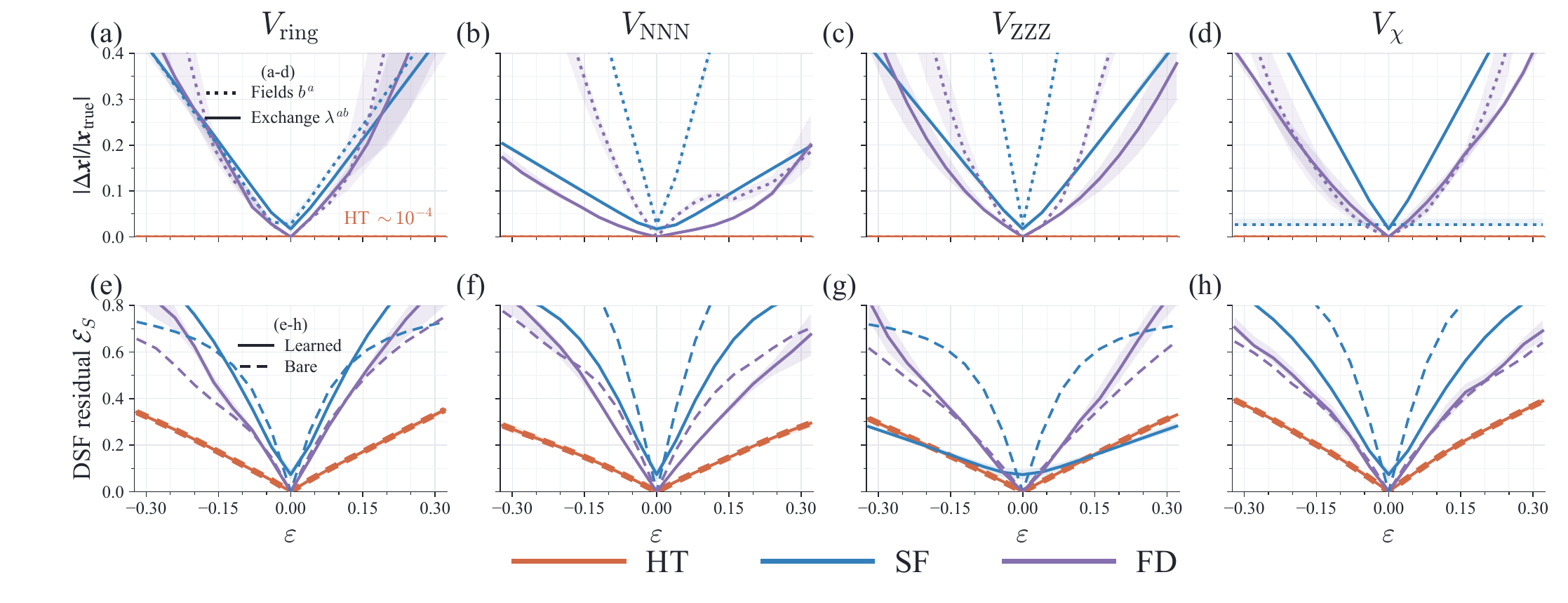}
    \caption{\textbf{Hamiltonian tomography with model mismatch.} We generate noiseless DSF data for eight-site periodic spin chains [Eq.~\eqref{eq:Hexpmismatch}; $N=8$] and learn the couplings assuming the model $H(\vec\lambda)$, using the HT, SF, and FD protocols. Columns show the four interactions defined in Eq.~\eqref{eq:mismatchVs}; the top row (a--d) plots the relative field (dotted) and exchange (solid) coupling errors, each normalized by the corresponding true parameter norm, and the bottom row (e--h) plots the DSF residual Eq.~\eqref{eq:DSFresidual}. In (e--h), solid curves show the DSF residual for the learned Hamiltonian $H(\vec\lambda_{\mathrm{pred}})$, while dashed curves show the residual obtained by simply omitting $\varepsilon V$ and keeping the original couplings, $H(\vec\lambda_{\mathrm{true}})$. Both residuals are evaluated against the DSF of $H_{\mathrm{exp}}$, at the temperatures and applied fields used by each protocol.}
    \label{fig:modelmismatch}
\end{figure}

\begin{figure}
    \centering
    \includegraphics[width=\linewidth]{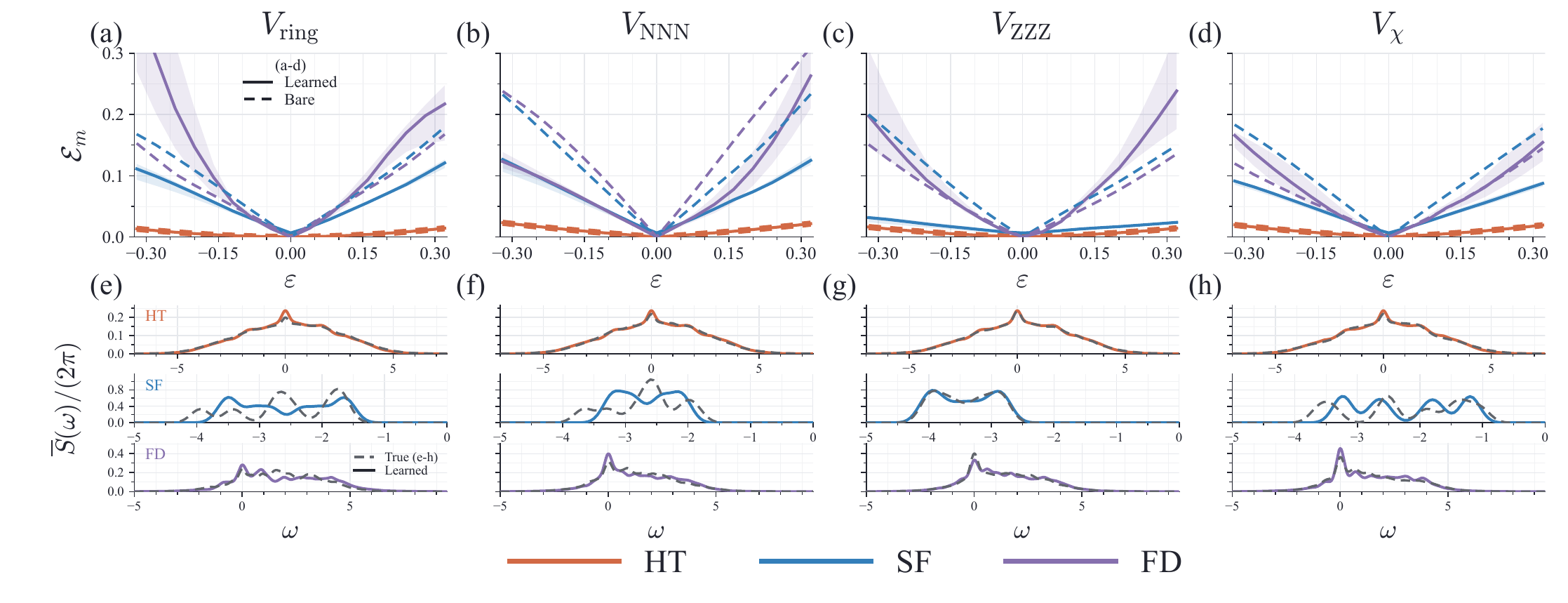}
    \caption{\textbf{Spectral moments and DSF spectra with model mismatch.} We use the same Hamiltonians and protocols as in Fig.~\ref{fig:modelmismatch}. (a--d) Moment residual $\mathcal{E}_m$ for $H(\vec\lambda_{\mathrm{pred}})$ (solid) and $H(\vec\lambda_{\mathrm{true}})$ (dashed), each compared with $H_{\mathrm{exp}}$. We plot medians across eight trial Hamiltonians and show ribbons spanning the 37.5th--62.5th percentiles. (e--h) True (gray dashed) and learned (colored solid) DSF for one example ($\varepsilon=0.2$). Each panel shows HT at $\beta = 0.1$, $\b=0$ (upper), SF at $\b=64$, $T=0$ (middle) and FD at $T=1,B=0$ (lower). Here $\overline S$ averages over momenta and three diagonal spin components of $S^{ab}(\vec q,\omega)$. We shift frequency by $2B$ for SF and use Gaussian broadening $\eta = 0.15$.}
    \label{fig:modelmismatch2}
\end{figure}

\par 
The results are shown in Figs.~\ref{fig:modelmismatch} and
\ref{fig:modelmismatch2}.
Remarkably, the HT protocol successfully learns the true couplings
(with relative errors of order $10^{-4}$) for all four interactions.
This accuracy follows from the leading-order HT protocol, which isolate
the coefficients of interactions in $H(\vec\lambda)$ without ``contamination" from any orthogonal (in the Hilbert-Schmidt inner product) interaction term, and is not expected to extend to the SF and FD protocols.
For the HT protocol, the DSF residual grows approximately linearly with
$|\varepsilon|$ as expected by this argument.
\par

For the SF protocol, we may point out the following patterns. The NNN and $ZZZ$
interactions mainly shift the learned fields $b^a$. For the $ZZZ$ interaction, we note that with spins polarized along $+\hat{\boldsymbol z}$, a single spin flip changes three $ZZZ$
terms, shifting its excitation energy by $-6\varepsilon$, which can be reproduced by a field correction
$\Delta b^z=3\varepsilon$, explaining how a large field error can
accompany a substantially improved DSF.

\par 
On the other hand, the scalar chirality interaction
mainly shifts exchange couplings $\lambda^{ab}$; we find opposite changes in
$\lambda^{ab}$ and $\lambda^{ba}$. Thus the scalar chirality interaction is mimicked by an effective antisymmetric ``renormalization" of the exchange parameters by the SF protocol, and the success of this renormalization is reflected in the comparative improvement of $\mathcal{E}_S$ and $\mathcal{E}_m$ for the learned Hamiltonians compared to the bare Hamiltonians [Fig.~\ref{fig:modelmismatch}h, Fig.~\ref{fig:modelmismatch2}d].

\par 

Finally, we note that the FD protocol offers less interpretable responses to model mismatch, though its performance is generally comparable with the SF protocol. 

\par 

These results distinguish two aims of Hamiltonian learning: identifying the couplings in a model and determining whether that model adequately describes the experimental data. The HT protocol can accomplish the former even when the latter fails, while SF and FD protocols may instead return effective couplings that partially capture the effects of omitted interactions. Since agreement with the spectral moments used for learning, even exactly, does not guarantee agreement with the full DSF, our learning guarantees remain conditional on the assumed Hamiltonian family. A more stringent test is to confront the learned Hamiltonian with additional spectral moments or DSF measurements, potentially at different temperatures and fields. If no choice of couplings can reproduce these independent data within experimental and approximation errors, the assumed Hamiltonian family can be ruled out--- for instance, in the noise-free setting, if the polynomial system $\F (\vec\lambda)= \FF$ has only spurious complex solutions. Conversely, systematically enriching the feature map and enlarging the Hamiltonian family can provide means of validating and refining the learned Hamiltonian.

\subsection{Spatial disorder }
\label{sec:spatialdisorder}

\noindent 
So far, we have always assumed translation-invariant Hamiltonians which are characterized by an $O(1)$ number of couplings $\vec \lambda$. In this section, we extend our high-temperature Hamiltonian learning framework to include quenched spatial disorder, i.e. non-translation-invariant couplings. In particular, we assume that the local coupling vectors $\vec\lambda_{\vec r}$ are independent and identically distributed (i.i.d.) across positions $\vec r$, with common distribution $\pi$, and we will seek to characterize $\pi$. We contrast this with the more difficult problem of characterizing each of the random couplings (e.g. as in Ref.~\cite{gu2024practical}), which we do not treat here. \par 

One motivating question is whether DSF data determines statistical properties of the disorder, such as the variance of a coupling distribution. We answer this in the affirmative for broad classes of Hamiltonians when high-temperature data is accessible. The disorder-averaged spectral moments depend on the disorder distribution only through its moments, and we show that these spectral moments produce a map that recovers the disorder distribution moments order by order. (Note the presence of two types of moments in this section, DSF spectral moments and moments of $\pi(\vec \lambda)$.) If we further assume the domain of $\pi$ is bounded, the distribution is fully determined by its moments, and hence $\pi$ is fully identifiable whenever a particular rank
condition holds. \par 

To set the stage, consider a \textit{statistically} translation-invariant disordered Hamiltonian on $\Lambda$,
\begin{equation}\label{eq:Hdisorder}
H_J = \sum_{\vec r\in\Lambda}\sum_{j=1}^p \lambda^j_{\vec r}\,h^j_{\vec r},
\end{equation}
where $J=\{\vec \lambda_{\vec r}\}$ labels a quenched disorder realization and the local coupling vectors $\vec\lambda_{\vec r}=(\lambda^1_{\vec r},\ldots,\lambda^p_{\vec r})\in[-1,1]^p$ are i.i.d. random variables drawn from $\pi$:
\begin{equation}
    \vec\lambda_{\vec r} \sim \pi,\qquad \pi \in \mathcal{P}([-1,1]^p),
\end{equation}
for each $\vec r\in \Lambda$. Each $h^j_{\vec r}$ is a fixed local term equal to $h^j_{\vec 0}$ up to translation, bounded by $\|h^j_{\vec r}\|_\infty\le h$ and $\mathrm{diam}(h^j_{\vec r})\le R$. We allow for coupling-space correlations, i.e. for $i\neq j$,
$\mathbb E_\pi[\lambda^i\lambda^j]$ need not equal $\mathbb E_\pi[\lambda^i]\mathbb E_\pi[\lambda^j]$. This allows for physically realistic scenarios; for example, a local change in the crystal field at $\vec r$ can affect multiple local couplings $\lambda^i_{\vec r}$. \par 

The measured DSF, accessed for instance by neutron scattering in a disordered crystal, involves a spatial average over $\pi$ in a manner which we make precise. We will quantify if and how this is self-averaging--- that is, how close a single quenched realization is to the true ensemble disorder average. The overall question we seek to answer is: which functionals of the disorder distribution $\pi$ are identifiable from a quenched-disorder DSF? For instance, can we recover the variance of the couplings, or even the full $\pi$?\par 

Remarkably, we will find the answer to be that any moment of $\pi$ is learnable in principle, i.e. with no measurement noise and otherwise ideal conditions, from the DSF. For spins $\sigma^a$ and a quenched disorder realization $J\coloneqq\{\vec\lambda_{\vec r}\}$, we define spatial harmonics of the DSF moments as
\begin{equation}\label{eq:mndi}
    m^{ab(J)}_{n,\vec d}
    =
    \frac{2\pi}{N}\sum_{{\vec r}\in\Lambda}\big\langle \mathrm{ad}_{H_J}^n (\sigma_{\vec r}^a)\sigma_{\vec r+\vec d}^b\big\rangle_{\beta,J}
    \coloneqq \sum_{i\ge 0} \m^{ab(J)}_{n,\vec d,i}\,\beta^i,
\end{equation}
where $\langle\cdot\rangle_{\beta,J}$ is the Gibbs expectation value in $H_J$ at inverse temperature $\beta$ and $\mathrm{ad}_H(A)=[A,H]$. We note that $\frac1N\sum_{\vec r}$ performs a spatial average. In what follows, we sometimes drop the $ab$.

\par

We illustrate the protocol with a simple worked example: finding the mean and variance of a disordered Ising model. Consider the periodic random-bond Ising model
 \begin{equation}
     H=\sum_jJ_jZ_jZ_{j+1}, \qquad J_j\sim\pi 
 \end{equation}
with i.i.d. couplings supported on $[-1,1]$. A simple calculation gives us
\begin{equation}
    \begin{aligned} \frac{\m_{0,1,1}^{zz(J)}}{2\pi} &= -\frac1N\sum_j \langle Z_jZ_{j+1}H\rangle_{0,J} = -\frac1N\sum_jJ_j,\end{aligned}
\end{equation}
and thus $-\m_{0,1,1}^{zz(J)}/2\pi$ is the spatial average of $J$. For the variance of $J$, it is natural to turn to the second spectral moment; using $(\mathrm{ad}_H^2X_j)X_j = 4(J_{j-1}^2+J_j^2)I + 8J_{j-1}J_jZ_{j-1}Z_{j+1}$, we find
\begin{equation}
    \begin{aligned}       \frac{\m^{xx(J)}_{2,0,0}}{2\pi} &=\frac4N\sum_j (J_{j-1}^2+J_j^2) = \frac{8}{N} \sum_j J_j^2.
    \end{aligned}
\end{equation}
We denote the disorder average of $\m^{ab(J)}_{n,\vec d,i}$ by $\overline{\m^{ab}_{n,\vec d,i}}$. Taking disorder averages, we have 
\begin{equation}
\label{eq:epiJmxx}
    \mathbb{E}_\pi[J]
    =-\frac{\overline{\m^{zz}_{0,1,1}}}{2\pi},
    \qquad
    \mathbb{E}_\pi[J^2]
    =\frac{\overline{\m^{xx}_{2,0,0}}}{16\pi},
\end{equation}
from which an expression for the variance $\mathbb{E}_\pi[J^2]-\mathbb{E}_\pi[J]^2$ follows. We note that the disorder average above is not \textit{a priori} equivalent to the spatial average $\langle \cdot \rangle = \frac1N\sum_j(\cdot)$. However, we will show that the spatial average concentrates around the disorder average in the thermodynamic limit: 
\begin{proposition}[Self-averaging]\label{self-average} Let $N = |\Lambda|$, $\m^{(J)}_{n,\d,i}$ be the spectral moment harmonics as above for quenched disorder $J = \{\vec\lambda_{\vec r}\}$, $R_h = \max_a \mathrm{diam}(h_{\vec r}^a)$, $R_O = \mathrm{diam}(O_{\vec 0})$ and $\rho_i = (n+i) R_h  + R_O+ |\vec d|$. Let $|B(\rho)|$ be the number of sites in a ball of radius $\rho$ and $G_i$ be independent of $N$ and satisfy
$|g_{\vec r}^{(J)}|\leq G_i$ for all $\vec r,J$, where $g_{\vec r}^{(J)}$ is defined below. Then we have
\begin{equation}
    \EE_\pi\big[ |\m_{n,\d,i}^{(J)} - \overline{\m_{n,\d,i}}|^2 \big] \leq \frac{|B(2\rho_i)|G_i^2}{N} \to 0
\end{equation}
as $N\to \infty$. Hence $\m_{n,\d,i}^{(J)}\to \overline{\m_{n,\d,i}}$ in the thermodynamic limit. 
\end{proposition}
\begin{proof}
    Let us write $\m_{n,\d,i}^{(J)} = \frac1N \sum_{\vec r\in \Lambda} g_{\vec r}^{(J)}$ where $g_\r^{(J)}\coloneqq \frac{2\pi}{i!}\pdv[i]{\beta}\Big[\langle\mathrm{ad}_{H_J}^n O_\r\,O'_{\r+\d}\rangle_{\beta,J}\Big]_{\beta=0}.$ Let $\mathrm{supp}(A)$ denote the support of the operator $A$ and let $B_{\vec r}(\rho)$ be the ball of radius $\rho$ centered at $\vec r$. Note that  $\mathrm{supp}(\mathrm{ad}_{H_J}(A))$ is contained in a ball slightly larger than $\mathrm{supp}(A)$ by an amount $R_h$. Hence $\mathrm{supp}(\mathrm{ad}_H^nO_{\vec r}O_{\vec r+\vec d})\subseteq B_{\vec r}(\rho_0)$ where $\rho_0 = n R_h+\mathrm{diam}(O)+|\vec d|$ and $\mathrm{supp}(g_\r^{(J)}) \subseteq B_{\vec r}(\rho_i)$
where $\rho_i = \rho_0 + i R_h$. We compute
\begin{multline}
    \EE_\pi[|\m_{n,\d,i}^{(J)} - \overline{\m_{n,\d,i}}|^2] = \frac{1}{N^2}\sum_{\vec r,\vec r'\in \Lambda} \EE_\pi [(g_{\vec r}^{(J)} - \EE_\pi[g_{\vec r}^{(J)}])(g_{\vec r'}^{(J)} - \EE_\pi[g_{\vec r'}^{(J)}])] \\
    \coloneqq \frac{1}{N^2}\sum_{\vec r,\vec r'\in \Lambda}\mathrm{Cov}(g_{\vec r}^{(J)},g_{\vec r'}^{(J)}).
\end{multline}
If $|\vec r-\vec r'| \geq 2\rho_i$ then $g_{\vec r}^{(J)}, g_{\vec r'}^{(J)}$ are functions of disjoint, independent couplings, so $\mathrm{Cov}(g_{\vec r}^{(J)},g_{\vec r'}^{(J)})=0$. Otherwise we can use $\mathrm{Cov}(g_{\vec r}^{(J)},g_{\vec r'}^{(J)})\leq \mathrm{Var}_\pi[g_{\vec r}^{(J)}]^{1/2} \mathrm{Var}_\pi[g_{\vec r'}^{(J)}]^{1/2}\leq G_i^2$. Hence
\begin{equation}
     \EE_\pi[|\m_{n,\d,i}^{(J)} - \overline{\m_{n,\d,i}}|^2]\leq \frac{1}{N^2}\sum_{\vec r}\sum_{\vec r' : |\vec r-\vec r'|\leq 2\rho_i} G_i^2 \leq \frac{N|B(2\rho_i)|G_i^2}{N^2} = \frac{B(2\rho_i)G_i^2}{N}.
\end{equation}
\end{proof}
\noindent In short, $\m_{n,\vec d,i}^{ab(J)}$ is the average of $N$ random variables, so it should obey central limit scaling, though some care is needed because these variables $g_{\vec r}^{(J)}$ are not all independent. However, using locality, we can show that they are sufficiently independen hence, in the thermodynamic limit, a single quenched measurement $\m^{ab}_{n,\vec d,i}$ concentrates to $\overline{\m^{ab}_{n,\vec d,i}}$ with central limit scaling. \par 

\begin{figure}[t]
    \centering
\includegraphics[width=\linewidth]{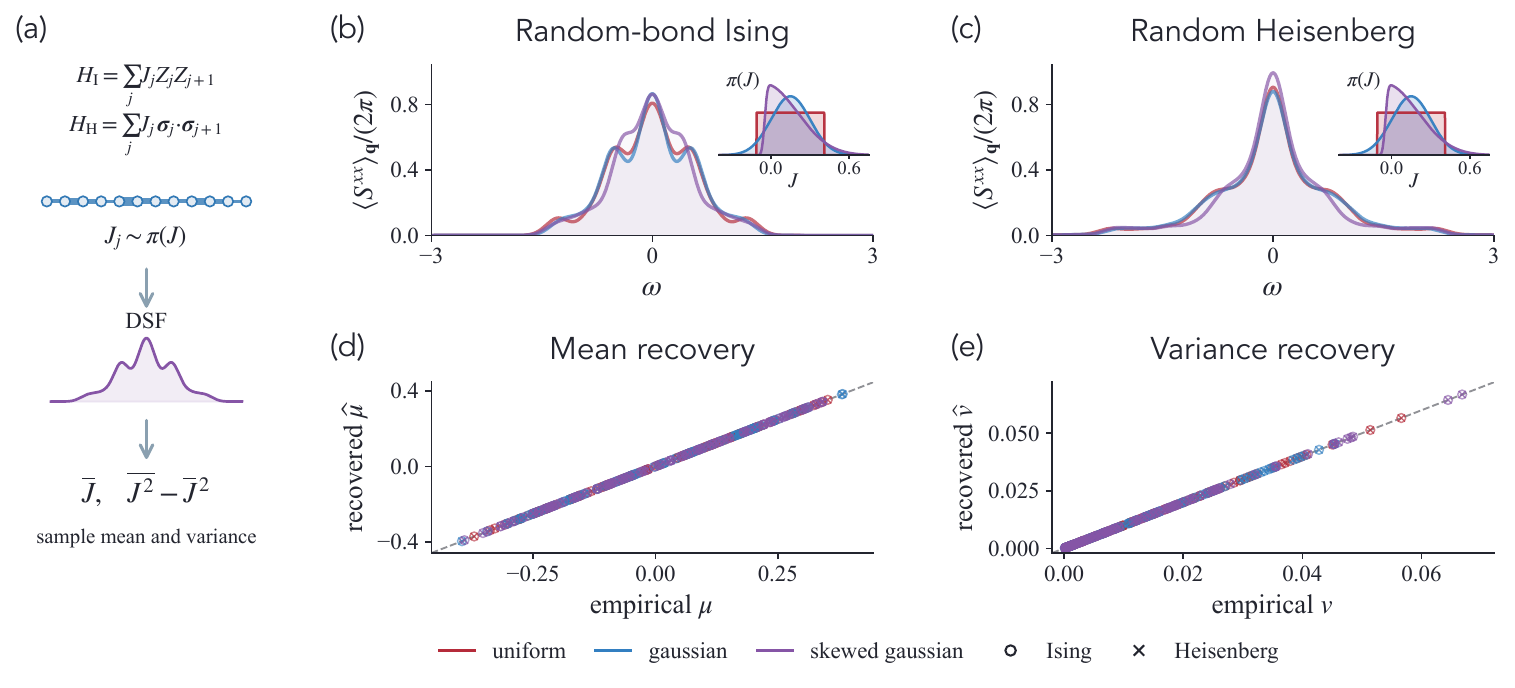}
\caption{\textbf{Recovering disorder statistics from the DSF.} We learn moments of the disorder distribution $\pi$ from the dynamical structure factor (DSF). We generate DSF data and spectral moments from the random-bond Ising and Heisenberg models ($N=12$) and recover the mean and variance $\overline{J}, \overline{J^2}-(\overline{J})^2$ with $J_i\sim \pi(J)$. (b,c) Momentum-averaged transverse DSF for random-bond Ising (b) and Heisenberg (c) models with uniform, Gaussian, and
skewed Gaussian disorder, drawn from distributions (insets) with mean and standard deviation
$0.15$. (d,e) Recovered means and variances across 8 realizations at each of 13 different means $\mu$ and variances $\nu$, for each distribution. The dashed line indicates perfect recovery.
}   \label{fig:disordertomography}
\end{figure}

We illustrate the characterization of disorder distributions using DSF spectral moments in Fig.~\ref{fig:disordertomography} for the random Ising and Heisenberg models. In Fig.~\ref{fig:disordertomography}b,c we show representative DSF components $\langle S^{xx}\rangle_{\vec q} = \frac1N\sum_{\vec q} S^{xx}(\vec q,\omega)$ in each model for couplings $J_i$ drawn randomly from three different distributions $\pi(J)$: a uniform distribution, a Gaussian distribution, and a skewed Gaussian distribution ($p(x) \propto \phi(x)\Phi(\alpha x)$ where $\phi(x)$ is a Gaussian and $\Phi(x)$ is the Gaussian cumulative distribution), observing informative variations arising from the different disorder distributions. In Fig.~\ref{fig:disordertomography}d,e, we show that the mean and variance can be successfully recovered from the DSF alone, following the preceding method for the random Ising model and its straightforward generalization to the Heisenberg case. We present the general method in the following section. 

\subsection{Hierarchical protocol for learning with disorder}

\noindent In this section, we extend the ideas of Eq.~\eqref{eq:epiJmxx} to the general case, presenting a high-temperature protocol for characterizing quenched disorder. Once again, rather than recovering the quenched couplings at each site $\vec r$, our goal will be to learn \textit{moments} of the distribution $\pi$ describing the local couplings $\vec\lambda_{\vec r}$. In contrast to the high-temperature and strong-field protocols, which both involved solving a polynomial system for the exact values of the couplings, the present protocol will instead involve solving a \textit{series} of \textit{linear} systems for these moments. To introduce this, we will need the following notation. First, we define a multi-index $\vec\alpha = (\alpha_1,\ldots, \alpha_p) \in \mathbb{Z}_{\geq 0}^p$ and define the order to be $|\vec \alpha | = \sum_a \alpha_a$. Next, we define the $\vec\alpha$-moment to be 
\begin{equation}
    \mu_{\vec \alpha} = \mathbb{E}_\pi\left[ \prod_a (\lambda^a)^{\alpha_a}\right].
\end{equation}
For example, we denote means as $\mu_{\vec e_a} = \mathbb{E}_\pi[\lambda^a]$ and  covariances as $\mu_{\vec e_a+\vec e_b} -\mu_{\vec e_a}\mu_{\vec e_b}$, where we use unit vectors $\vec e_a$ satisfying $[\vec e_a]_b = \delta_{ab}$. Recall, without disorder, the coefficient $\m_{n,\vec d,i}$ is a homogeneous polynomial of degree $n+i$ in the couplings. Disorder-averaging acts on each monomial, and by independence across sites this reduces to a product of moments $\mu_{\vec \alpha}$ with total order $n+i$. Hence disorder averaging conserves this degree. We establish the structure of these disorder averages in the Proposition below. 

 \begin{proposition}[Structure of disorder-averaged spectral moments] Let $\m_{n,\vec d,i}^{(J)}$ denote a high-temperature coefficient of a spectral moment with quenched disorder. We have
\begin{equation}
    \overline{\m_{n,\vec d,i}} = \sum_{|\vec \alpha|=n+i} Q_{\vec \alpha}^{(n,\vec d,i)} \mu_{\vec \alpha} + R_{n,\vec d,i}\left( \{\mu_{\vec \alpha} \mid 1\leq |\vec\alpha|\leq n+i-1\}\right),
\end{equation}
where $Q_{\vec \alpha}^{(n,\vec d,i)}$ are coefficients and $R_{n,\vec d,i}$ is a polynomial in lower-order moments. In particular, disorder moments of highest order $n+i$ enter linearly.
\end{proposition}
\begin{proof}
    We may expand $\m_{n,\d,i}$ into monomials $\prod_{(\r,a)}(\lambda^a_\r)^{e_{a,\r}}$ with
$\sum_{(\r,a)}e_{a,\r}=n+i$. Each averaged monomial becomes a product $\prod_{\vec r}\mu_{\vec\alpha(\vec r)}$ with $\sum_{\vec r}|\alpha(\vec r)| = n+i$. A term carrying an order $n+i$ moment leaves no room for any other factor, so the formula above follows. 
\end{proof}
The preceding proposition implies the following useful structure of the disorder-averaged spectral moments. Let us collect all order-$k$ moments of the disorder
distribution into $\boldsymbol{\mu}^{(k)}$, and all
disorder-averaged spectral moment coefficients with
$n+i=k$ into $\overline{\boldsymbol{\m}}^{(k)}$.
Then
\begin{equation}
    \overline{\boldsymbol{\m}}^{(k)}
    =\mathsf{Q}^{(k)}\boldsymbol{\mu}^{(k)}
    +\boldsymbol{b}^{(k)},
\end{equation}
where $\vec b^{(k)}$ depends only on moments with order less than $k$ and $\mathsf{Q}^{(k)}$ is a coefficient matrix. At $k=1$, this recovers the means, and at $k=2$, it recovers the second moments and hence variances and covariances. For instance, if disorder is present in two couplings $(J,h)$, $k=1$ determines $\mathbb E[J]$ and $\mathbb E[h]$, while $k=2$ determines $\mathbb E[J^2], \mathbb E[Jh]$, and $\mathbb E[h^2]$. Thus the $k=2$ system determines not only both variances but also $\operatorname{Cov}(J,h)$. Higher moments are successively recoverable from higher orders in this hierarchy. Furthermore, it is clear that if $\mathsf Q^{(k)}$ has full column rank for every $k\leq K$, then the equation above uniquely determines all moments of $\pi$ through order $K$. \par 

We have assumed that the disorder is independent across spatial positions. For spatially correlated disorder, averages such as $\mathbb E[J_{\mathbf r}J_{\mathbf r+\mathbf d}]$ no longer factor into products of single-site means, and the hierarchy naturally contains spatial disorder correlation functions, suggesting a route to learning quantities such as disorder correlation lengths from the momentum dependence of the DSF. We leave this extension to future work.

\end{document}